\documentclass[12pt]{article}

\usepackage{verbatim}
\usepackage{amsmath}
\usepackage{amssymb}
\usepackage{graphicx}
\usepackage{cite}
\usepackage{subfig}
\usepackage{setspace}
\usepackage{url}
\usepackage[percent]{overpic}
\usepackage{slashed}
\usepackage{authblk}

\usepackage{xspace}

\usepackage{fullpage}
\usepackage{hyperref}
\def\tev{\,{\rm TeV}}
\def\gev{\,{\rm GeV}}
\def\ie{{\it i.e.}}
\def\eg{{\it e.g.}}

\def\to{\rightarrow}

\title{pMSSM Studies at the 7, 8 and 14 TeV LHC\footnote{White Paper contributed to the Snowmass Community Summer Study 2013,
Minneapolis, MN July 29 - August 6, 2013}}
\date{September 30, 2013}
\author{M. Cahill-Rowley}
\author{J.L. Hewett}
\author{A. Ismail}
\author{T.G. Rizzo}
\affil{SLAC National Accelerator Laboratory, Menlo Park, CA, USA\footnote{mrowley, hewett, aismail, rizzo@slac.stanford.edu}}

\begin{document}

\rightline{\vbox{\halign{&#\hfil\cr
&SLAC-PUB-15554\cr
}}}


{\let\newpage\relax\maketitle}

\begin{abstract}
The 19/20-parameter p(henomenological)MSSM with either a neutralino or gravitino LSP offers a very flexible 
framework for the study of a wide variety of R-parity conserving MSSM SUSY phenomena at the 7, 8 and 14 TeV 
LHC. Here we present the results of a study of SUSY signatures at these facilities obtained via a fast 
Monte Carlo `replication' of the ATLAS SUSY analysis suite. In particular, we show the ranges of the sparticle 
masses that are either disfavored or remain viable after all of the various searches at the 7 and 8 TeV runs are 
combined.  We then extrapolate to 14 TeV with both 300 fb$^{-1}$ and 3 ab$^{-1}$ of integrated luminosity and 
determine the sensitivity of a jets + MET search to the pMSSM parameter space.  We find that the high-luminosity
LHC performs especially well in probing natural SUSY models. 
\end{abstract}

\section{Introduction and Overview of the pMSSM Model Sets}

An important issue for new physics of any kind is whether it can be discovered or excluded in collider searches given backgrounds 
arising from the Standard Model (SM). In particular, within a specific model, it is crucial to know how well a given set of 
experimental analyses can probe the full parameter space of interest. With the lack of any experimental 
evidence for new physics so far, this is certainly true in the case of Supersymmetry (SUSY), which is the most widely studied theory beyond
the SM. However, even in the simplest SUSY scenario, the MSSM, the number of free parameters ($\sim$ 100) is too large 
to study in complete generality. The traditional approach 
is to assume the existence of some high-scale theory with only a few parameters (such as mSUGRA{\cite {SUSYrefs}) from which all 
the properties of the sparticles at the TeV scale can be determined and studied in detail. While such an approach is often quite 
valuable~\cite{Cohen:2013kna}, these scenarios are somewhat phenomenologically limiting and are under increasing tension with a 
wide range of experimental data including, in some cases, the $\sim 126$ GeV mass of the recently discovered Higgs boson{\cite {ATLASH,CMSH}}.

One way to circumvent such limitations is to examine the more general 19/20-parameter pMSSM{\cite{Djouadi:1998di}}. The 
increased dimensionality of the parameter space not only allows for a more unprejudiced study of SUSY, but can also yield valuable
information on `unusual' scenarios, identify weaknesses in the current LHC analyses and can be used to combine results obtained from
many independent SUSY-related searches. To these ends, we have recently embarked on a detailed study of the signatures for the pMSSM at the 7 and 8 TeV LHC, 
supplemented by input from Dark Matter (DM) experiments as well as from precision electroweak and flavor 
measurements{\cite {us1,us2,us4}}. The pMSSM is the most general version of the R-parity conserving MSSM when it is subjected to  
a minimal set of experimentally-motivated guiding principles: ($i$) CP conservation, ($ii$) Minimal Flavor Violation at the 
electroweak scale so that flavor physics 
is controlled by the CKM mixing matrix, ($iii$) degenerate 1\textsuperscript{st} and 2\textsuperscript{nd} generation sfermion masses,
and ($iv$) negligible Yukawa couplings 
and A-terms for the first two generations. In particular, no assumptions are made about physics at high scales, e.g., the nature of SUSY 
breaking, in order to capture electroweak scale phenomenology for which a UV-complete theory may not yet exist. Imposing these principles 
($i$)-($iv$) decreases the number of free parameters in the MSSM at the TeV-scale from 105 to 19 for the case of a neutralino LSP, or to 
20 when the gravitino mass is included as an additional parameter when it plays the role of the LSP.  We have not assumed that the LSP relic density 
necessarily saturates the WMAP/Planck value{\cite{Komatsu:2010fb}} in order to allow for the possibility of multi-component 
DM. For example, the axions introduced to solve the strong CP problem may may make up a substantial amount of DM. The 19/20 pMSSM parameters 
and the ranges of values employed in our scans are listed in Table~\ref{ScanRanges}. Like throwing darts, to study the pMSSM   
we generate many millions of model points in this space (using SOFTSUSY{\cite{Allanach:2001kg}} and checking for consistency with 
SuSpect{\cite{Djouadi:2002ze}}), with each point then corresponding to a specific set of values for these parameters. These individual models 
are then subjected to a large set of collider, flavor, precision measurement, dark matter and theoretical constraints~\cite{us1}.  
Roughly $\sim$225k models with either type of LSP survive this initial selection and can then be used for further physics studies. Decay 
patterns of the SUSY partners and the extended Higgs sector are calculated using privately modified versions of SUSY-HIT~\cite{Djouadi:2006bz}, CalcHEP~\cite{calchep}, and MadGraph~\cite{madgraph}. 
Since our scan ranges include sparticle masses up to 4 TeV, an upper limit chosen to enable phenomenological studies at the 14 TeV LHC, 
the neutralinos and charginos in either of our model sets are typically very close to being in a pure 
electroweak eigenstate as the off-diagonal elements of the corresponding mass matrices are at most $\sim M_W$.

In addition to these two large pMSSM model sets, we have recently generated a smaller, specialized neutralino LSP set of $\sim$ 10.2k 
`natural' models, all of which predict $m_h = 126 \pm 3$ GeV, have an LSP that {\it does} saturate the WMAP relic density and produce 
values of fine-tuning (FT) better than $1\%$ using the Ellis-Barbieri-Giudice measure~\cite{Ellis:1986yg, Barbieri:1987fn}. This low-FT model 
set will also be used as part of the present study. In order to obtain this model set we modified the parameter scan ranges listed in Table~\ref{ScanRanges} to 
greatly increase the likelihood that a chosen point will satisfy the combined relic density, higgs mass, and FT constraints.  Amongst other things, satisfying these requirements necessitates a bino at the bottom of the spectrum as well as light Higgsinos and highly-mixed stops. We generated $\sim 3.3 \times 10^8$ low-FT points 
in this 19-parameter space and subjected them to updated precision, flavor, DM and collider constraints as before. Since our requirements 
were much stricter here than for our two larger model sets only $\sim$ 10.2k low-FT models survive for further study. 

We now subject these three sets of pMSSM models to the SUSY searches performed at the 7 and 8 TeV LHC, as well as planned searches at 14 TeV.

\begin{table}
\centering
\begin{tabular}{|c|c|} \hline\hline
$m_{\tilde L(e)_{1,2,3}}$ & $100 \gev - 4 \tev$ \\ 
$m_{\tilde Q(q)_{1,2}}$ & $400 \gev - 4 \tev$ \\ 
$m_{\tilde Q(q)_{3}}$ &  $200 \gev - 4 \tev$ \\
$|M_1|$ & $50 \gev - 4 \tev$ \\
$|M_2|$ & $100 \gev - 4 \tev$ \\
$|\mu|$ & $100 \gev - 4 \tev$ \\ 
$M_3$ & $400 \gev - 4 \tev$ \\ 
$|A_{t,b,\tau}|$ & $0 \gev - 4 \tev$ \\ 
$M_A$ & $100 \gev - 4 \tev$ \\ 
$\tan \beta$ & $1 - 60$ \\
$m_{3/2}$ & 1 eV$ - 1 \tev$ ($\tilde{G}$ LSP)\\
\hline\hline
\end{tabular}
\caption{Scan ranges for the 19 (20) parameters of the pMSSM with a neutralino (gravitino) LSP. The gravitino mass is scanned with 
a log prior. All other parameters are scanned with flat priors, though we expect this choice to have little qualitative impact on 
our results~\cite{us}.}
\label{ScanRanges}
\end{table}

\section{7 and 8 TeV LHC Searches}

We begin with a short overview of the searches for the pMSSM at the 7 and 8 TeV LHC; the same overall approach will carry over to our 
14 TeV study. In general, we follow the suite of ATLAS SUSY analyses as closely as possible employing fast Monte Carlo, however these are also 
supplemented by several searches performed by CMS. The specific analyses applied to the neutralino model set are briefly summarized in 
Tables~\ref{SearchList7} and ~\ref{SearchList8}. We further augment the MET-based SUSY searches by including a search for heavy neutral 
SUSY Higgs $\to \tau^+\tau^-$ performed by CMS~\cite{Chatrchyan:2012vp} and measurements of the rare decay mode $B_s\to \mu^+\mu^-$ as discovered by 
CMS and LHCb~\cite{BSMUMU}.  Both of these play distinct but important roles in restricting the pMSSM parameter space. Presently, we have 
implemented every relevant ATLAS SUSY search publicly available as of the beginning of March 2013. This list is currently being 
expanded to include more recent ATLAS (and some CMS) analyses with results to be expected in the not too distant future.

\begin{table}
\centering
\begin{tabular}{|l|l|c|c|c|} \hline\hline
Search & Reference & Neutralino & Gravitino & Low-FT   \\
\hline
2-6 jets & ATLAS-CONF-2012-033  & 21.2\% &  17.4\% & 36.5\% \\
multijets & ATLAS-CONF-2012-037 & 1.6\%  & 2.1\% & 10.6\% \\
1-lepton & ATLAS-CONF-2012-041 & 3.2\%  & 5.3\% & 18.7\%  \\

HSCP      &  1205.0272  & 4.0\% & 17.4\% & $<$0.1\%  \\
Disappearing Track  & ATLAS-CONF-2012-111 & 2.6\%  & 1.2\% & $<$0.1\% \\
Muon + Displaced Vertex  & 1210.7451 & - & 0.5\% & - \\
Displaced Dilepton & 1211.2472 & - & 1.1\% & - \\

Gluino $\to$ Stop/Sbottom   & 1207.4686 & 4.9\% &  3.5\% & 21.2\% \\
Very Light Stop  & ATLAS-CONF-2012-059 & $<$0.1\% & $<$0.1\% & 0.1\%  \\
Medium Stop  & ATLAS-CONF-2012-071 & 0.3\% & 5.1\% & 2.1\% \\
Heavy Stop (0l)  & 1208.1447 & 3.7\% & 3.0\% & 17.0\% \\
Heavy Stop (1l)   & 1208.2590 & 2.0\% & 2.2\% & 12.6\% \\
GMSB Direct Stop  & 1204.6736 & $<$0.1\% & $<$0.1\% & 0.7\% \\
Direct Sbottom & ATLAS-CONF-2012-106 & 2.5\% & 2.3\% & 5.1\% \\
3 leptons & ATLAS-CONF-2012-108 & 1.1\% & 6.1\% & 17.6\% \\
1-2 leptons & 1208.4688 & 4.1\% & 8.2\% & 21.0\% \\
Direct slepton/gaugino (2l)  & 1208.2884 & 0.1\% & 1.2\% & 0.8\% \\
Direct gaugino (3l) & 1208.3144 & 0.4\% & 5.4\% & 7.5\% \\
4 leptons & 1210.4457 & 0.7\% & 6.3\% & 14.8\% \\
1 lepton + many jets & ATLAS-CONF-2012-140 & 1.3\% & 2.0\% & 11.7\% \\
1 lepton + $\gamma$ & ATLAS-CONF-2012-144 & $<$0.1\% & 1.6\% & $<$0.1\% \\
$\gamma$ + b & 1211.1167 & $<$0.1\% & 2.3\% & $<$0.1\% \\
$\gamma \gamma $ + MET & 1209.0753 & $<$0.1\% & 5.4\% & $<$0.1\% \\

$B_s \to \mu \mu$ & 1211.2674 & 0.8\% & 3.1\% & * \\
$A/H \to \tau \tau$ & CMS-PAS-HIG-12-050 & 1.6\% & $<$0.1\% & * \\

\hline\hline
\end{tabular}
\caption{7 TeV LHC searches included in the present analysis and the corresponding fraction of the neutralino, gravitino and low-FT pMSSM 
model sets excluded by each search. Note that in the case of the last two rows the experimental constraints have already been included 
in the model generation process for the low-FT model set and therefore are not shown here.}
\label{SearchList7}
\end{table}

\begin{table}
\centering
\begin{tabular}{|l|l|c|c|c|} \hline\hline
Search & Reference & Neutralino & Gravitino & Low-FT    \\
\hline

2-6 jets   & ATLAS-CONF-2012-109 & 26.7\% & 21.6\% & 44.9\% \\
multijets   & ATLAS-CONF-2012-103 & 3.3\% & 3.8\% & 20.9\% \\
1-lepton     & ATLAS-CONF-2012-104 & 3.3\% & 6.0\% & 20.9\% \\
SS dileptons & ATLAS-CONF-2012-105 & 4.9\% & 12.4\% & 35.5\% \\

Medium Stop (2l) & ATLAS-CONF-2012-167 & 0.6\% & 8.1\% & 4.9\% \\
Medium/Heavy Stop (1l) & ATLAS-CONF-2012-166 & 3.8\% & 4.5\% & 21.0\% \\
Direct Sbottom (2b) & ATLAS-CONF-2012-165 & 6.2\% & 5.1\% & 12.1\% \\
3rd Generation Squarks (3b) & ATLAS-CONF-2012-145 & 10.8\% & 9.9\% & 40.8\% \\
3rd Generation Squarks (3l) & ATLAS-CONF-2012-151 & 1.9\% & 9.2\% & 26.5\% \\
3 leptons & ATLAS-CONF-2012-154 & 1.4\% & 8.8\% &32.3\% \\
4 leptons & ATLAS-CONF-2012-153 & 3.0\% & 13.2\% & 46.9\% \\
Z + jets + MET & ATLAS-CONF-2012-152 & 0.3\% & 1.4\% &6.8\% \\

\hline\hline
\end{tabular}
\caption{Same as in the previous table but now for the 8 TeV ATLAS MET-based SUSY searches. Note that when all the searches from this table and the previous table are combined for the neutralino 
(gravitino, low-FT) model set we find that $\sim 37~(52,~70)\%$ of these models are excluded by the LHC.}
\label{SearchList8}
\end{table}

Briefly stated our procedure is as follows: We generate SUSY events for each model for all relevant (up to 85) production channels in PYTHIA 
6.4.26~\cite{Sjostrand:2006za}, and then pass the events through fast detector simulation using PGS 4~\cite{PGS}. Both programs have been modified to, 
e.g., correctly deal with gravitinos, multi-body decays, hadronization of stable colored sparticles, and ATLAS b-tagging. We then scale our 
event rates to NLO by calculating the relevant K-factors using Prospino 2.1~\cite{Beenakker:1996ch}. The individual searches are then 
implemented using our customized analysis code{\cite {us}}, which follows the published cuts and selection criteria as closely as possible. 
This analysis code is validated for each of the many search regions in every analysis employing the benchmark model points provided by ATLAS 
(and CMS). Models are then excluded using the 95\% $CL_s$ limits as obtained by ATLAS (and CMS). For the two large model sets these analyses are 
performed {\it without} requiring the Higgs mass constraint, $m_h = 126 \pm 3$ GeV (combined experimental and theoretical errors) so that we 
can understand its influence on the search results. Note that roughly $20(10)\%$ of models in the neutralino (gravitino) model set predict a Higgs mass in the above range. While there is some variation amongst the individual searches themselves, we find that, 
once combined, the total fraction of our models surviving the set of all LHC searches is to an excellent approximation 
{\it independent} of whether or not the Higgs mass constraint has been applied. Conversely, the $\sim 20(10)\%$ fraction of 
neutralino (gravitino) models predicting the correct Higgs mass is also found to be approximately independent of whether  
the SUSY searches have been applied. These results can best be seen explicitly in Fig.~\ref{figm1}, which shows the 
predicted Higgs mass distribution in the neutralino model set both before and after the LHC SUSY searches have been applied. This result is 
very powerful and demonstrates the approximate decoupling of SUSY search results from the discovery of the Higgs boson which allows 
us to continue examining the entire model set for either LSP type with some reasonable validity.  To a similar approximation, we find that
the results for the distribution of Higgs branching fractions in the pMSSM are also insensitive to the LHC SUSY searches as 
demonstrated in detail in our companion Higgs White Paper{\cite {Higgs}}.

\begin{figure}[htbp]
\centerline{\includegraphics[width=5.0in]{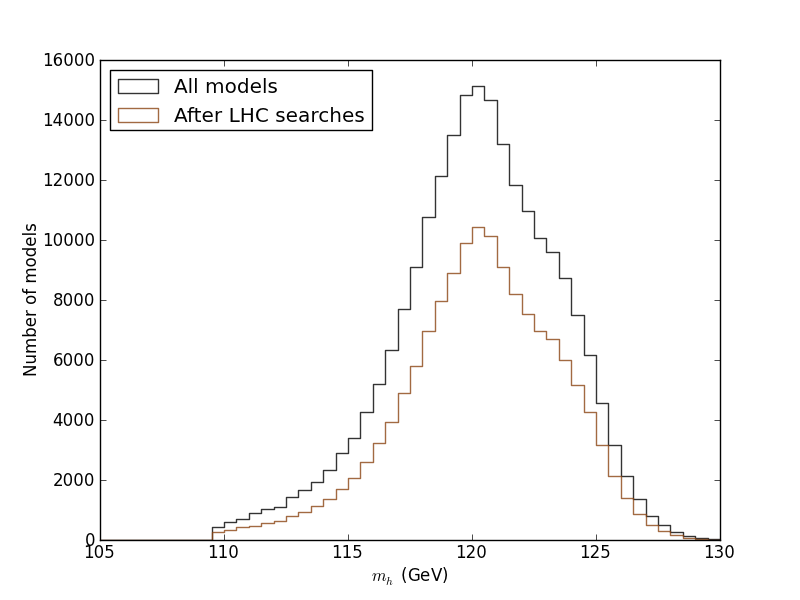}}
\vspace*{-0.10cm}
\caption{The distribution of predicted Higgs mass before and after the LHC search constraints are applied to the neutralino LSP set as indicated. 
Note that the LHC search efficiency is essentially independent of the Higgs mass.}
\label{figm1}
\end{figure}

\subsection{Neutralino Model Set}
\label{sec:neutlsp}

We first discuss the results of our analysis for the case of the neutralino LSP model set. The first important question to address is how well 
the combination of LHC searches cover the pMSSM parameter space.  One way to address this is to project  
2-dimensional slices in the multi-dimensional space of sparticle masses and show the exclusion efficiency of the combined LHC searches within 
specified mass ranges.  The results of this procedure for some of the sparticles in the general neutralino pMSSM model set are
shown in Figs.~\ref{fig00}, ~\ref{fig1}, ~\ref{fig2} and ~\ref{fig3}. In addition, Tables~\ref{SearchList7} 
and~\ref{SearchList8} provide further information by listing the fraction of the neutralino pMSSM set (as well as for the corresponding 
gravitino and low-FT model sets) that is excluded for each of the individual LHC searches.  Combining all of the searches we find that 
$\sim 37\%$ of these neutralino LSP models are currently excluded.   Clearly this implies that a large fraction of the excluded models are eliminated
by more than one search.

\begin{figure}[htbp]
\centerline{\includegraphics[width=3.5in]{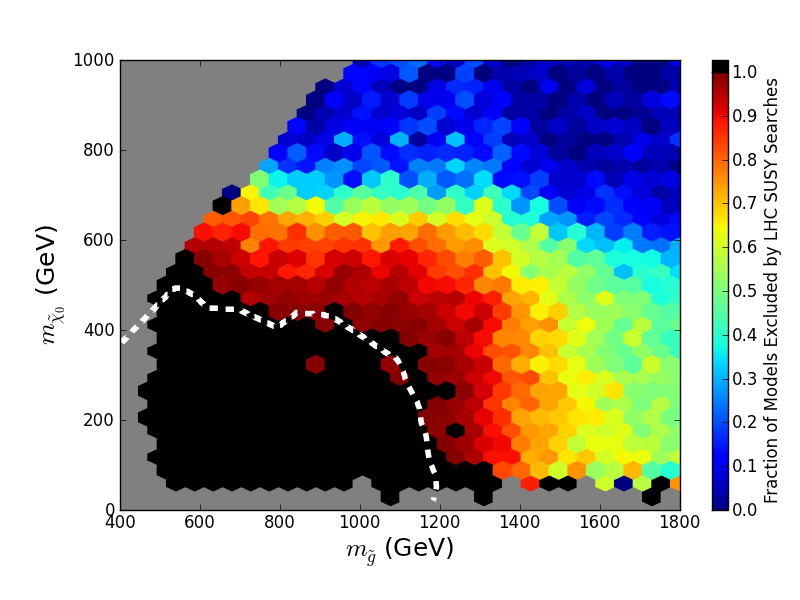}
\hspace{-0.50cm}
\includegraphics[width=3.5in]{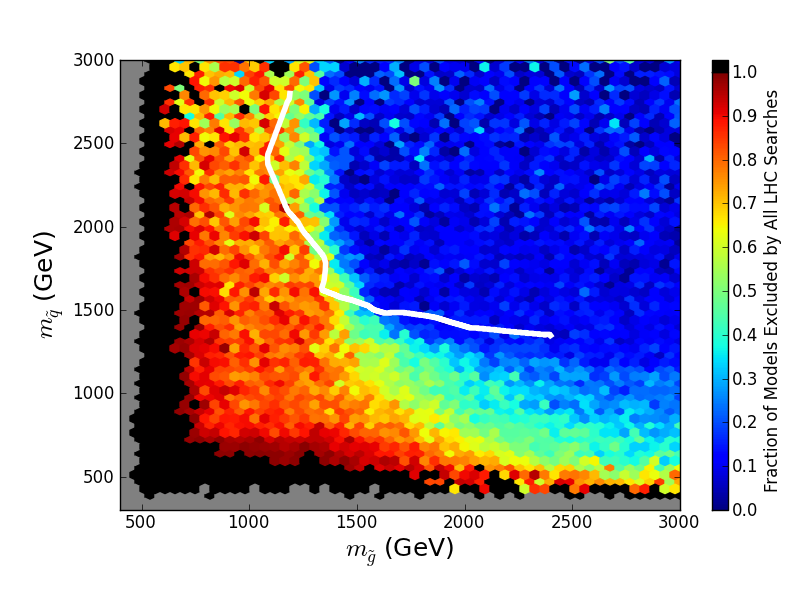}}
\vspace*{-0.10cm}
\caption{Projections of the pMSSM model coverage efficiencies for the neutralino LSP set from the combined 7 and 8 TeV LHC searches shown in the gluino-LSP (left) 
and the lightest squark-gluino (right) mass planes. The color 
code provides the total search efficiency in a specific mass bin. In this and all subsequent figures, 7 (8) TeV simplified model analysis results from ATLAS are shown as solid (dashed) white curves in the various LSP-sparticle mass planes, while the solid white line in the squark-gluino mass plane is from the 8 TeV 5.8 fb$^{-1}$ 2-6 Jets + MET search, assuming degenerate squarks and a massless LSP.}
\label{fig00}
\end{figure}

Figure~\ref{fig00} shows the combined LHC search efficiencies projected onto both the gluino-LSP and the lightest (1\textsuperscript{st}/2\textsuperscript{nd} generation) 
squark-gluino mass planes together with the corresponding $95\%$ CL limits from the ATLAS simplified model analysis. Here we see that the 
region excluded by the ATLAS simplified model analysis (below and to the left of the white line) in the gluino-LSP mass plane roughly encircles 
the all-black region which is excluded by our combination of analyses. This is interesting as while the ATLAS simplified model result 
is based solely on a jets + MET analysis under the assumption of decoupled squarks in the left panel, 
ours is a combination of {\it many} analyses, making 
no additional assumptions about the sparticle spectra under consideration. As can be seen here, most of the surviving light gluino models 
have relatively compressed mass spectra although a few of them evade detection by having rather complex decay patterns. The lightest squark-gluino 
panel shows that many models survive that are far below the ATLAS simplified model exclusion line (where degenerate squarks and a massless LSP have been 
assumed) as might be expected from the rather more complex spectra in the pMSSM. It is important to note the rather large set of models, 
particularly when the gluino is quite heavy, where rather light squarks are allowed in comparison to the simplified treatment. 

\begin{figure}[htbp]
\centerline{\includegraphics[width=3.5in]{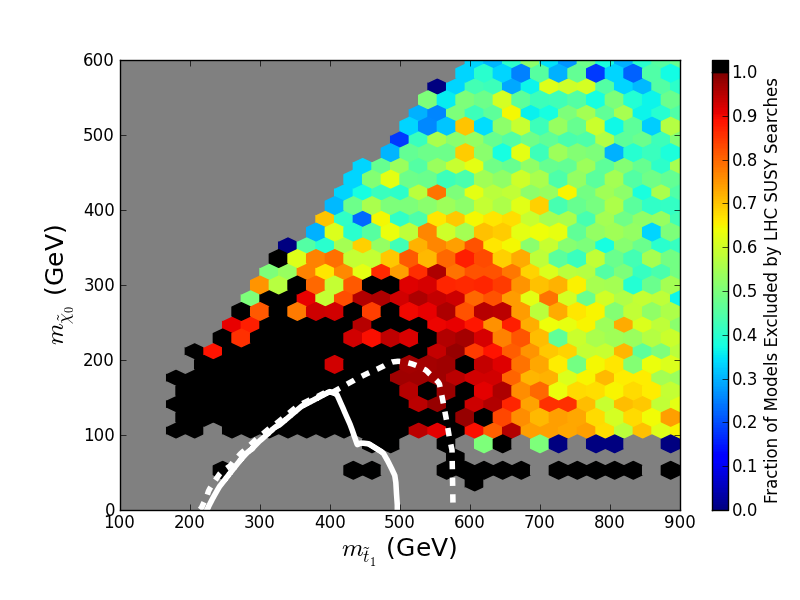}
\hspace{-0.50cm}
\includegraphics[width=3.5in]{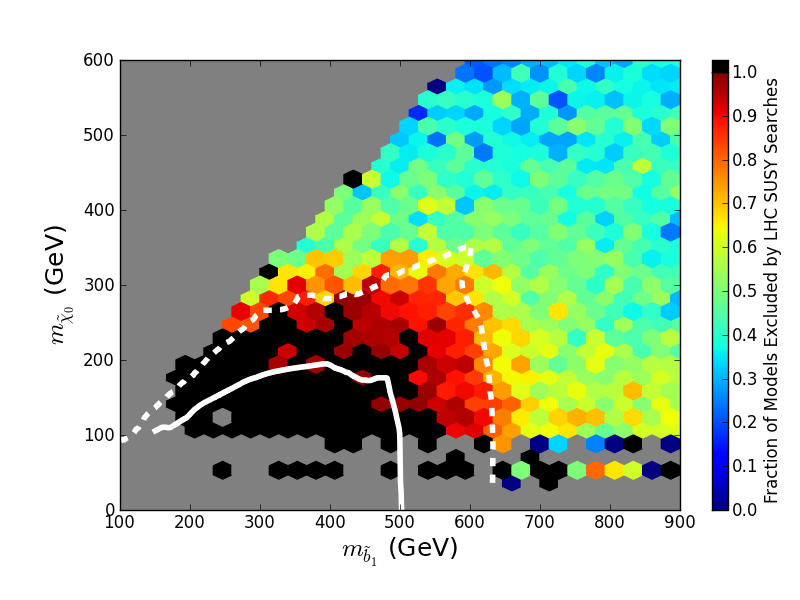}}
\vspace*{0.50cm}
\centerline{\includegraphics[width=3.5in]{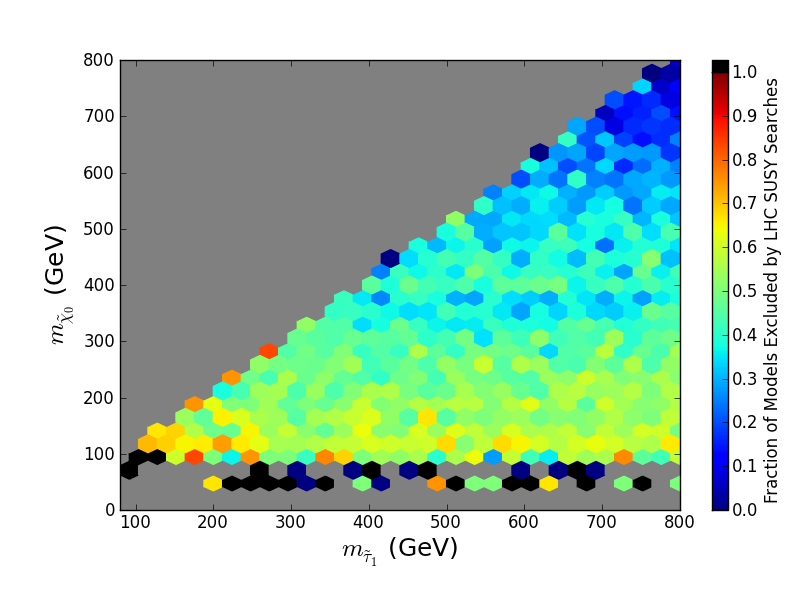}}
\vspace*{-0.10cm}
\caption{Projections of the pMSSM model coverage efficiencies for the neutralino LSP set from the 7 and 8 TeV LHC searches shown in the lightest stop-LSP mass plane 
(top left), the lightest sbottom-LSP mass plane (top right) and the lightest stau-LSP mass plane (bottom). The solid and dashed
white lines represent the corresponding 
$95\%$ CL limit results obtained by ATLAS at 7 and 8 TeV, respectively, in the simplified model limit as discussed in the text.}
\label{fig1}
\end{figure}

Searches for 3\textsuperscript{rd} generation sparticles are of particular importance since these sparticles couple more strongly to the Higgs and are most responsible 
for solving the `naturalness' and fine-tuning problems associated with the Higgs mass quadratic divergence. At least one of the stops is expected 
to be reasonably light and if it is mostly left-handed it will likely bring along with it a light sbottom as well. Figure~\ref{fig1} shows the 
impact of the LHC searches in the lightest stop-, lightest sbottom- and lightest stau-LSP mass planes. Note that whereas the simplified model 
treatment by ATLAS arising from searches at 7 (solid) and 8 (dashed) TeV qualitatively describes the coverage in the sbottom-LSP mass 
plane (though this is again somewhat accidental), it is entirely inadequate for stop searches. 
One reason for this is that many of the searches for sbottoms 
(in particular that for two b-jets, zero leptons + MET inclusive) can also strongly constrain models with light stops. As we will see below, even  
non-3\textsuperscript{rd}-generation searches play an important role in obtaining the exclusion regions shown here. 

We note here that we have not incorporated ATLAS searches involving taus in the final state as PGS has a strong tendency to lead to a large mis-tag 
rate while having a simultaneously low tau tagging efficiency. Thus the exclusions we see in the lower panel of this figure (which is relatively 
uniform in density as might be expected from these arguments) is actually the result of the non-tau searches. 

\begin{figure}[htbp]
\centerline{\includegraphics[width=3.5in]{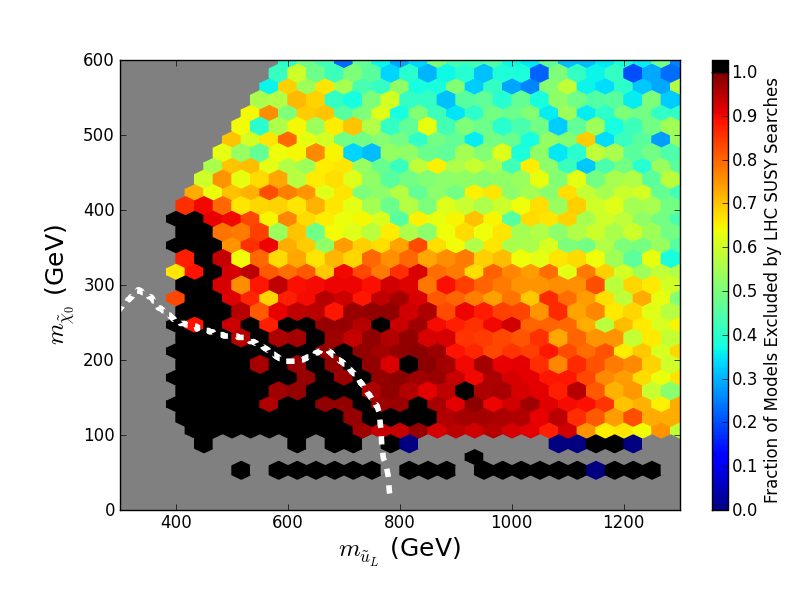}
\hspace{-0.50cm}
\includegraphics[width=3.5in]{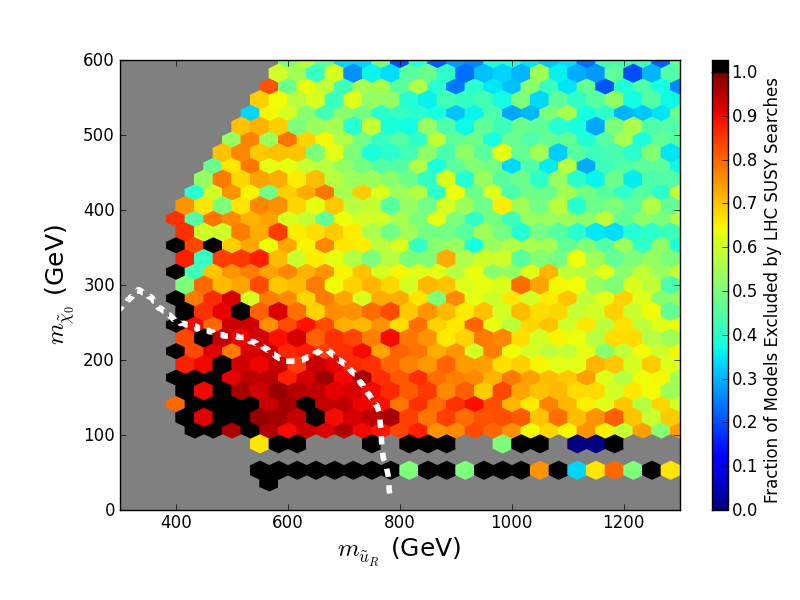}}
\vspace*{0.50cm}
\centerline{\includegraphics[width=3.5in]{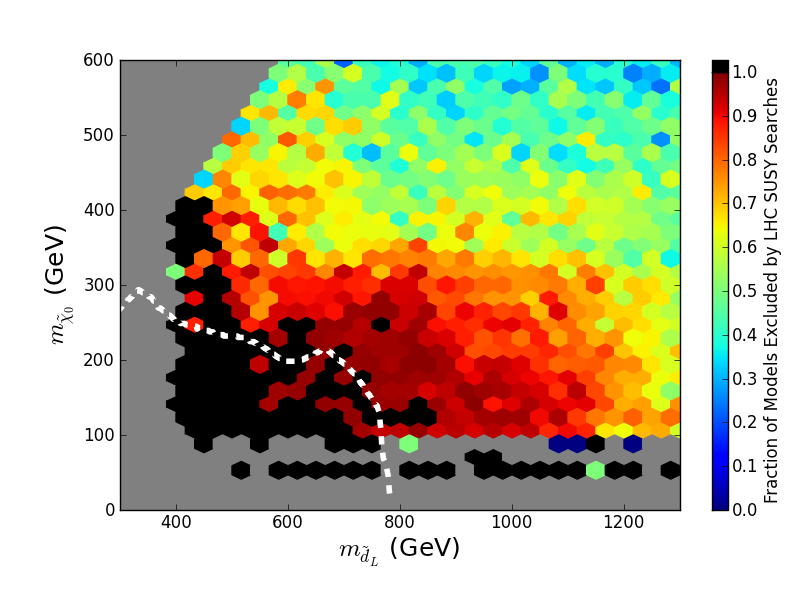}
\hspace{-0.50cm}
\includegraphics[width=3.5in]{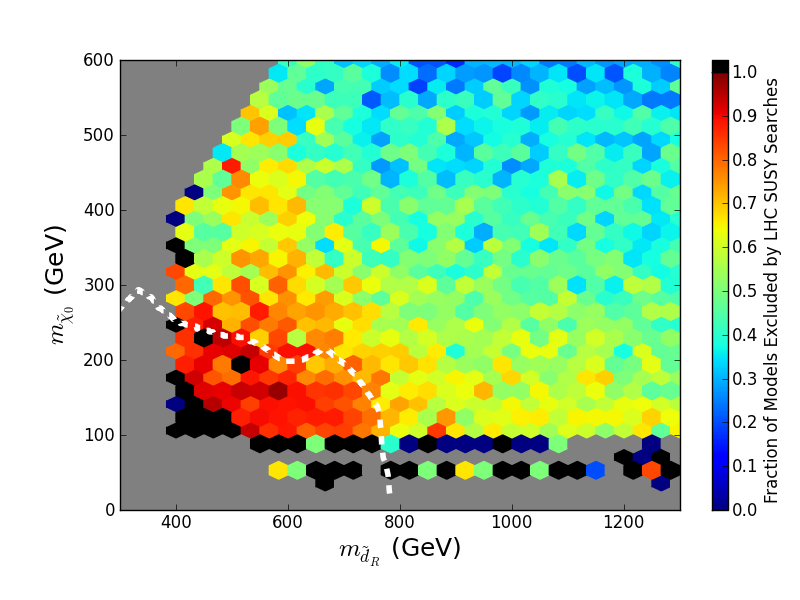}}
\vspace*{-0.10cm}
\caption{Same as the previous figure but now for $\tilde u_L$ (top left), $\tilde u_R$ (top right), $\tilde d_L$ (bottom left) and 
$\tilde d_R$ (bottom right). The ATLAS simplified model result assuming degenerate squarks is shown for comparison.}
\label{fig2}
\end{figure}

As observed above, there is a reasonably large set of models, particularly when the gluino is heavy, where rather light squarks are 
still allowed in comparison to the expectation provided by the simplified model treatment. It is informative to examine this aspect in 
a bit more detail. The search coverage for the first and second generation squarks are shown individually in Fig.~\ref{fig2}. Conventionally, LHC searches assume that these 4 squark states are degenerate, but here in the pMSSM where their masses can be different (except for the 
symmetry breaking relationship between the masses of $\tilde u_L$ and $\tilde d_L$) we see that the searches lead to quite different coverage 
for these sparticles. These differences are generally easy to understand and are essentially related to the relative sizes of the production 
cross sections for these sparticles. Since $\tilde u_L$ and $\tilde d_L$ are relatively degenerate they are produced simultaneously with somewhat  
similar rates (although that for $\tilde d_L$ is somewhat suppressed due to the smaller $d$-quark parton densities) 
and, one might expect similar, though not completely identical, exclusion rates. Indeed we see that is the case as the exclusions for $\tilde u_L$ 
and $\tilde d_L$ are quite similar, with the one for $\tilde u_L$ being slightly stronger.  In the case of $\tilde u_R$ and $\tilde d_R$, their masses are uncorrelated 
so that they will produce a smaller signal than the corresponding left-handed states; again, for identical masses, $\tilde d_R$ will have a smaller 
production rate due to the PDFs. In the figure we see that exclusions for either of these right-handed squarks is rather poor even though all they 
can do is decay to the bino components available in the lighter gauginos.  In particular, we see that $\tilde d_R$ masses as low as $\sim 450-500$ 
GeV remain possible with LSP masses in the range of $\sim 150$ GeV.  Additional work at the LHC will be needed to close the light squark mass 
loophole.

\begin{figure}[htbp]
\centerline{\includegraphics[width=3.5in]{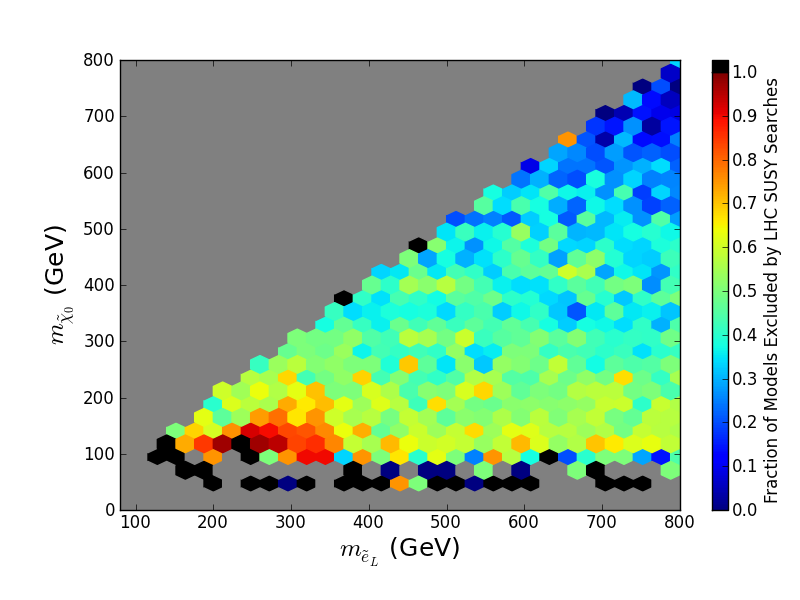}
\hspace{-0.50cm}
\includegraphics[width=3.5in]{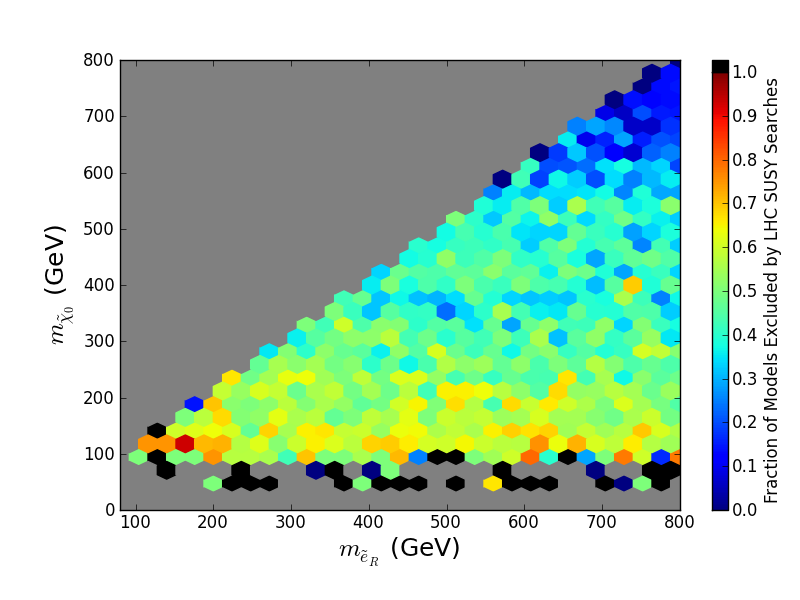}}
\vspace*{0.50cm}
\centerline{\includegraphics[width=3.5in]{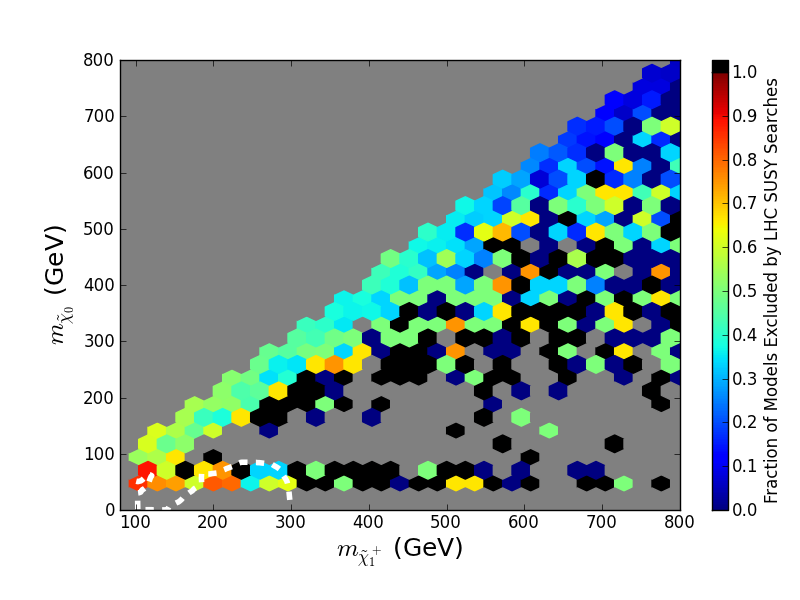}
\hspace{-0.50cm}
\includegraphics[width=3.5in]{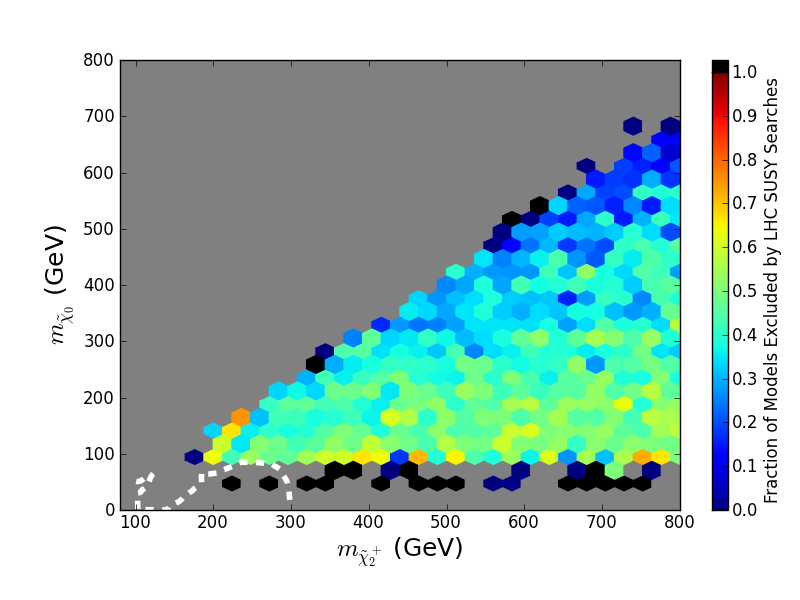}}
\vspace*{-0.10cm}
\caption{Same as the previous figure but now for $\tilde e_L$ (top left),  $\tilde e_R$ (top right), $\tilde \chi_1^\pm$ (bottom left), 
and $\tilde \chi_2^\pm$ (bottom right).}
\label{fig3}
\end{figure}

Figure \ref{fig3} shows the coverage in the LSP - left- and right-handed selectron and light/heavy chargino mass planes.  As expected, we see that the coverage is rather poor and very light selectrons and charginos are still allowed.

Having available the individual results of many 
different SUSY search analyses we next examine and compare their various strengths and weaknesses. Figure~\ref{fig4} provides an example of 
this where we compare the impact of the ``vanilla'' jets (+ leptons) + MET analyses (entries 1-3 in Table~\ref{SearchList7} and 1-4 in Table~\ref{SearchList7}) and the combination of the 3\textsuperscript{rd} generation searches in the lightest stop- and 
lightest sbottom-LSP mass plane along with the corresponding simplified model results obtained by ATLAS at both 7 (solid) and 8 TeV (dashed). 
Here the red (blue) bins indicate regions in the parameter space where most of the exclusion arises from the ``vanilla'' (3\textsuperscript{rd} 
generation) searches while green indicates a balance between these two extremes. In fact, most of the regions in both panels are 
green, indicating that both types of searches are having comparable impact in excluding models. However, in the upper portion of the plots 
near the kinematic boundary we see that the ``vanilla'' searches are more powerful, probably because of low b-tagging efficiencies for soft b-jets, while for small LSP masses we see that the 3\textsuperscript{rd} generation searches are 
dominating the exclusion. Clearly these types of results may change substantially as we add additional searches using the full luminosity 
available at 8 TeV.

\begin{figure}[htbp]
\centerline{\includegraphics[width=3.5in]{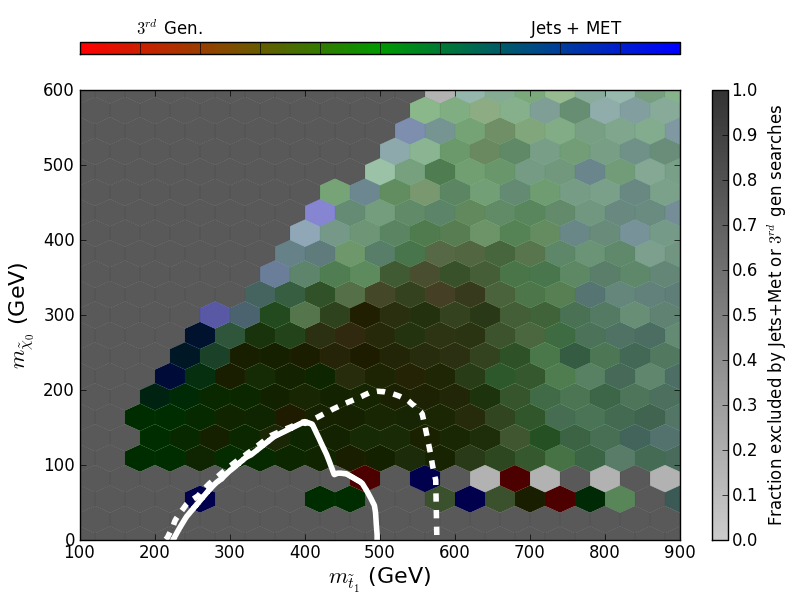}
\hspace{-0.50cm}
\includegraphics[width=3.5in]{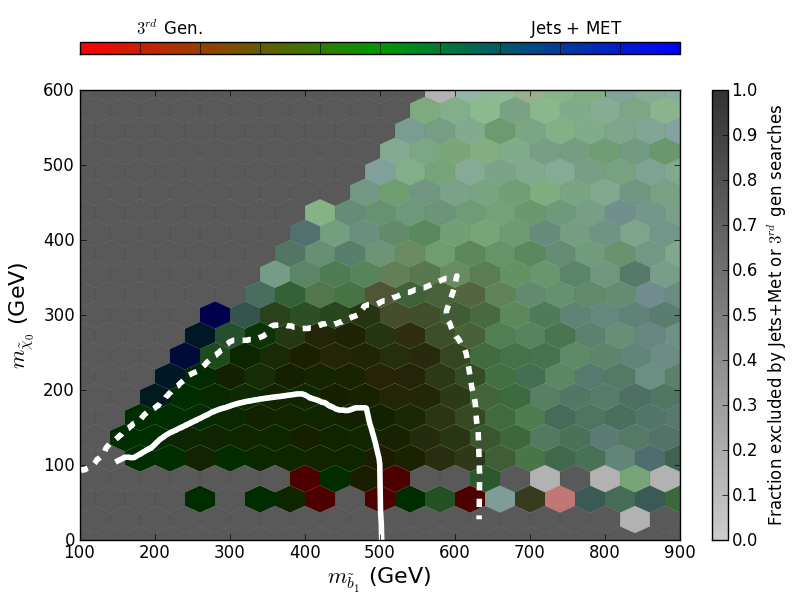}}
\vspace*{-0.10cm}

\caption{Comparison of the contributions to model exclusion arising from jets (+ leptons) + MET and 3\textsuperscript{rd} generation searches for light stops (left) 
and light sbottoms (right) in the neutralino model set. While the intensity in each mass bin indicates the fraction of models
that are excluded, the color indicates the 
relative importance of the two types of searches as described in the text.}
\label{fig4}
\end{figure}

\subsection{Gravitino Model Set}

We now examine the gravitino LSP model set. The preliminary results found here differ from that for the neutralino 
LSP model set for numerous reasons as outlined in Ref.~\cite {us4}. The distinctive 
phenomenology of gravitino LSP models mostly arises from the fact that the NLSP has a strongly suppressed coupling to the gravitino, with the result that the NLSP decay is displaced by a millimeter or more for most gravitino masses we consider. This has important implications for the standard SUSY searches; visible products from displaced decays may be rejected or may cause the whole event to be rejected by failing quality cuts designed to suppress large backgrounds from pileup. Alternatively, if a charged NLSP lives long enough to traverse the detector, it will not be counted as missing energy, making the standard MET-based searches ineffective. Of course, long-lived NLSPs also create new signals which can be even more striking than the original MET-based signature, as occurs when the NLSP lives long enough to be seen in searches for heavy stable charged particles (HSCPs). Another key difference from the neutralino LSP case is that the gravitino LSP is effectively massless in nearly all of our models. As a result, compressed spectra (which provide one of the most effective mechanisms for hiding light sparticles when the LSP is a neutralino) are only possible when the NLSP is a sneutrino (which decays invisibly to a gravitino and a neutrino) or a detector-stable neutralino.

A summary of the results of the 7 and 8 TeV LHC searches applied to the gravitino model set can be seen in Tables~\ref{SearchList7} and \ref{SearchList8}. We see that the ATLAS SUSY searches have a very different sensitivity to models with a gravitino LSP than to neutralino LSP models, as we anticipated. In particular, we see significant
differences in the fraction of models excluded by each search, and the fraction of models excluded by the combination of these searches is substantially higher for the gravitino LSP model set. Specifically, we see that $\sim 52\%$ of the gravitino LSP models are excluded by the combination of all 7 and 8 TeV searches, a result which is found to be rather insensitive to the 
$m_h=126\pm 3$ GeV mass requirement. 

The differences in the exclusion power of the combined ATLAS searches for models with neutralino or gravitino LSPs are displayed in the Figures to be discussed below, which show the results analogous to those presented in the previous subsection, but now shown for the 
gravitino LSP model set. First, we must note that the phenomenological differences between the two model sets make the results difficult to compare directly. The most important difference is that the neutralino LSP is not only a source of MET, but also forms a `kinematic boundary' for particle decays where the NLSP has a mass in the few hundred GeV range, since the LSP and NLSP generally have masses of comparable magnitudes. Additionally, the NLSP does not necessarily appear in cascade decays of heavier sparticles. For gravitino LSP models, on the other hand, the NLSP plays a critical role in the phenomenology, since it is almost invariably produced as the penultimate step in cascade decays, due to the very weak gravitino-sparticle interactions. If the NLSP decays within the detector to visible decay products, they generally have large transverse momentum (because the NLSP mass splitting with a massless gravitino is just the NLSP mass), and as a result these decay products are often the most important handle for distinguishing SUSY production from backgrounds. Alternatively, if the NLSP is detector-stable, its identity determines whether production of any sparticle will ultimately result in a SM + MET or a SM + heavy charged track final state. The 
nature of the NLSP is therefore crucial for determining the collider signature for a given gravitino LSP model. 

Based on this 
discussion, we can divide gravitino models into several categories based on the following properties of the NLSP: ($i$) whether of not it produces a charged track or visible decay products (invisibly-decaying NLSPs and neutral NLSPs which are detector stable are recorded only as MET) and ($ii$) the decay length of the NLSP, \ie, whether its decay is prompt $(c\tau < 200~\mu\mathrm{m})$ or displaced  $(200~\mu\mathrm{m} <c\tau < 2~\mathrm{m})$, or 
the NLSP is detector stable $(c\tau > 2~\mathrm{m})$. (The values of $200~\mu\mathrm{m}$ and $2~\mathrm{m}$ are chosen as the approximate length scales for displacements from the primary vertex allowed by quality cuts and the decay length necessary for the NLSP to be seen by the HSCP search, respectively.) Clearly the sensitivity of the ATLAS SUSY searches will depend strongly on how the model is categorized according to these criteria.

\begin{figure}[htbp]
\centerline{\includegraphics[width=3.5in]{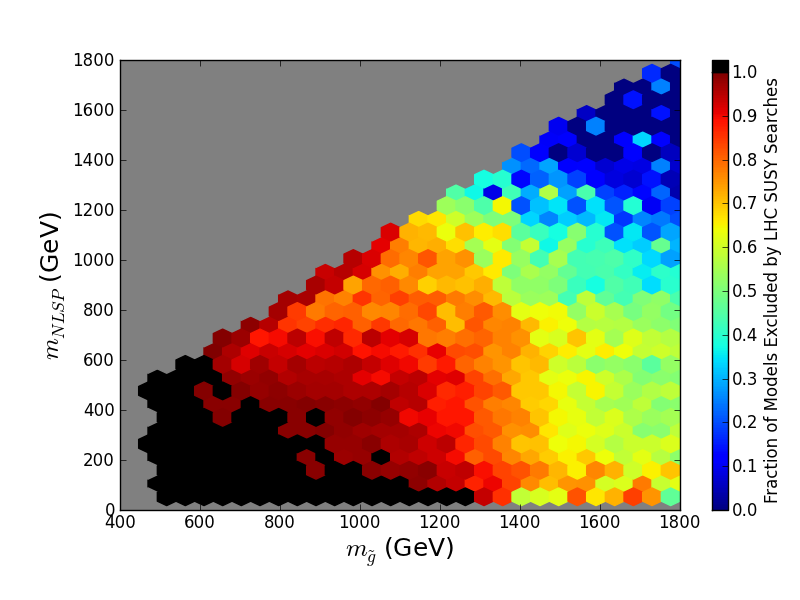}
\hspace{-0.50cm}
\includegraphics[width=3.5in]{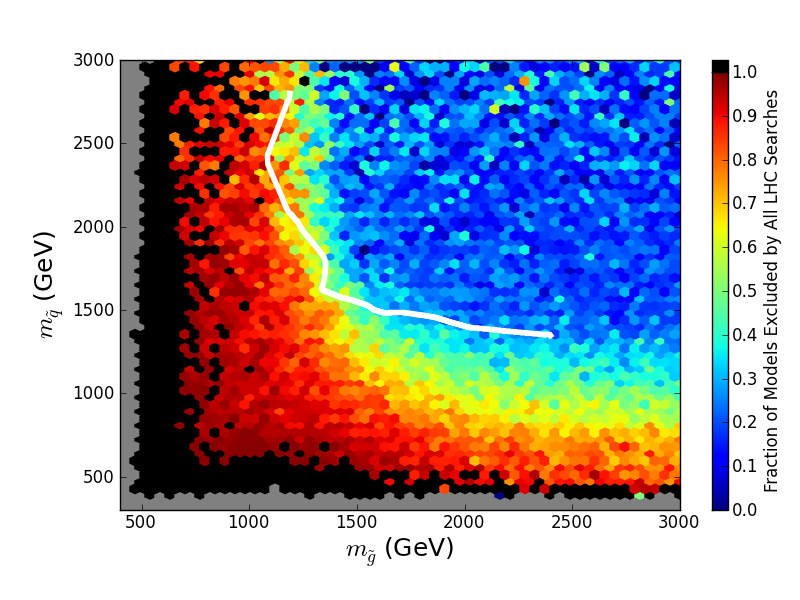}}
\vspace*{-0.10cm}
\caption{Projections of the pMSSM model coverage efficiencies from the 7 and 8 TeV LHC searches shown in the gluino-NLSP (left) and the lightest squark-gluino (right) 
mass planes for the gravitino LSP models, including all NLSP types and lifetimes. As usual the color code provides the total search efficiency in a specific mass bin.}
\label{fig1g}
\end{figure}

Fig.~\ref{fig1g} provides a summary of the impact of LHC searches on models with gravitino LSPs, analogous to Fig.~\ref{fig00} above for the neutralino LSP model set. For the left panel, it is important 
to note that here the mass on the $y-$axis is that of the NLSP and not that of the gravitino and that here we have summed over all the NLSP scenarios described above. 
Overall, the searches perform quite favorably compared to what we found for the neutralino model set, although the excluded regions are qualitatively similar. On the gluino-NLSP mass plot we see that the area of complete coverage is somewhat smaller than that found in the corresponding gluino-LSP plot 
for the neutralino set. However, in the gravitino LSP case we see that the ``mostly excluded'' region of orange and red bins includes all NLSP masses for gluinos below 1200 GeV. For the neutralino model set, there are bins in which most models remain viable (indicated by a green or blue color) with gluino masses as light as 800 GeV. Again, this is because energetic NLSP decay products (or a massive charged track) may be seen regardless of the sparticle-NLSP splitting. In the lightest (1\textsuperscript{st}/2\textsuperscript{nd} generation) squark-gluino mass plane shown in the second panel we again see somewhat similar coverage to that for neutralino LSP models, but once again the ``mostly excluded'' region ranges up to much higher gluino and squark masses than in the neutralino LSP models. Of course we reemphasize that the two model sets differ in many quantitative aspects.

\begin{figure}[htbp]
\centerline{\includegraphics[width=3.5in]{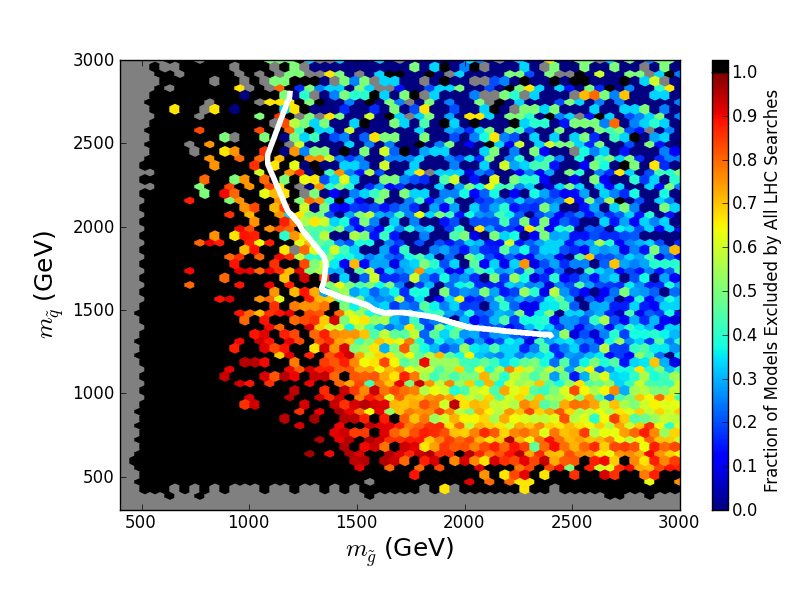}
\hspace{-0.50cm}
\includegraphics[width=3.5in]{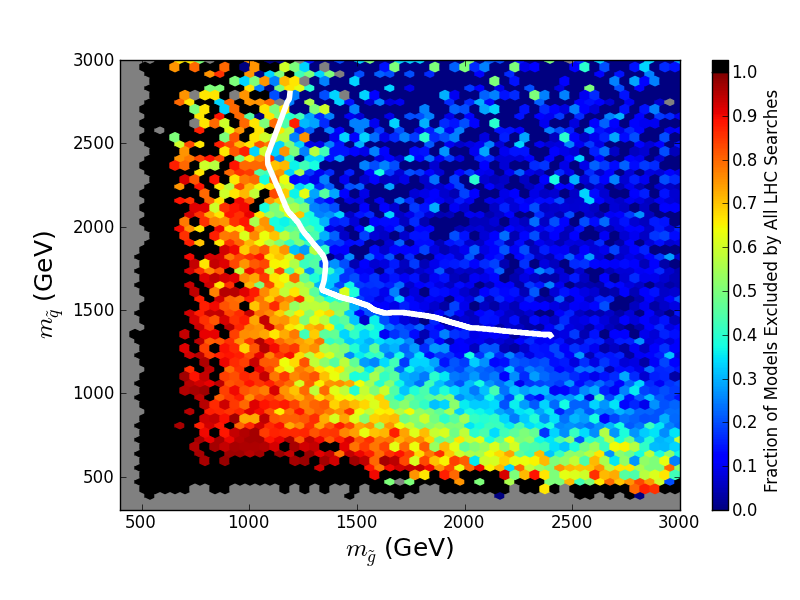}} 
\vspace*{0.50cm}
\centerline{\includegraphics[width=3.5in]{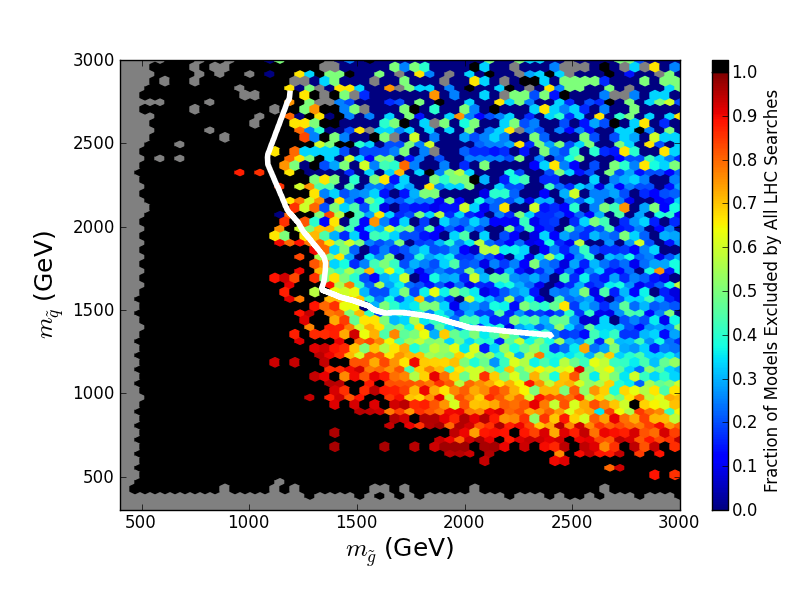}
\hspace{-0.50cm}
\includegraphics[width=3.5in]{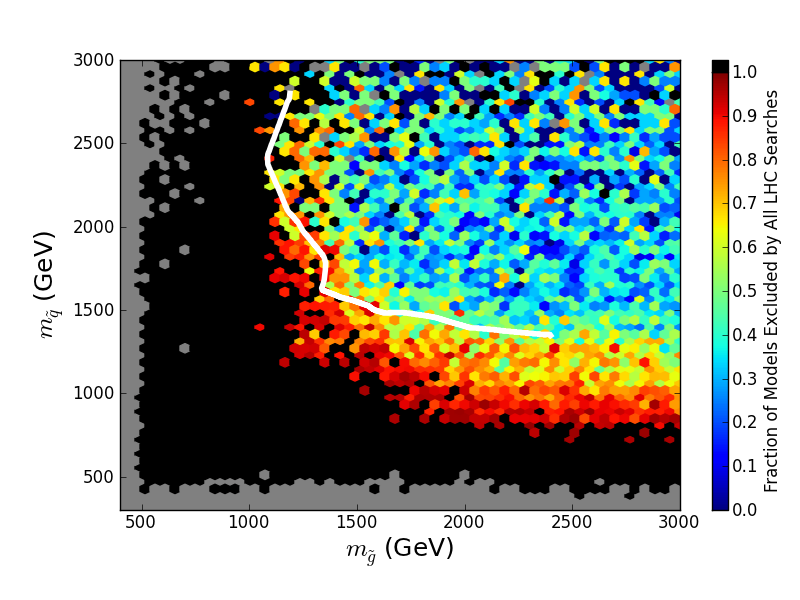}}
\vspace*{-0.10cm}
\caption{Projections of the gravitino pMSSM model coverage efficiencies from the 7 and 8 TeV ATLAS searches shown in the lightest squark-gluino mass plane for various 
model subsets: (top left) models with NLSPs that have displaced decays yielding observable decay products, (top right) models with NLSPs that are detector stable and invisible or have invisible decay products, (lower left) 
models with NLSPs that decay promptly, producing visible decay products and (bottom right) models in which the NLSP is detector-stable and charged.}
\label{fig2g}
\end{figure}

It is interesting to see how these gravitino LSP search results change when we select various subsets of models according to the NLSP categories we defined above; 
these results are shown projected onto the familiar lightest squark-gluino mass plane in Fig.~\ref{fig2g}. Comparing the upper right panel with the other three panels, we see that models wherein the NLSP or its decay products can be seen in the detector are far more strongly constrained than models in which the NLSP produces only MET, in which case the standard compressed spectrum scenario arises when the NLSP is closely split with the sparticle of interest. Not surprisingly, the results for models with invisible NLSPs are quite similar to the results for the neutralino LSP model set (compare the upper right panel with  the right panel of Fig.~\ref{fig00}). We can also see the reduced sensitivity to models with displaced NLSP decays, shown by the presence of surviving models with lighter squarks and gluinos in the upper left panel (showing models with displaced NLSP decays) compared with the lower left panel (showing models with prompt NLSP decays). Finally, the lower right panel shows that the exclusion limits for models with detector-stable charged or colored NLSPs are particularly strong, which is what we would expect given the low backgrounds and high sensitivity that characterize searches for heavy stable charged particles (HSCPs). One important difference between HSCP and MET based searches is that the former are sensitive to electroweak production of fairly heavy sparticles (e.g. 600 GeV charginos), whereas the MET-based searches have little direct sensitivity to particles with only electroweak interactions. This sharp distinction is demonstrated by the fact that the fraction of models with HSCPs that are excluded by the current searches is still $\sim 33 \% $ even when the squarks and gluino are kinematically inaccessible; by comparison, only $ \sim 15 \%$ of models with inaccessible squarks and gluinos and a decaying or invisible NLSP are excluded.

We can further investigate the effect of quality cuts removing or vetoing on displaced objects by comparing the subset of models with visible NLSP decay products with the subset of models in which the NLSP decay products are both visible {\it and} prompt. We see that the search reach in the gluino-NLSP mass plane, shown in Fig.~\ref{fig3g}, is significantly better for models with prompt decays. This results from the application of quality cuts that remove displaced objects, or even reject events with displaced objects, in the standard SUSY searches, combined with the fact that the current ATLAS searches for displaced decays are not optimized for NLSPs decaying to gravitinos (the disappearing track search has strong isolation requirements which veto the hard NLSP decay products, the displaced dilepton search requires the dilepton pair momentum to be aligned with the parent particle momentum, which is unlikely since the gravitino shares some of the parent momentum, and the muon + displaced vertex search requires a jet and a high-$p_T$ muon from the same vertex, which is not a standard NLSP decay signature). However, we note that despite the stringent quality cuts, the search reach for models with displaced NLSP decays yielding visible decay products is still better than that for models with invisible NLSPs. One possible reason for the remaining sensitivity to models with metastable NLSPs is that quality cuts that remove displaced jets from an event don't remove their contribution to the event's missing energy, meaning that the missing energy does not suffer from the same suppression as it would in the case of an invisible NLSP near the gluino mass.

\begin{figure}[htbp]
\centerline{\includegraphics[width=3.5in]{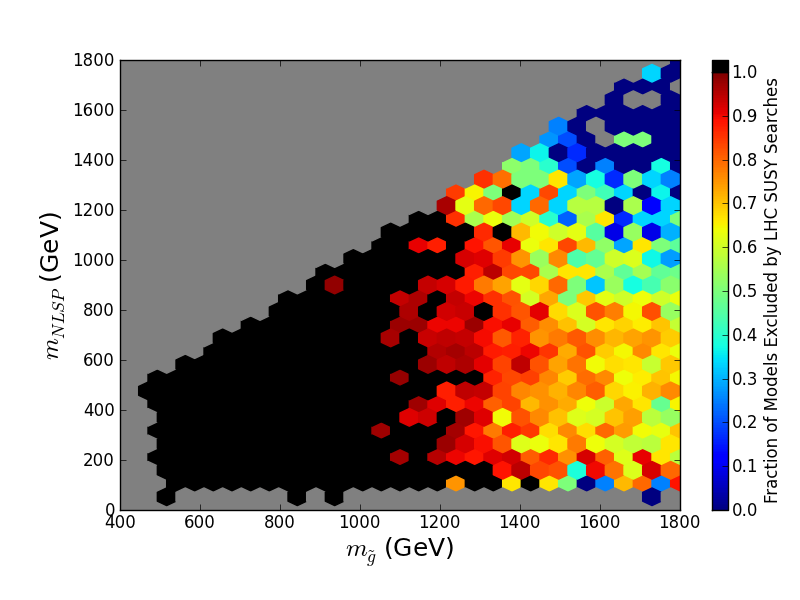}
\hspace{-0.50cm}
\includegraphics[width=3.5in]{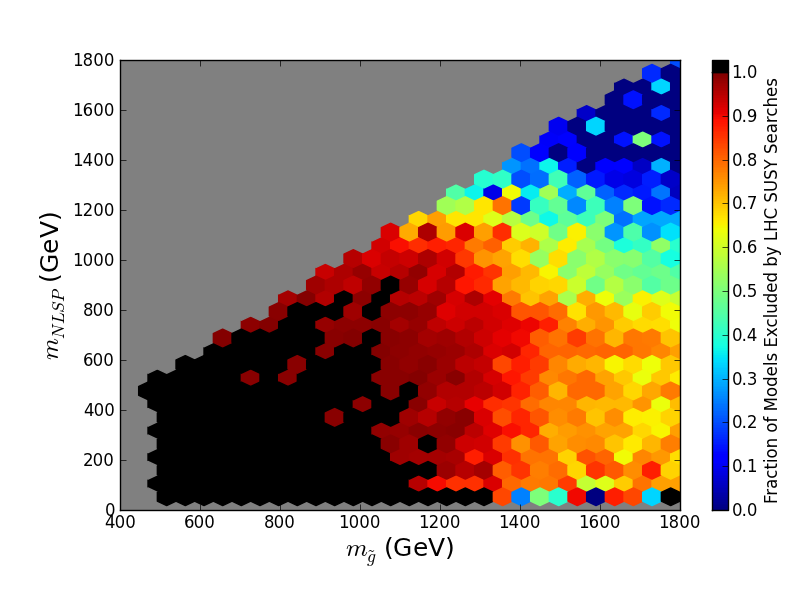}}
\vspace*{-0.10cm}
\caption{Projections of the pMSSM model coverage efficiencies from the 7 and 8 TeV LHC searches shown in the gluino-NLSP mass plane for the gravitino LSP models 
comparing the subsets of models where the NLSP produces a prompt, visible decay (left) compared to the larger subset where the NLSP decay need not be prompt (right) but 
still produces a visible final state.}
\label{fig3g}
\end{figure}

We saw above that the sensitivity of ATLAS searches to individual squarks is greatly reduced compared to the case where the squarks are degenerate. This lack of sensitivity to individual light squarks is even more striking for models with a gravitino LSP. Fig.~\ref{fig4g} shows the SUSY search coverages for the various 
1\textsuperscript{st}/2\textsuperscript{nd} generation squarks in the squark-NLSP mass plane; note that since $\tilde u_L$ and $\tilde d_L$ are produced together the coverage in these two cases 
is quite similar, as we saw for the neutralino LSP model set in Section~\ref{sec:neutlsp}. Here we see that the search reaches appear to be substantially reduced for the gravitino LSP models in all cases when compared with their reaches in the corresponding planes in the neutralino LSP model set. In particular, we see that it remains relatively easy for squarks in the gravitino LSP model set to have masses as low as 400 GeV without a near-degeneracy between any of the squarks and the NLSP. However, for very light NLSPs we find that the coverage is seen 
to extend out to somewhat larger squark masses, since in this case the LSP may be excluded through direct production. The limits on light squarks improve significantly when we consider only the subset of models in which the NLSPs produce one or more visible decay products (top panels of Fig.~\ref{fig6g}).

\begin{figure}[htbp]
\centerline{\includegraphics[width=3.5in]{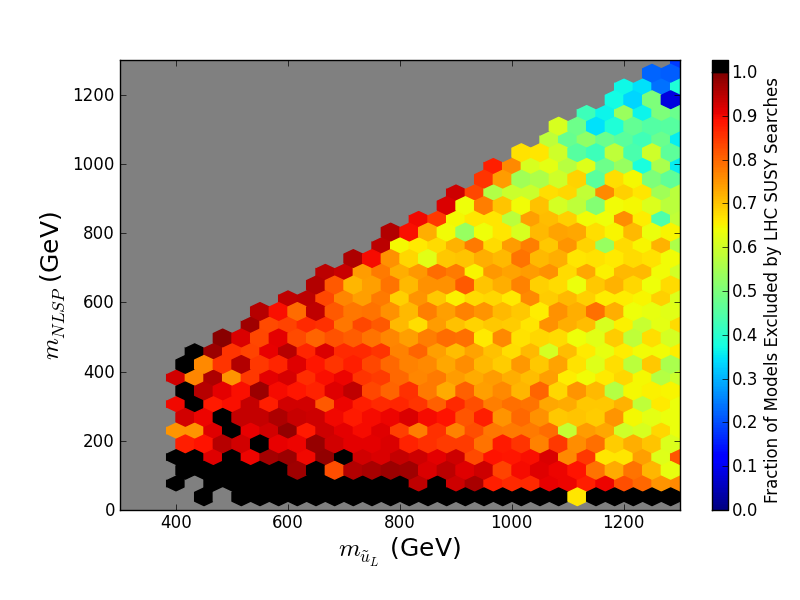}
\hspace{-0.50cm}
\includegraphics[width=3.5in]{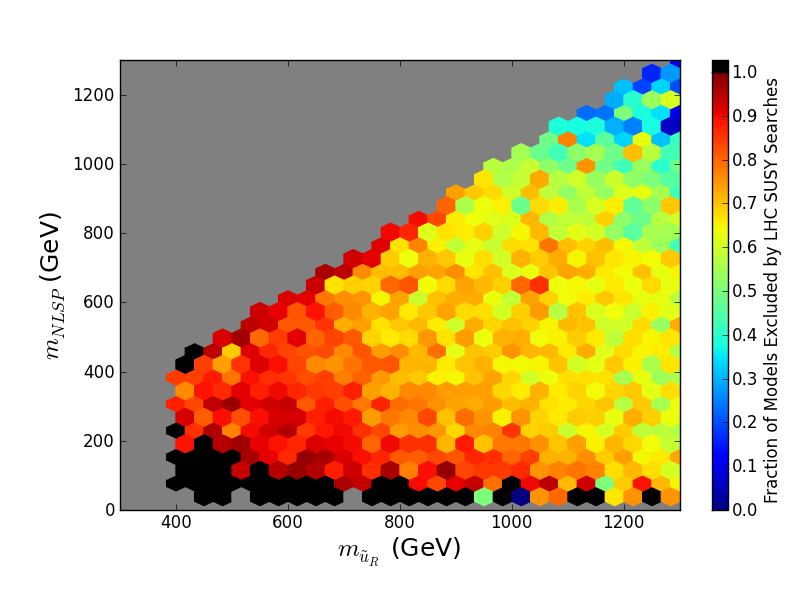}}
\vspace*{0.50cm}
\centerline{\includegraphics[width=3.5in]{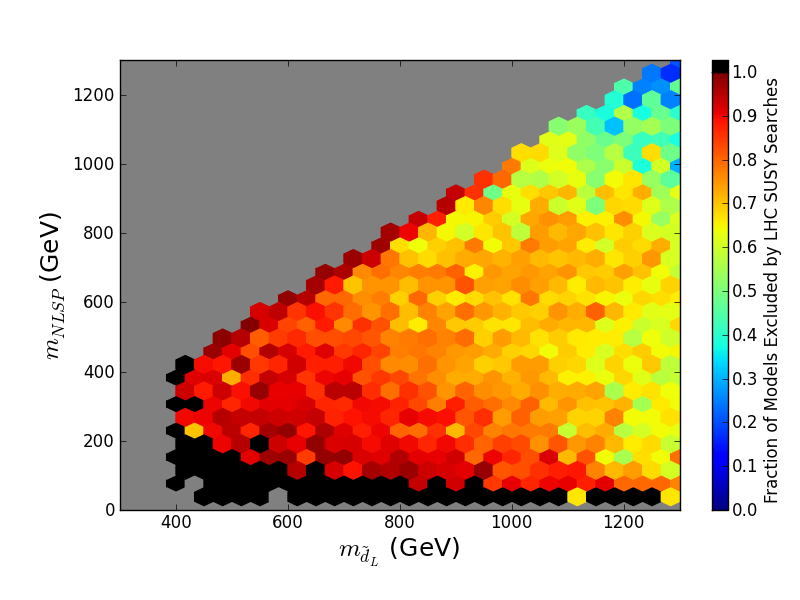}
\hspace{-0.50cm}
\includegraphics[width=3.5in]{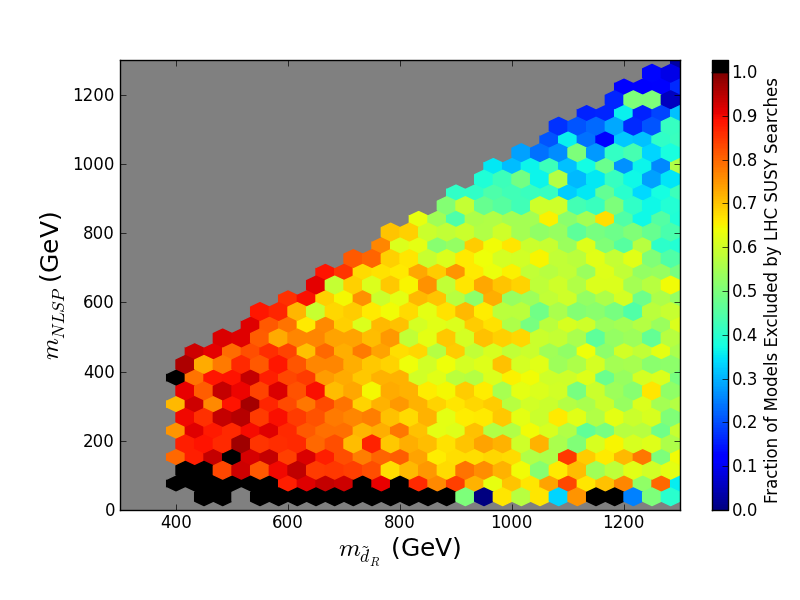}}
\vspace*{-0.10cm}
\caption{Search efficiencies for $\tilde u_L$ (top left), $\tilde u_R$ (top right), $\tilde d_L$ (bottom left) and $\tilde d_R$ (bottom right) in the gravitino LSP model set as a function of their 
masses and that of the NLSP.} 
\label{fig4g}
\end{figure}
\begin{figure}[htbp]
\centerline{\includegraphics[width=3.5in]{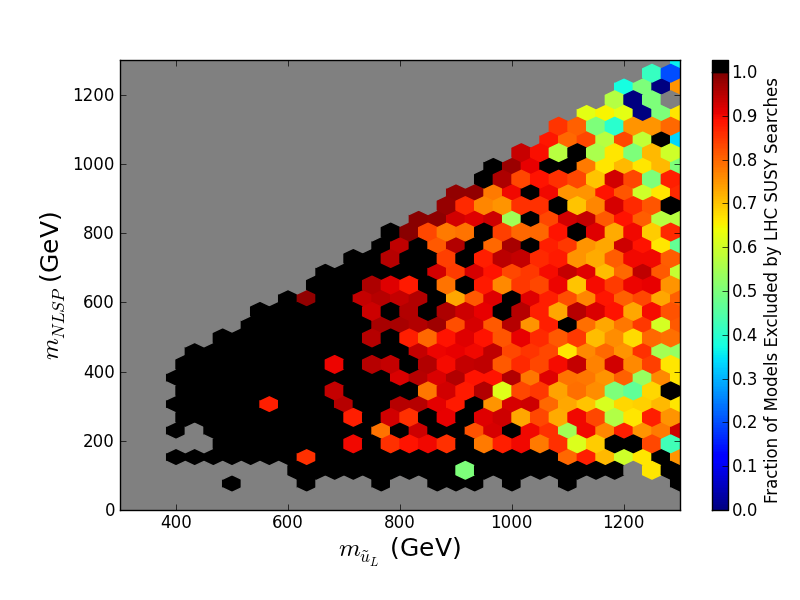}
\hspace{-0.50cm}
\includegraphics[width=3.5in]{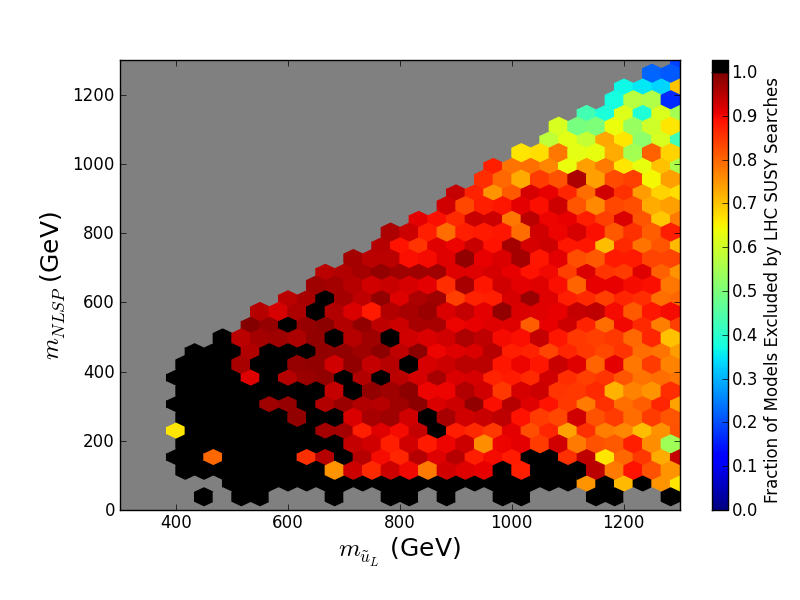}} 
\vspace*{0.50cm}
\centerline{\includegraphics[width=3.5in]{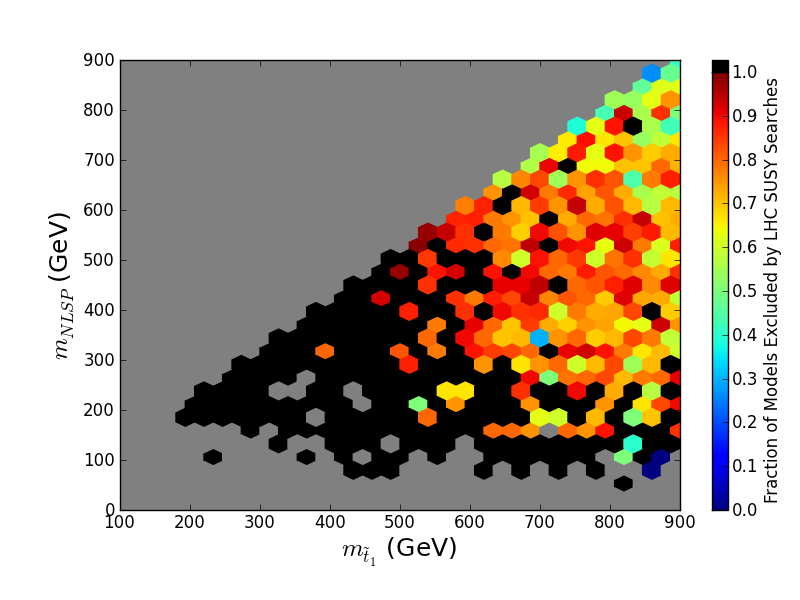}
\hspace{-0.50cm}
\includegraphics[width=3.5in]{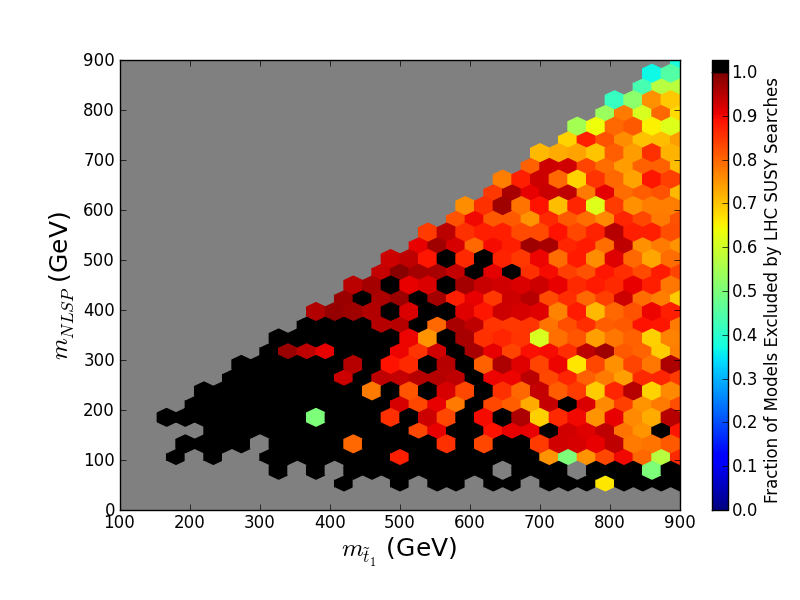}}
\vspace*{-0.10cm}
\caption{Search efficiencies for $\tilde u_L$ (top) and light stops (bottom) in the gravitino LSP model set as a function of their masses and that 
of the NLSP. Left panels are for prompt and visible decays while those on the right include all NLSPs which are visible in the detector or produce visible decay products.}
\label{fig6g}
\end{figure}

Fig.~\ref{fig5g} shows the resulting search efficiencies for the lightest stops and sbottoms in the gravitino LSP model set, shown in the sparticle-NLSP mass plane. The search efficiencies here are seen to be superior to those for the light squarks, as was the case in the neutralino set, but are still seen to be rather poor 
especially for the stop case. Overall, the results are similar to those for the light squarks - the search reach is somewhat weaker than that found 
for the neutralino LSP models but extends to higher masses when the NLSP mass is small. However, moderate coverage at the $\sim 60\%$ level extends out to significantly 
larger stop/sbottom masses in the gravitino case.  As in the case of light squarks, when we limit our discussion to the subset of models wherein the NLSPs produce a visible 
signal (prompt or not), the search efficiencies are found to be similarly improved. This is explicitly shown for the case of light stops as well as for the $\tilde u_L$ 
in Fig.~\ref{fig6g}. 

\begin{figure}[htbp]
\centerline{\includegraphics[width=3.5in]{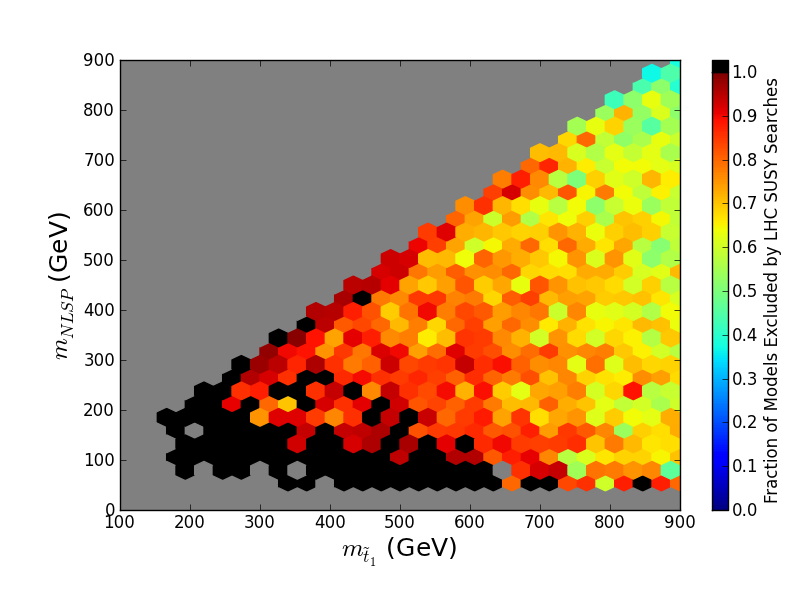}
\hspace{-0.50cm}
\includegraphics[width=3.5in]{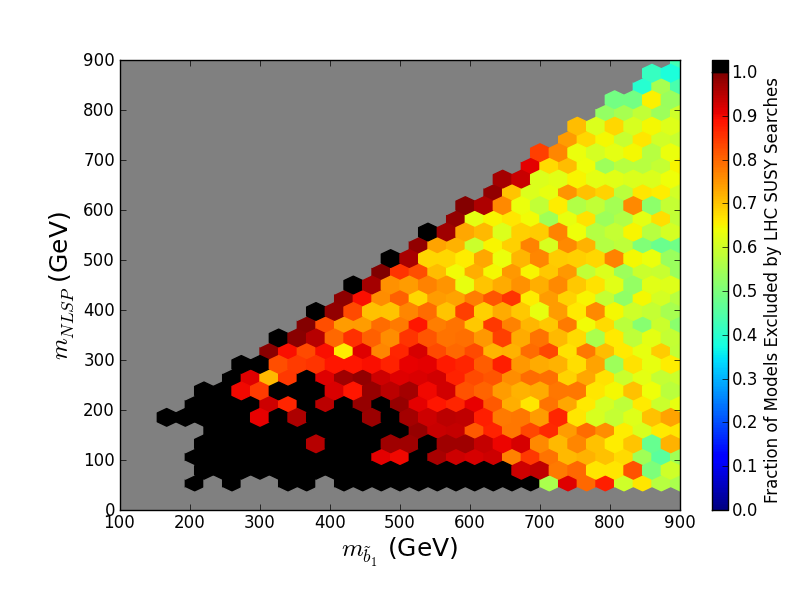}}
\vspace*{-0.10cm}
\caption{Search efficiencies for light stops (left) and sbottoms (right) in the gravitino LSP model set as a function of their masses and that of the NLSP.}
\label{fig5g}
\end{figure}

\subsection{Low Fine-Tuning Model Set}

As discussed above, we also have generated a small ($\sim 10.2$k) set of models with low Fine-Tuning where the neutralino LSP saturates the thermal relic density (with a Higgs mass 
of $126 \pm 3$ GeV); this low-FT set was selected 
from an initial sample of $3.3 \times 10^8$ points. This shows that satisfying the additional constraints of the `correct' 
relic density and the observed Higgs mass (in addition to all of the standard collider, precision electroweak, DM search and flavor constraints) is non-trivial to accomplish. One reason for 
this is that while $\sim 20\%$ of the original neutralino LSP models gave the `correct' Higgs mass of $126\pm 3$ GeV, the range we now allow for the relic density around its central value 
($\Omega h^2 = 0.1153 \pm 0.095$) is quite narrow compared to the range of values allowed for the full neutralino model set, which extends over several orders of magnitude \cite{us1}.  
Figure~\ref{fig1xx} displays the 
resulting distributions of the Higgs mass, relic density and amount of fine-tuning ($\Delta$, the Ellis-Barbieri-Giudice parameter \cite{Ellis:1986yg,Barbieri:1987fn}) 
for this model set. Here we see that the set is dominated by models 
which have larger values of $\Delta$ and somewhat smaller Higgs masses as we might expect. The smallest value of $\Delta$ we obtain is $\sim 30$ and to go much lower 
would likely require a dedicated Markov chain Monte Carlo study using our lower $\Delta$ points as seeds. 

\begin{figure}[htbp]
\centerline{\includegraphics[width=3.5in]{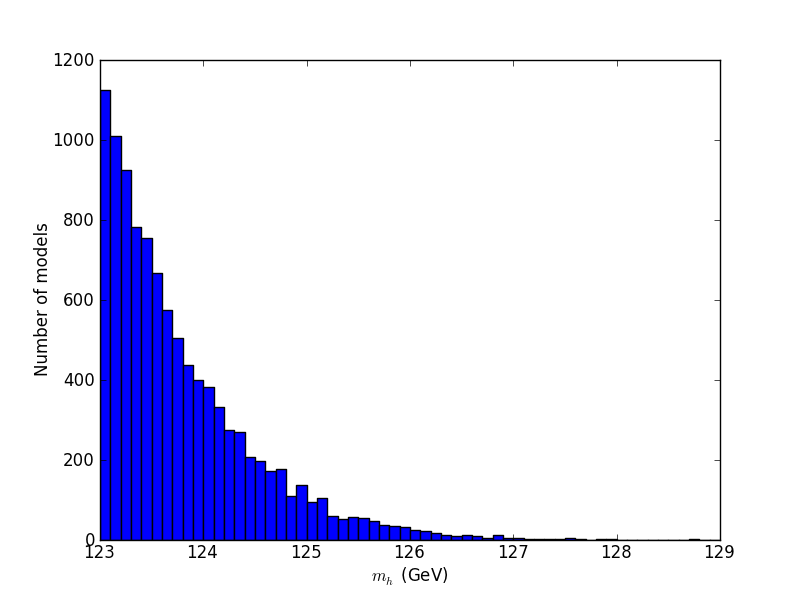}
\hspace{-0.50cm}
\includegraphics[width=3.5in]{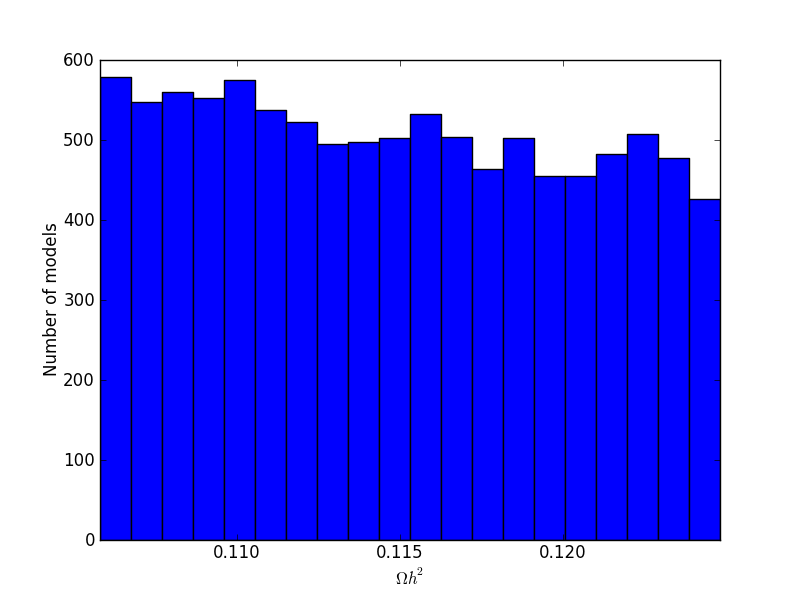}}
\vspace*{0.50cm}
\centerline{\includegraphics[width=3.5in]{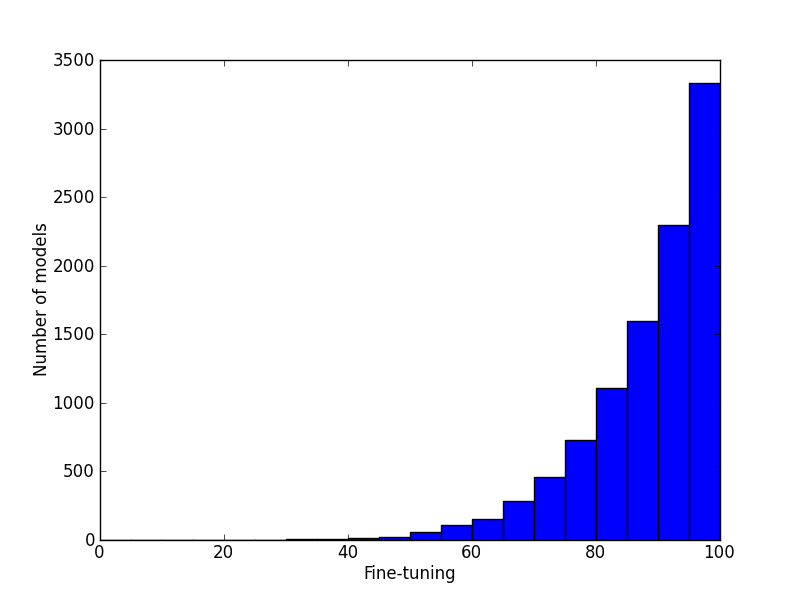}}
\vspace*{-0.10cm}
\caption{The distribution of Higgs masses (top left), thermal relic density (top right) and the amount of fine-tuning $\Delta$ (bottom) are shown for the 
low-FT model set.}
\label{fig1xx}
\end{figure}
\begin{figure}[htbp]
\centerline{\includegraphics[width=4.50in]{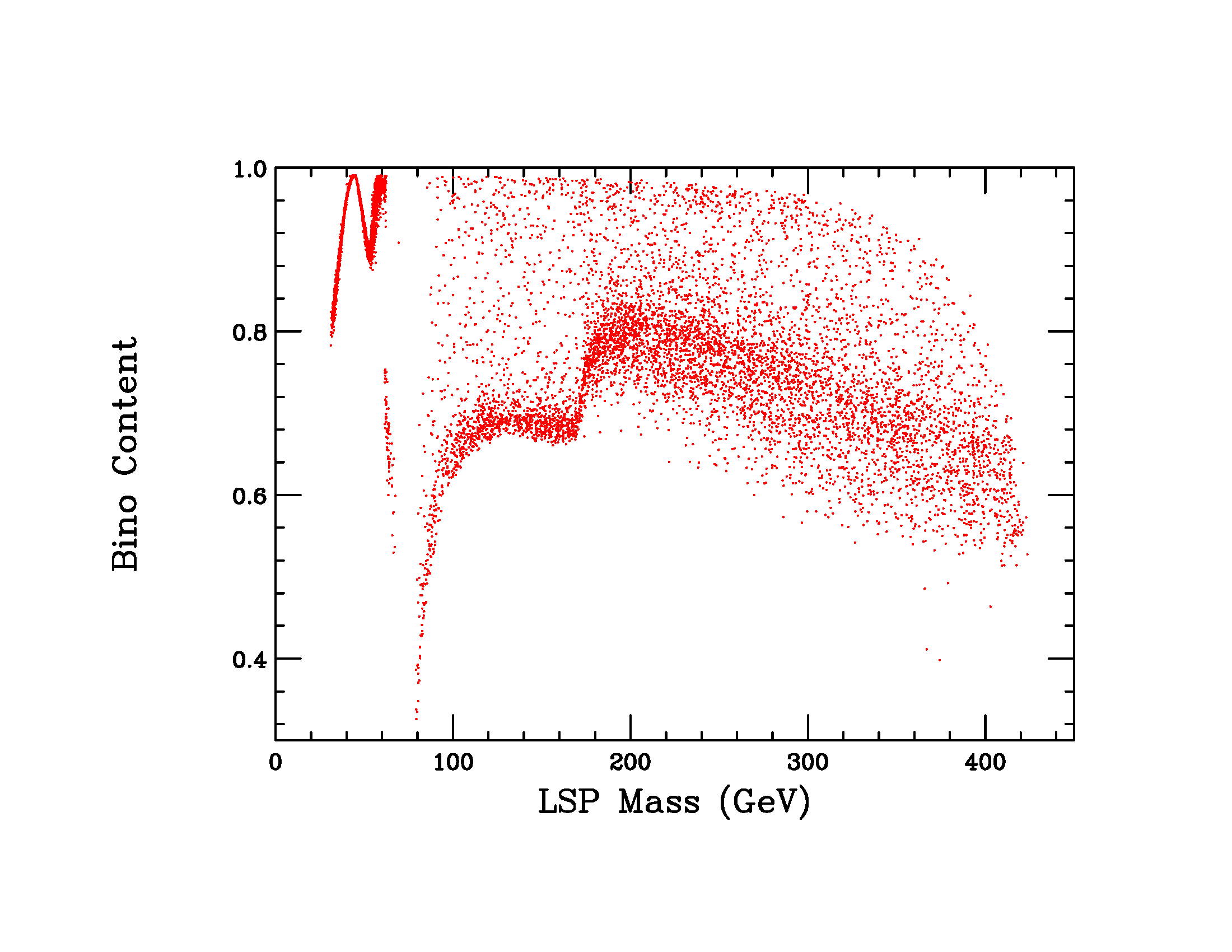}
\hspace{-0.70cm}
\includegraphics[width=4.50in]{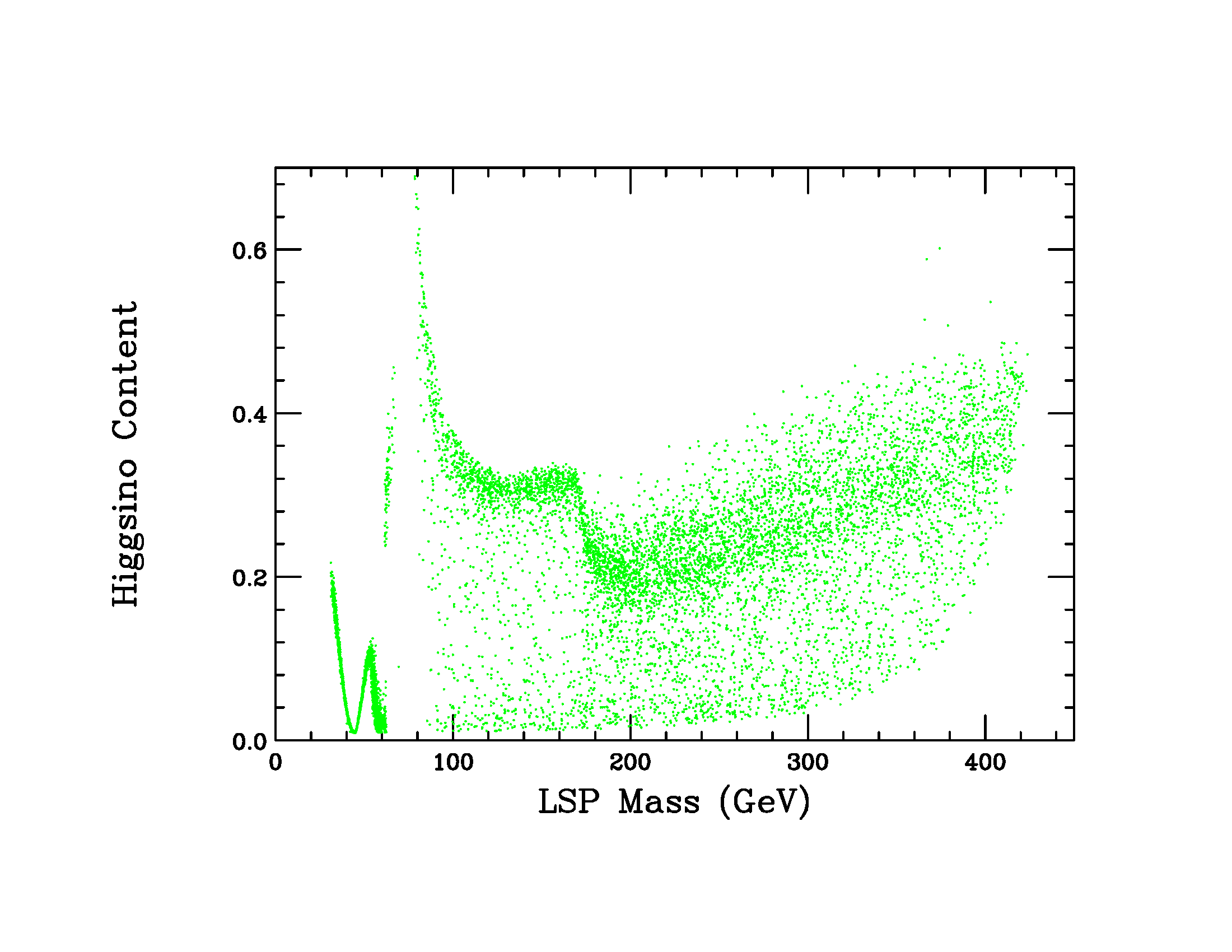}}
\vspace*{0.30cm}
\centerline{\includegraphics[width=4.50in]{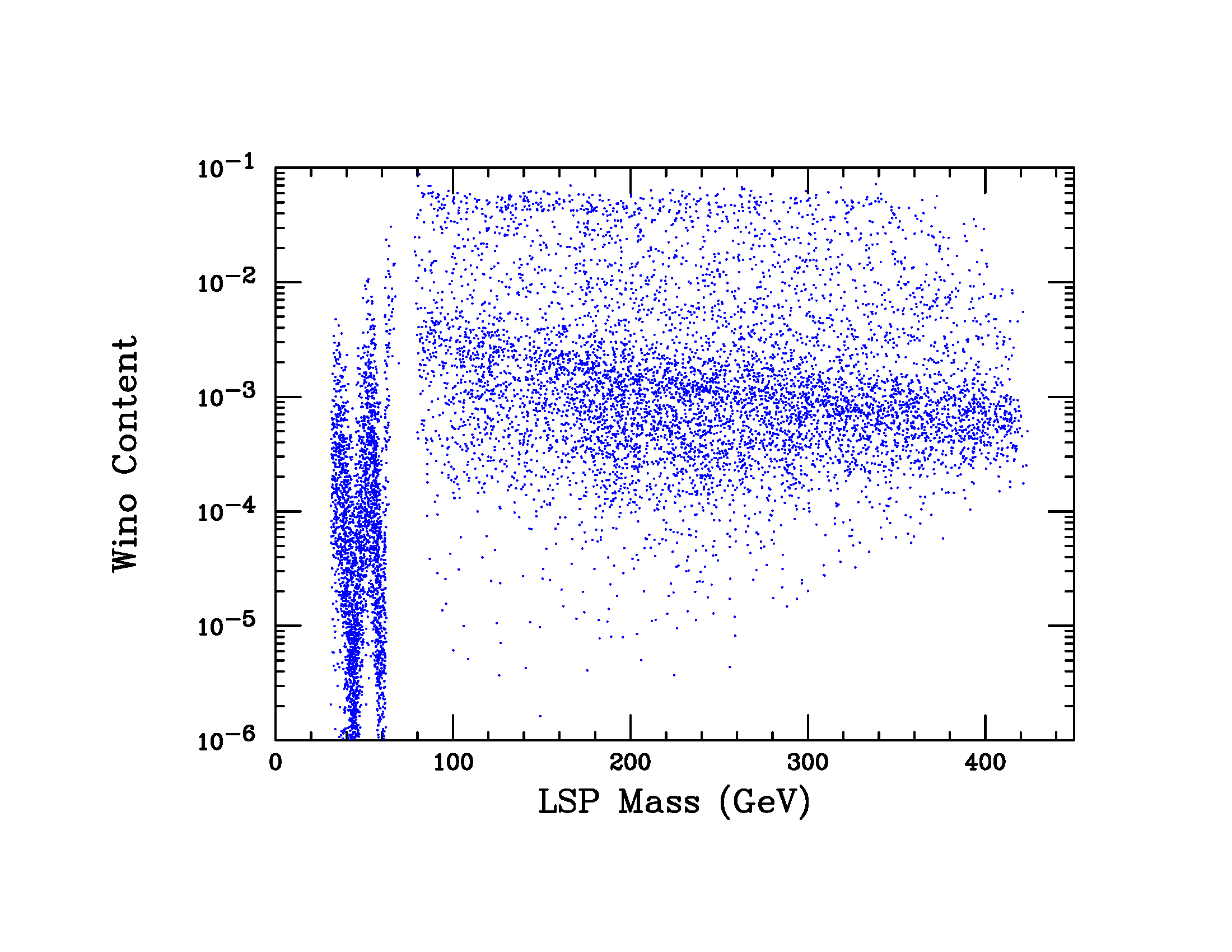}}
\vspace*{-0.3cm}
\caption{The bino (top left), Higgsino (top right) and wino (bottom) content of the LSP in the low-FT model set as a function of its mass.}
\label{fig1yy}
\end{figure}

These low-FT models necessarily have a relatively light stop and a mostly bino-like LSP along with Higgsinos with masses below $\sim 450$ 
GeV. Well-tempered bino-Higgsino mixing is mostly responsible for achieving the correct relic density in this model set, although co-annihilation (frequently with a 
light slepton or stop) or annihilation through either the $Z$ or Higgs funnel is also rather common. 
Figure~\ref{fig1yy} 
shows the electroweak content of the LSP as a function of its mass for all the models in the low-FT set and much of the structure associated with this physics 
is directly observed here. Note that for 
rather light LSPs, co-annihilation is not possible given the constraints from LEP on chargino and slepton masses so that the LSP must be a bino-Higgsino 
admixture in this case. 
Although our scan ranges allow for somewhat lighter LSPs, we find that all such sparticles must have masses greater than $\sim 30$ GeV in 
order to satisfy the constraint on the invisible decay width of the $Z$: $\Gamma(Z\to \chi \chi)< 3$ MeV as shown in Fig.~\ref{fig1xxx}{\footnote {The 
invisible width of the {\it Higgs} can also constrain the light neutralino spectrum. However, the model-independent limit on this quantity, $\sim 50-60\%$, is not yet 
sufficiently strong to be meaningful as can be seen in the figure.}}. 

We can, in fact, make an even stronger statement based on our study of both the neutralino and low-FT model sets. Under the following assumptions: ($i$) $m_h = 126 \pm 3$ 
GeV,  ($ii$) $\Gamma(Z\to \chi \chi)< 3$ MeV, ($iii$) the LSP produces a thermal relic density that either saturates or is below the WMAP/Planck value and ($iv$) the LEP 
constraints on {\it charged} sparticles are trivially satisfied (\ie, their masses can't `tunnel' to values below $\sim 90-100$ GeV for any reason),  {\it then} the mass 
of the LSP must exceed $\simeq 30$ GeV.

Returning now to the low-FT models, we note that since mostly left-handed $\tilde t_1$'s are common, light $\tilde b_1$'s are as well; furthermore, 
$\sim 11\%$ of the the time we find the sbottom to be lighter than the stop. Interestingly, since $|M_2|<2$ TeV to satisfy the low-FT requirement we find that 
$\sim 60\%$ of the models will also have winos below the stop/sbottom. This makes for a rather complex spectrum and even more complex decay patterns for the 
stops and sbottoms.

\begin{figure}[htbp]
\centerline{\includegraphics[width=3.7in]{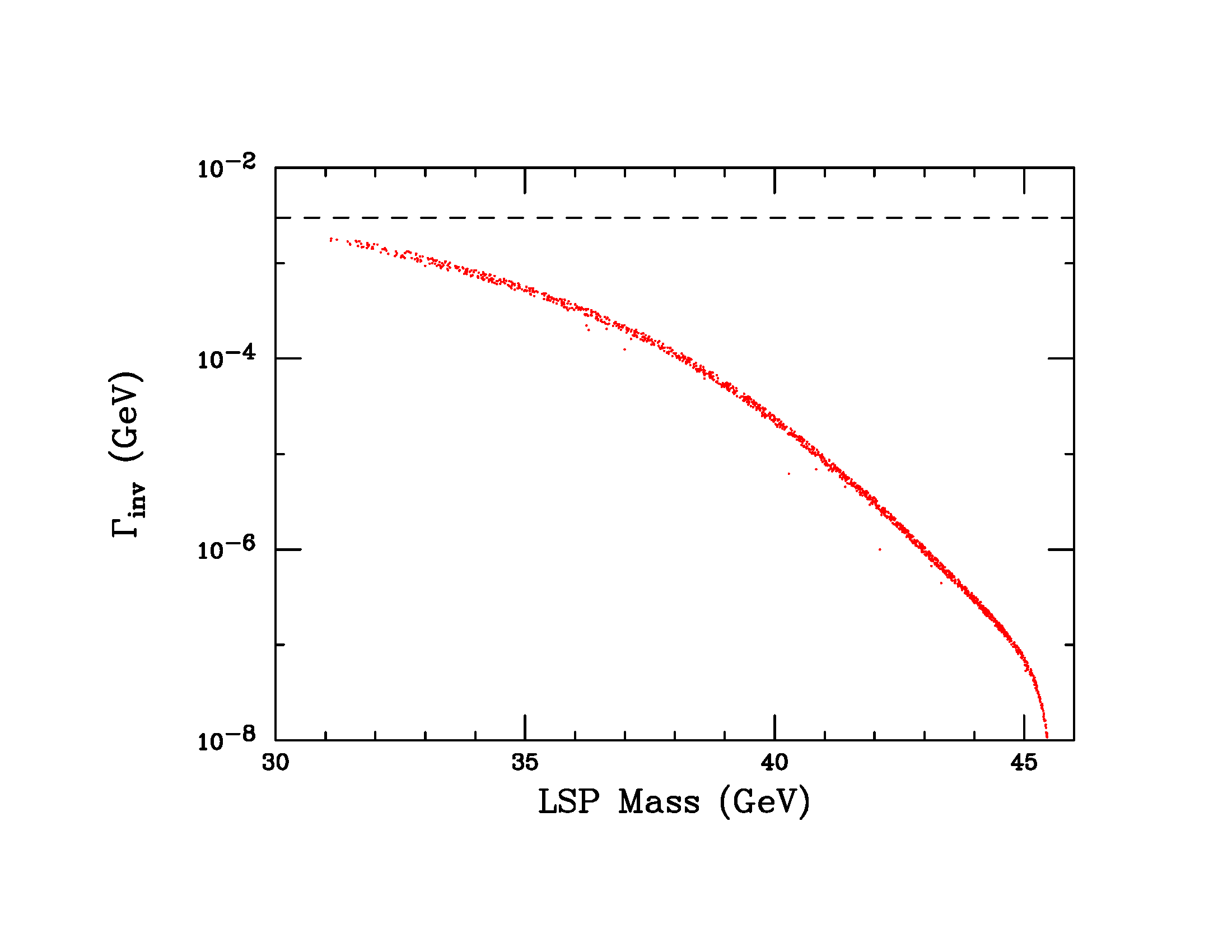}
\hspace{-0.50cm}
\includegraphics[width=3.5in]{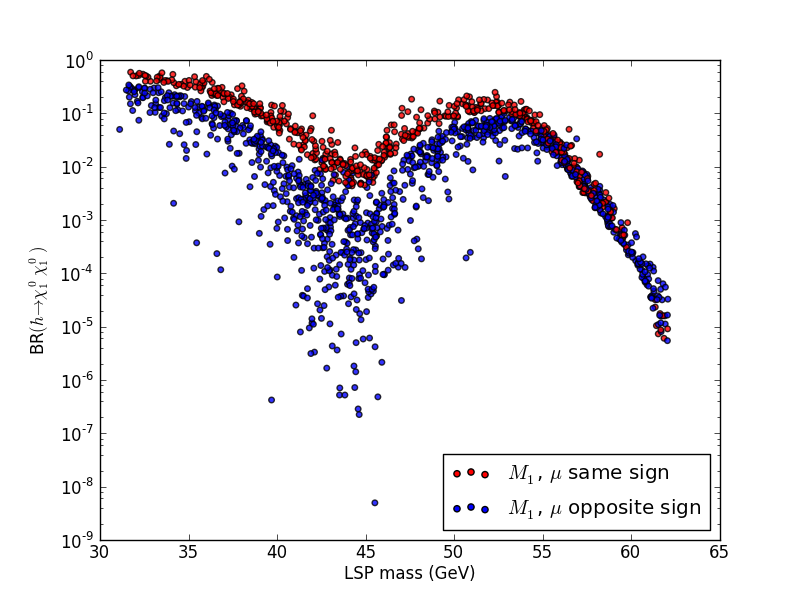}}
\vspace*{-0.10cm}
\caption{Invisible width of the $Z$ (left) and the Higgs (right) for kinematically accessible LSPs in the low-FT model set. In the left panel the LEP upper 
bound is also shown.}
\label{fig1xxx}
\end{figure}
\begin{figure}[htbp]

\subfloat{
  \centerline{\includegraphics[width=5.5in]{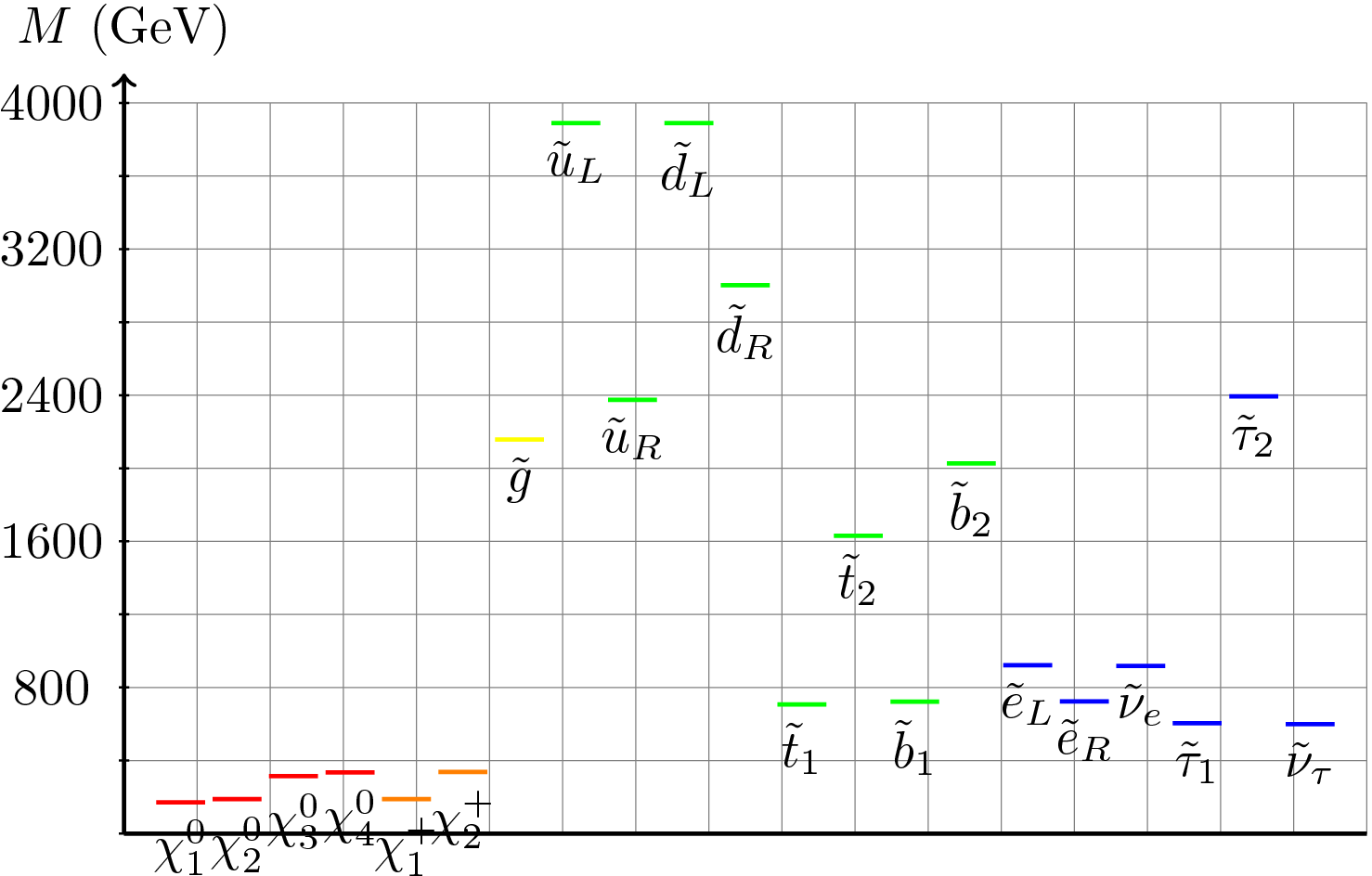}}
  \hspace{-0.50cm}

} // 

\subfloat{
  \vspace*{10cm}
}

\subfloat{\centerline{\includegraphics[width=3.5in]{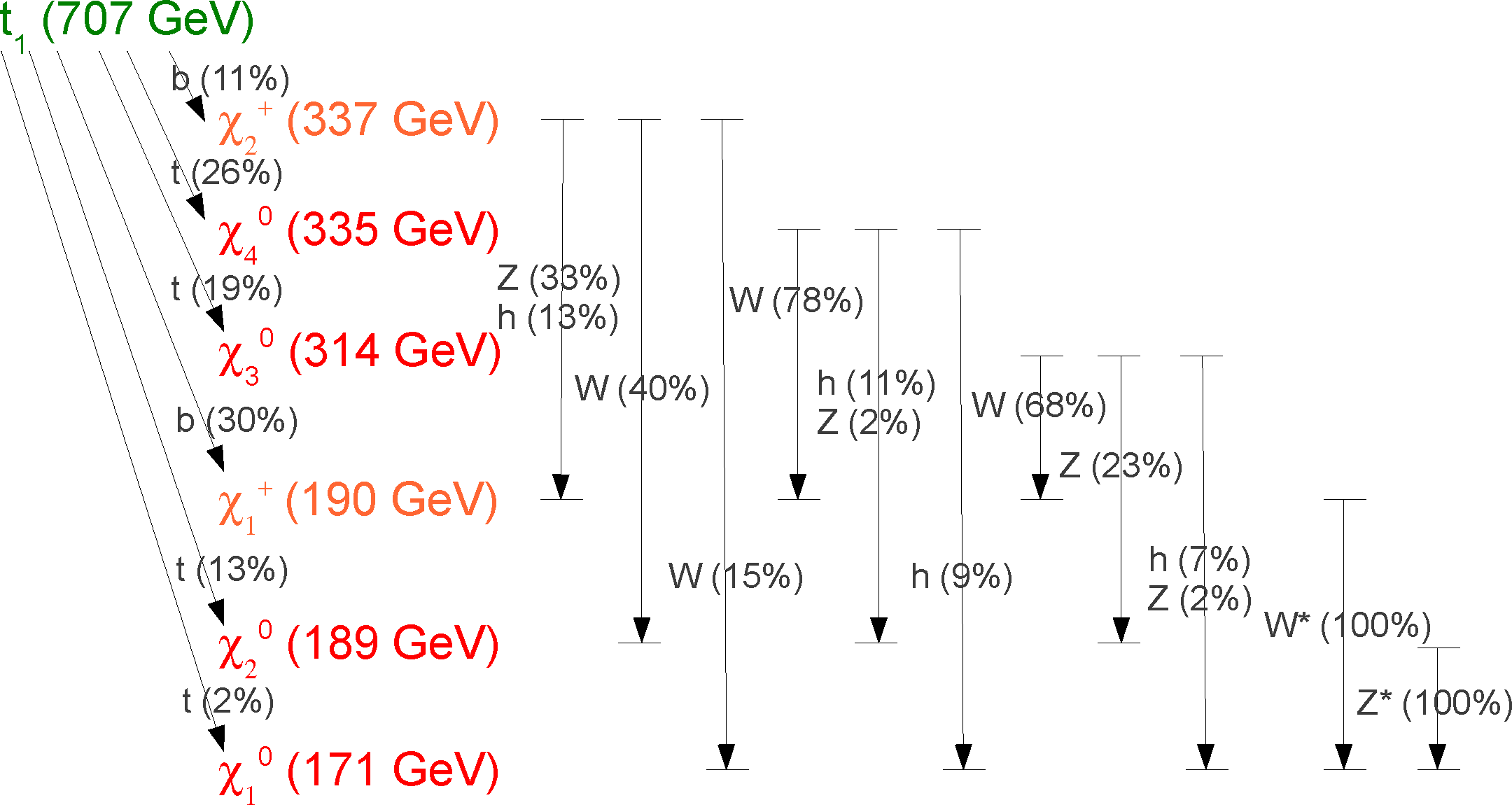}
\vspace*{0.5cm}
\includegraphics[width=3.5in]{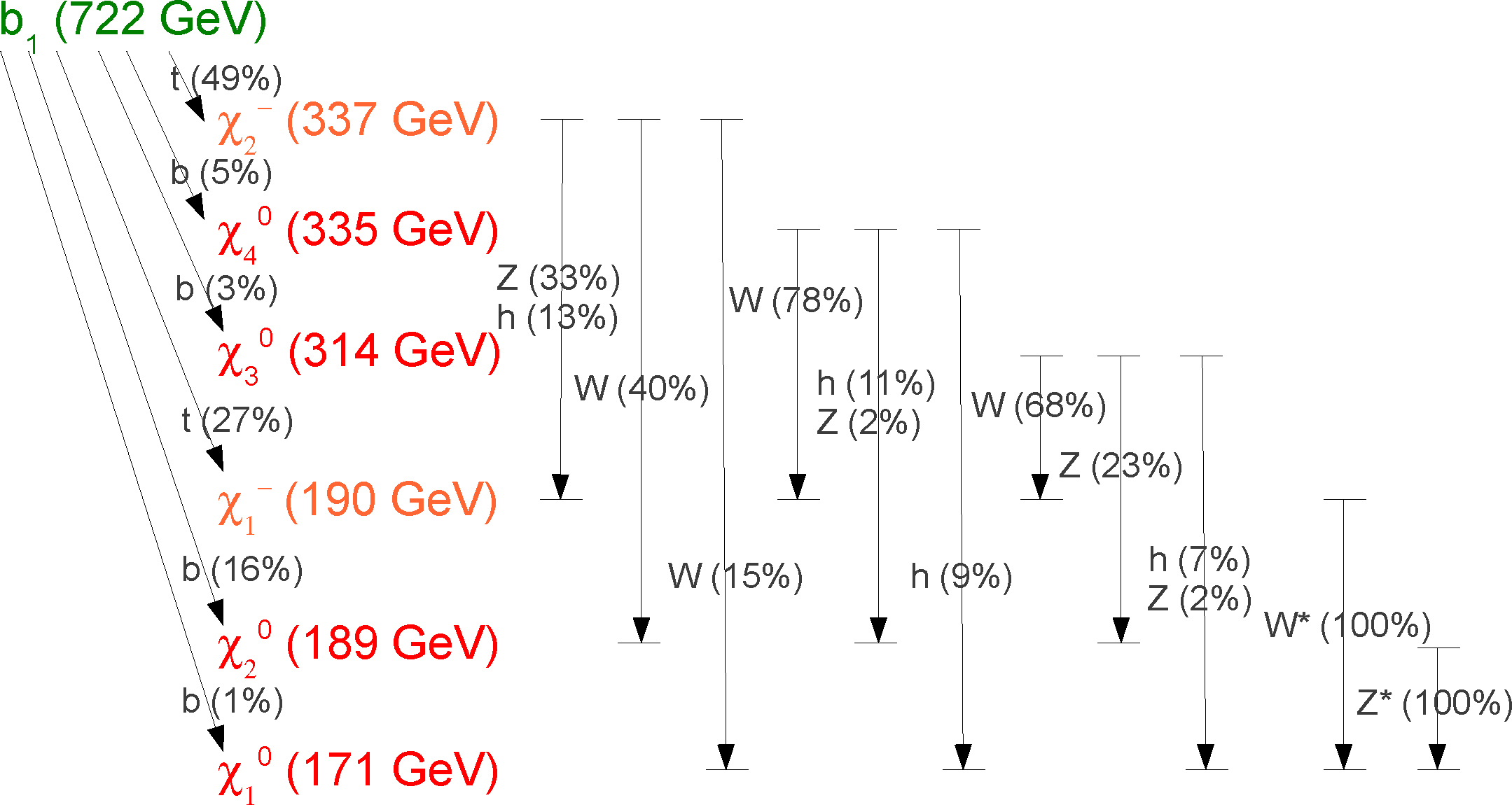}}
}
\vspace*{0.5cm}
\caption{Typical spectrum (top) and decay patterns for stops (bottom left) and sbottoms (bottom right) in a low-FT model.}
\label{fig1x}
\end{figure}

Figure~\ref{fig1x} shows a typical spectrum for one of the low-FT models with very heavy 1\textsuperscript{st}/2\textsuperscript{nd} generation squarks, reasonably heavy gluinos and 
a light stop/sbottom. Here we see that all the electroweakinos lie below the lighter stop/sbottom. This results in the rather complex decays for both of these sparticles, which are 
shown in the lower two panels of this Figure. Note that the light stop/sbottom can decay to any of the lighter electroweakinos with comparable branching fractions; 
these states then cascade down to lighter ones producing, \eg,  $W$'s and $Z$'s so we might expect multi-lepton searches to be useful here. Given these 
decay patterns, it is clear that searches for any {\it one} particular final state in stop/sbottom decay will not be very useful. However, by combining all 
of the 3\textsuperscript{rd} generation and other searches, we expect to find that very significant model coverage can be obtained.

Tables~\ref{SearchList7} and \ref{SearchList8} above show the ATLAS/CMS SUSY search analyses applied to this model set and the resulting fractions of models excluded by each 
of the individual searches; when combined we find that $\sim 70\%$ of the low-FT models are already excluded by the 7 and 8 TeV results. 
We note that many of the individual searches perform significantly better than in the case of the general neutralino model set, and as a result, the fraction of models excluded by the combined set of searches is nearly twice as large for the low-FT model set. This is clearly observed in Fig.~\ref{fig00ft} which shows the search efficiencies in both the gluino-LSP and lightest 
squark-gluino mass planes. Comparing this figure to the general neutralino LSP results shown above demonstrates the stark contrast between the two exclusion efficiencies. Since most of the time only the 
3\textsuperscript{rd} generation squarks are lighter than the gluino, the production of gluinos automatically leads to final states with a profusion of top and 
bottom quarks for which the 3\textsuperscript{rd} generation ATLAS searches were designed. Figure~\ref{fig00ft} shows, given our level of statistics, that this results 
in the exclusion of all of our low-FT models with gluinos below $\sim 1.2$ TeV! As in the general neutralino set we see that 1\textsuperscript{st}/2\textsuperscript{nd} generation squarks can be 
relatively light provided the gluinos are heavy.

\begin{figure}[htbp]
\centerline{\includegraphics[width=3.5in]{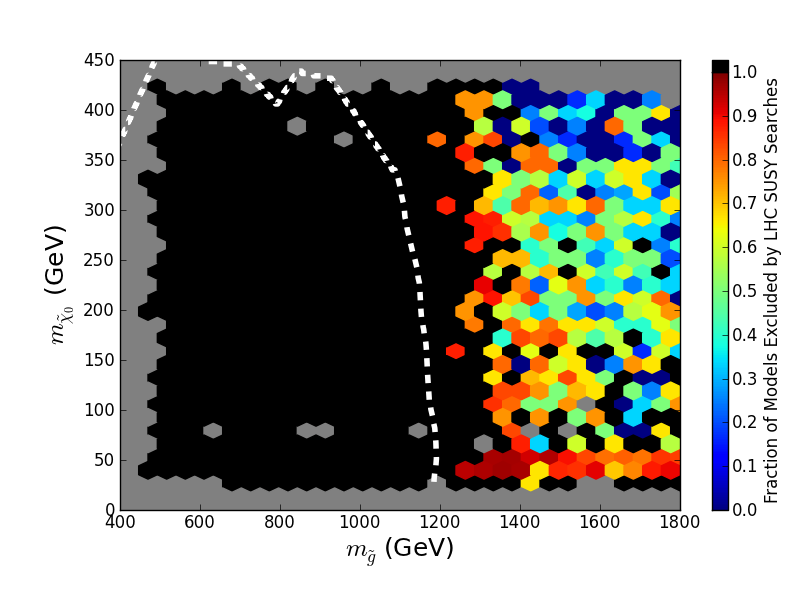}
\hspace{-0.50cm}
\includegraphics[width=3.5in]{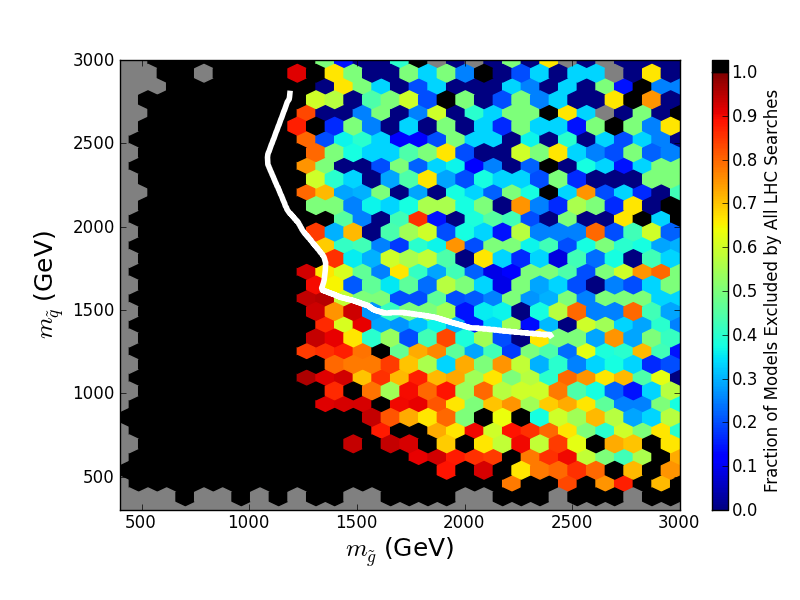}}
\vspace*{-0.10cm}
\caption{Projections of the pMSSM low-FT model coverage efficiencies from the 7 and 8 TeV LHC searches shown in the gluino-LSP (left) 
and the lightest squark-gluino (right) mass plane. The simplified model analysis results from ATLAS are also shown for comparison as the white lines. 
The grey holes in these 
panels arise from the rather low statistics of the low-FT model sample.}
\label{fig00ft}
\end{figure}

Based on this discussion, we might also expect the coverage of the stop-LSP and sbottom-LSP mass planes to be significantly improved for the low-FT set compared with the 
general neutralino model set results discussed above, and this is indeed the case as shown in Fig.~\ref{fig1ft}. The combination of ATLAS searches is clearly found to be particularly 
powerful for the low-FT model set. For the case of light staus, also shown in this Figure, we see that while the coverage is far more complete in the low-FT set than in the 
standard neutralino set, it remains rather uniform as no searches involving taus have been implemented.  Again, we see that
the simplified model limit does not accurately depict the pMSSM coverage for stops.

\begin{figure}[htbp]
\centerline{\includegraphics[width=3.5in]{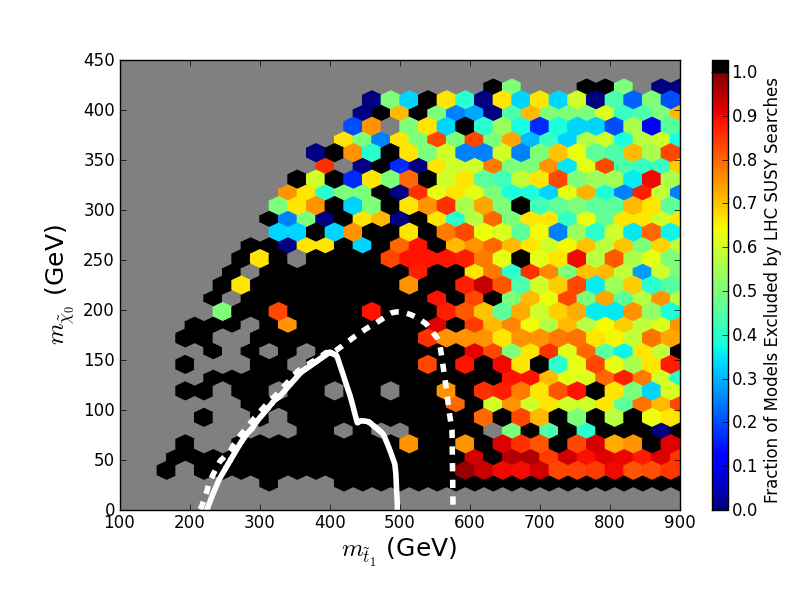}
\hspace{-0.50cm}
\includegraphics[width=3.5in]{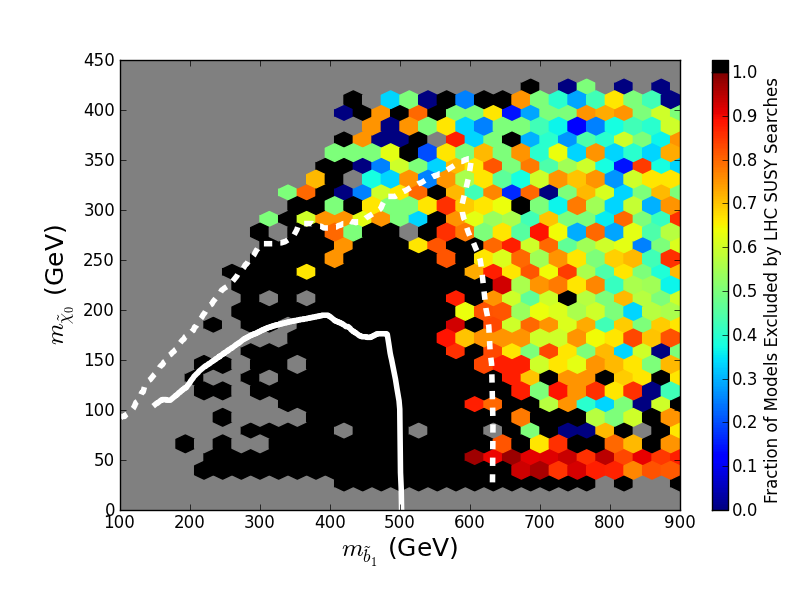}}
\vspace*{0.50cm}
\centerline{\includegraphics[width=3.5in]{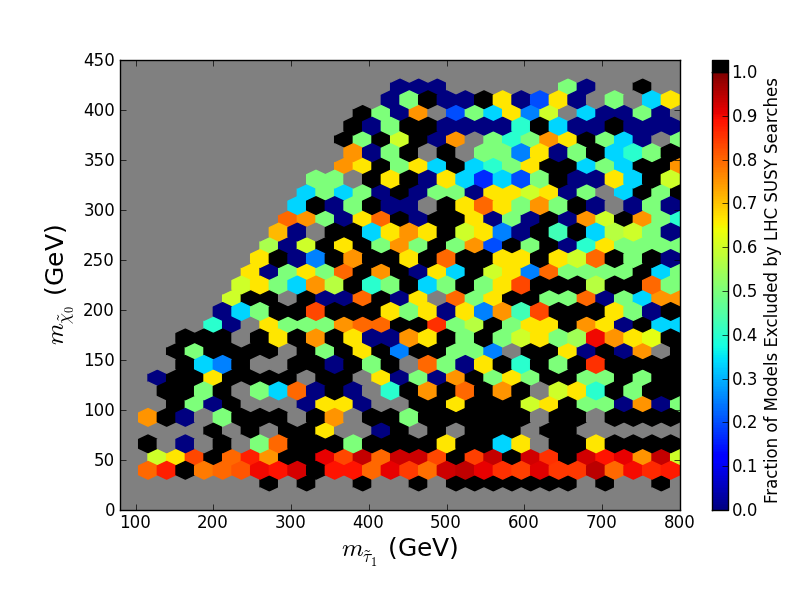}}
\vspace*{-0.10cm}
\caption{Projections of the pMSSM low-FT model coverage efficiencies from the 7 and 8 TeV LHC searches shown in the lightest stop-LSP mass plane 
(top left), the lightest sbottom-LSP mass plane (top right) and for the lightest stau-LSP mass plane (bottom). 
The white lines represent the corresponding 
$95\%$ CL limit results obtained by ATLAS in the simplified model limit as discussed in the text.}
\label{fig1ft}
\end{figure}

Figure~\ref{fig2ft} shows the coverage of the 1\textsuperscript{st}/2\textsuperscript{nd} generation squark-LSP mass plane for the low-FT set which should be compared with the analogous results for 
the standard neutralino model sample in Fig.~\ref{fig2}, shown above. As before, we see that the coverage is greatest for $\tilde u_L$ and $\tilde d_L$, then followed by 
$\tilde u_R$ with the least coverage for $\tilde d_R$. However, in all cases, we see that the coverage is far more complete for the low-FT set while also simultaneously 
being generally more uniform across the mass plane than for the general neutralino model set (even though lower mass regions are somewhat more disfavored).

\begin{figure}[htbp]
\centerline{\includegraphics[width=3.5in]{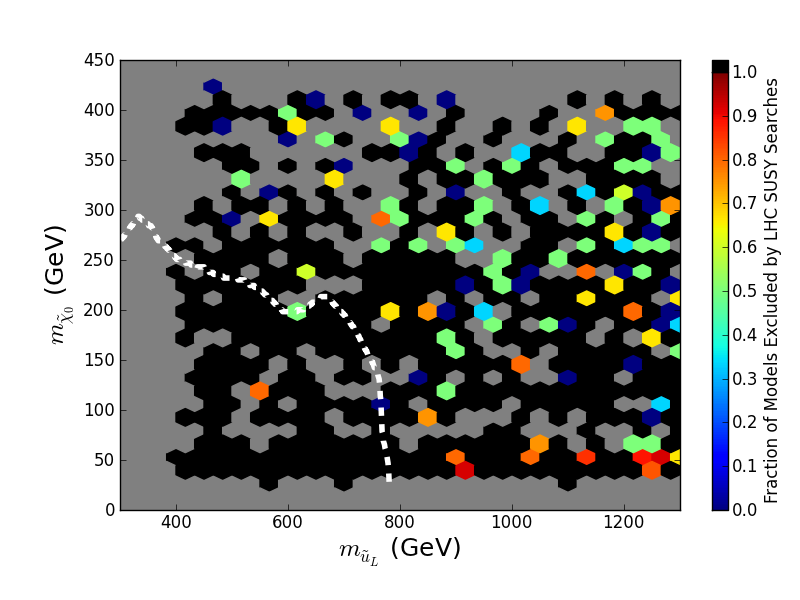}
\hspace{-0.50cm}
\includegraphics[width=3.5in]{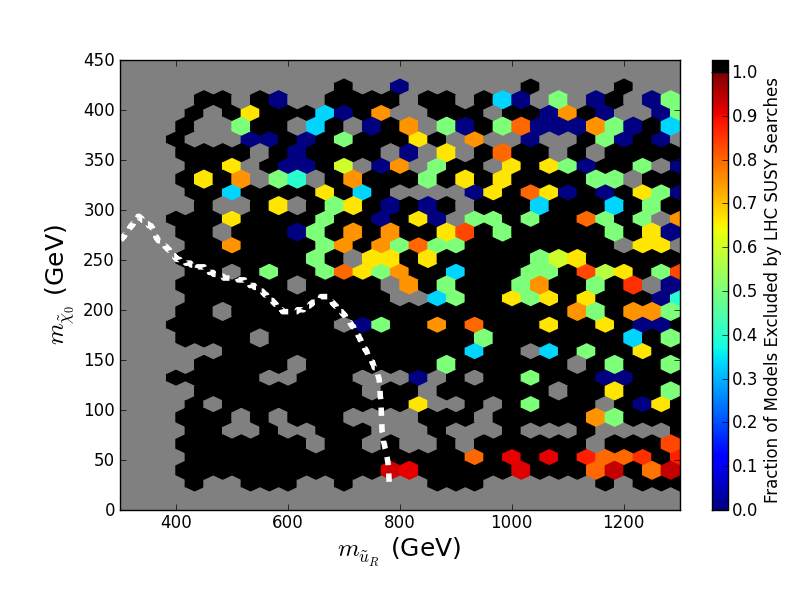}}
\vspace*{0.50cm}
\centerline{\includegraphics[width=3.5in]{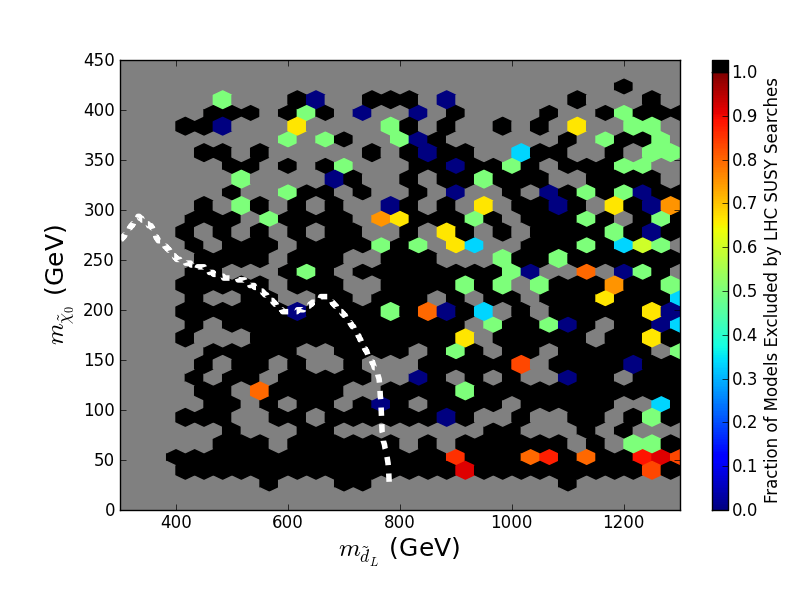}
\hspace{-0.50cm}
\includegraphics[width=3.5in]{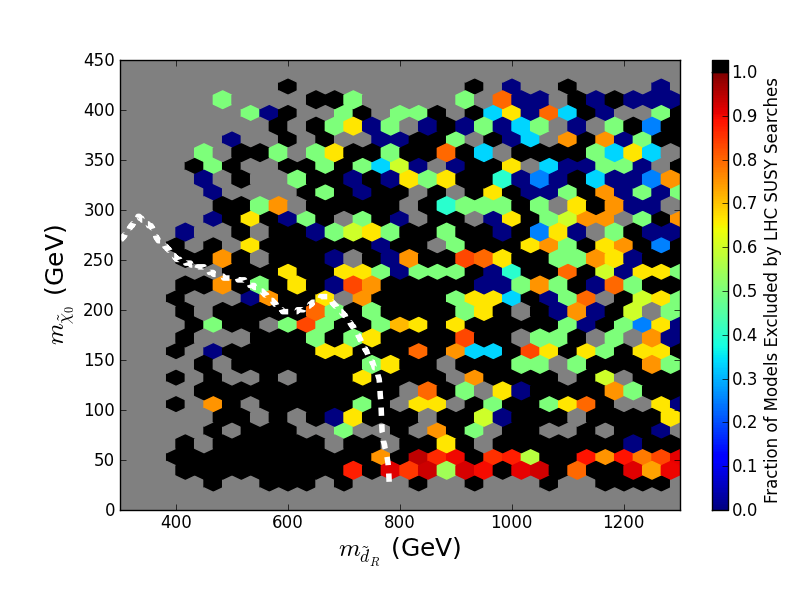}}
\vspace*{-0.10cm}
\caption{Same as the previous figure but now for $\tilde u_L$ (top left),  $\tilde u_R$ (top right), $\tilde d_L$ (bottom left), 
and $\tilde d_R$ (bottom right).}
\label{fig2ft}
\end{figure}

Figure~\ref{fig3ft} shows the analogue of Fig.~\ref{fig3} for the low-FT model set. In all cases the 7 and 8 TeV LHC coverage is, of course, 
more complete. In the upper panels, we see that the sensitivity to light sleptons is improved in the low-FT set (although the mass regions that are completely excluded remain small). The enhanced sensitivity to light sleptons most likely arises from the ubiquitous presence of a light chargino (with a mass below $\sim 460$ GeV). Having a light chargino means that light sleptons can be excluded not only via slepton pair production, but also by enhancing the detectability of the chargino. The latter possibility occurs when the slepton is an intermediate in the chargino decay cascade, producing a much more distinctive signature (hard leptons) than the soft gauge bosons typically produced in electroweakino cascades. In the bottom panels of Figure~\ref{fig3ft}, we see that the exclusion efficiency for models with light charginos has also improved somewhat; part of this improvement may result from an increased frequency of light sleptons (which are more common for the Low-FT model set because of their role as co-annihilators) enhancing the chargino visibility through the mechanism described above. The LHC searches are particularly sensitive to models containing light second charginos, in which case all 6 electroweak gauginos are light (in contrast with the general neutralino model set, in which the bino is frequently heavier than both charginos).{\footnote {The second chargino 
is always found to be at least $\sim 100$ GeV heavier than the lighter one but the distribution also peaks near this value due to the nature of the parameter scan.}} In this case, the 4 lepton search is highly effective, since a large number of leptons are frequently produced in cascades between the gaugino multiplets; although some of these leptons may be rather soft, they can still pass the low $p_T$ thresholds allowed by the high multiplicity lepton searches. Recall that in many cases, the charginos may be produced dominantly through decays of light stops and sbottoms, boosting their production cross-section and making them even more accessible to searches for multi lepton final states.

\begin{figure}[htbp]
\centerline{\includegraphics[width=3.5in]{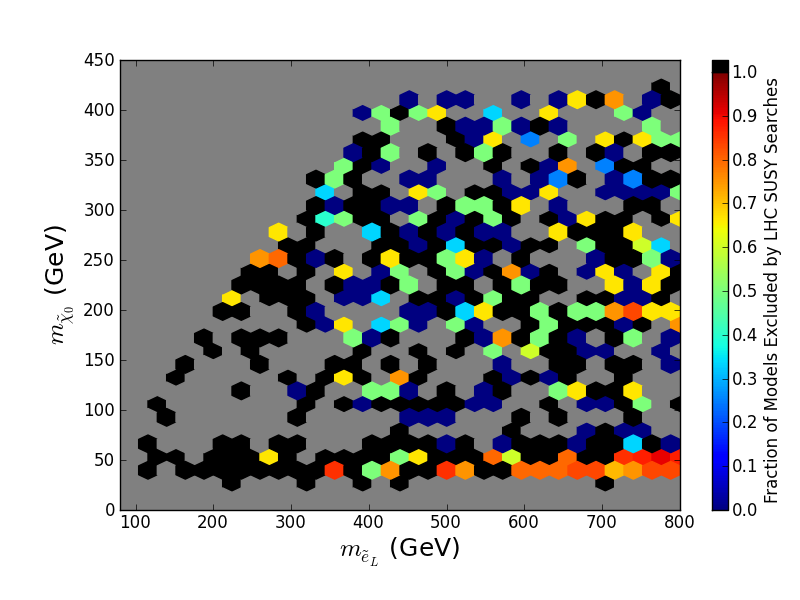}
\hspace{-0.50cm}
\includegraphics[width=3.5in]{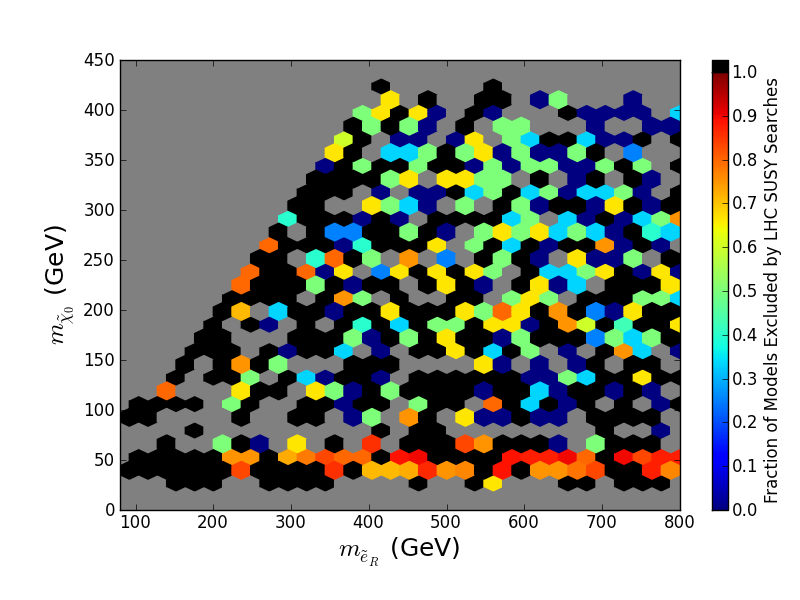}}
\vspace*{0.50cm}
\centerline{\includegraphics[width=3.5in]{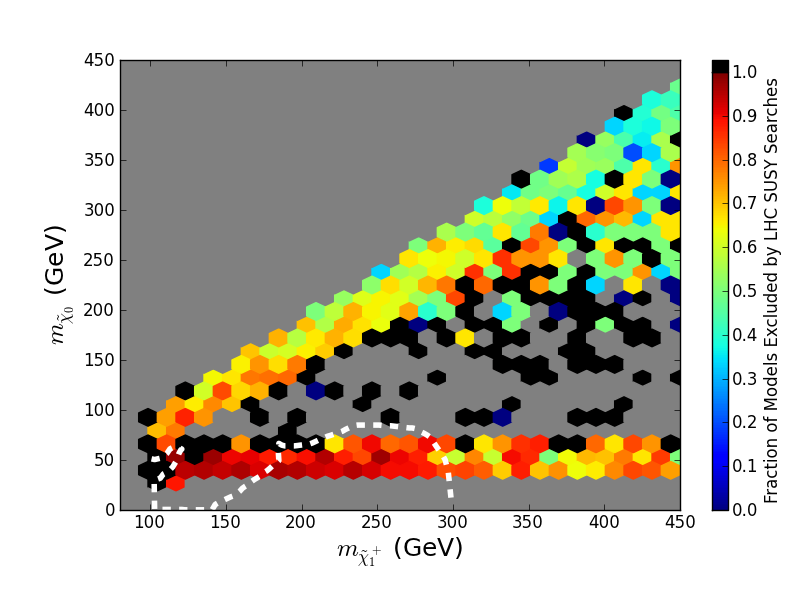}
\hspace{-0.50cm}
\includegraphics[width=3.5in]{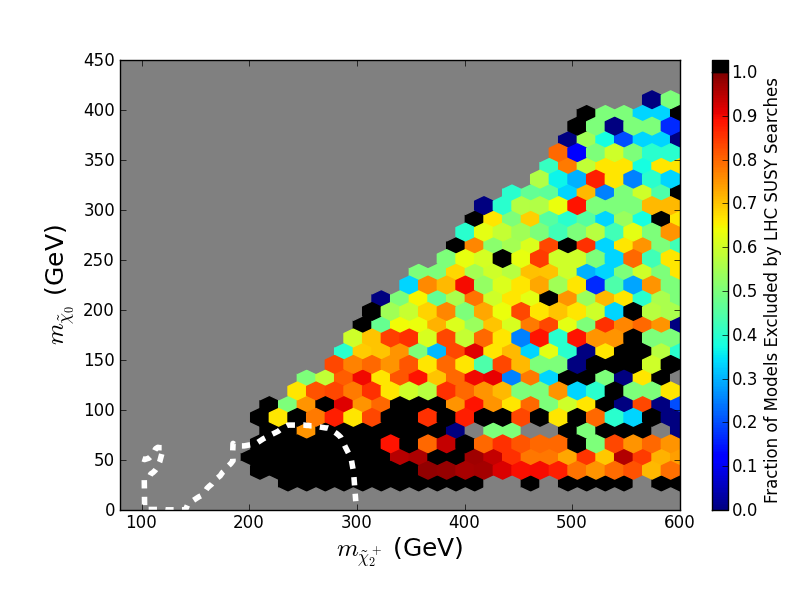}}
\vspace*{-0.10cm}
\caption{Same as the previous figure but now for $\tilde e_L$ (top left),  $\tilde e_R$ (top right), $\tilde \chi_1^\pm$ (bottom left), 
and $\chi_2^\pm$ (bottom right).}
\label{fig3ft}
\end{figure}

Next, we note in Fig.~\ref{fig4ft} the analogues of the results shown for the neutralino set above in Fig.~\ref{fig4}, comparing the effectiveness of the 
3\textsuperscript{rd} generation searches with the ``vanilla'' jets (+ leptons) + MET analyses (entries 1-3 in Table~\ref{SearchList7} and 1-4 in Table~\ref{SearchList7}). As we saw in Tables~\ref{SearchList7} and~\ref{SearchList8}, both sets of searches are significantly more effective in the low-FT model set. This Figure shows that both search categories have similar exclusion reaches for models with light 3rd generation squarks, although for light stops the ``vanilla'' searches are again slightly more important in the compressed region.

\begin{figure}[htbp]
\centerline{\includegraphics[width=3.5in]{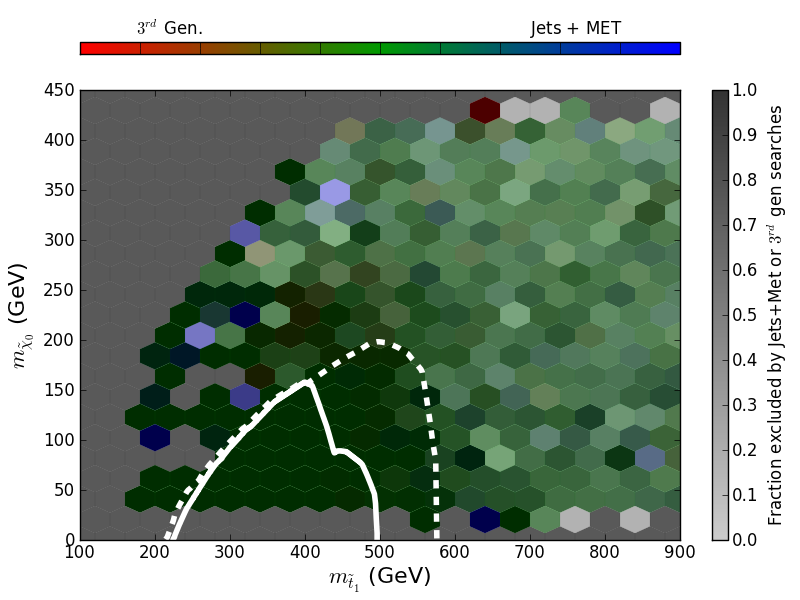}
\hspace{-0.50cm}
\includegraphics[width=3.5in]{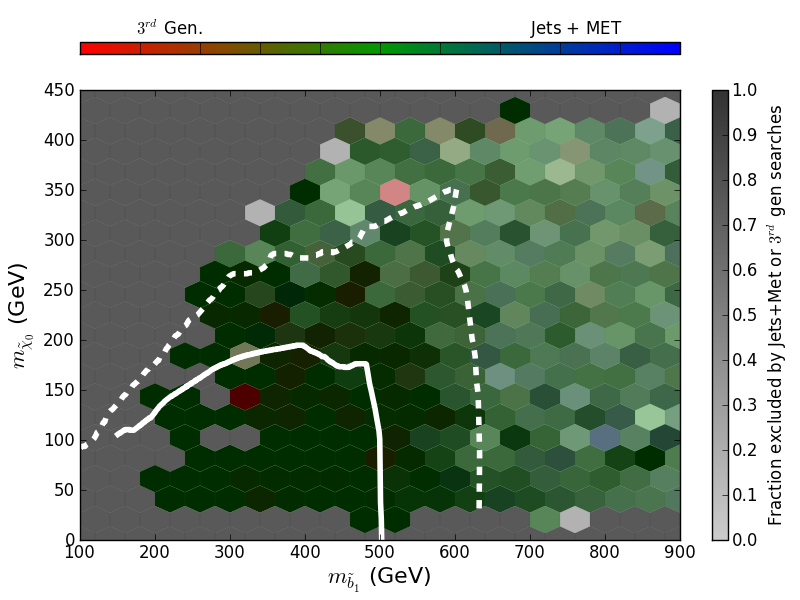}}
\vspace*{-0.10cm}
\caption{Comparison of the contributions to model exclusion arising from jets (+ leptons) + MET and 3\textsuperscript{rd} generation searches for light stops (left) and 
light sbottoms (right) in the low-FT model set. The color coding seen here is as described above.}
\label{fig4ft}
\end{figure}

\section{14 TeV Results}

In addition to the 7 and 8 TeV LHC searches, future data taking and enhanced analyses at $\sim$ 14 TeV will greatly extend the expected coverage of the 
pMSSM parameter space for both LSP types. In this section, we consider the impact of one of the most powerful of these searches to be performed by ATLAS, namely the zero-lepton jets + 
MET final state, as presented in their contribution to the Update of the European Strategy for Particle Physics~\cite {ATLAS-EP}}. We have performed our own version of this analysis 
in a manner identical to that employed above for the 7 and 8 TeV LHC by following ATLAS as closely as possible. We note that in this analysis ATLAS has somewhat underestimated 
the effect of systematic errors, so that it is likely that our results will correspondingly overestimate the efficiency of the pMSSM model 
coverage for this search. We further note that in extrapolating from 300 fb$^{-1}$ to 3 ab$^{-1}$ luminosity scaling of the required signal rate has been employed to obtain 
the results shown below.

In order to simplify our analysis and to obtain the results presented here in a relatively timely fashion, we consider only the $\sim 30.7 (10.2)$k 
neutralino (gravitino) LSP models that survive the 7 and 8 TeV LHC analyses above and also predict a Higgs mass of $126\pm 3$ GeV, as well as the 
corresponding subset of $\sim 3.1$k surviving low-FT models. Given the high luminosities, these subsets of models alone required $\sim 2 \cdot 10^6$ core-hrs of CPU 
to generate 14 TeV signal events and perform the necessary analysis{\footnote {Note that since this represents less than $\sim 10\%$ of our total set of models implying that a 
study of these sets in their entirety would have required $\sim 20-25 \cdot 10^6$ core-hrs of CPU which is far beyond our current capabilities.}}.
Note that since the dominant direct impact of this search is on the production of colored sparticles, 
we have restricted the discussion of our results below to the impact these searches have in the gluino-squark-(N)LSP sector.  We note that since the results of the 7 and 
8 TeV analyses are essentially independent of the Higgs mass, it is quite likely that the results presented here for this narrow Higgs mass range would in fact be applicable, 
at least to a very good approximation, to the entire neutralino, gravitino and low-FT model sets.

\begin{figure}[htbp]
\centerline{\includegraphics[width=3.5in]{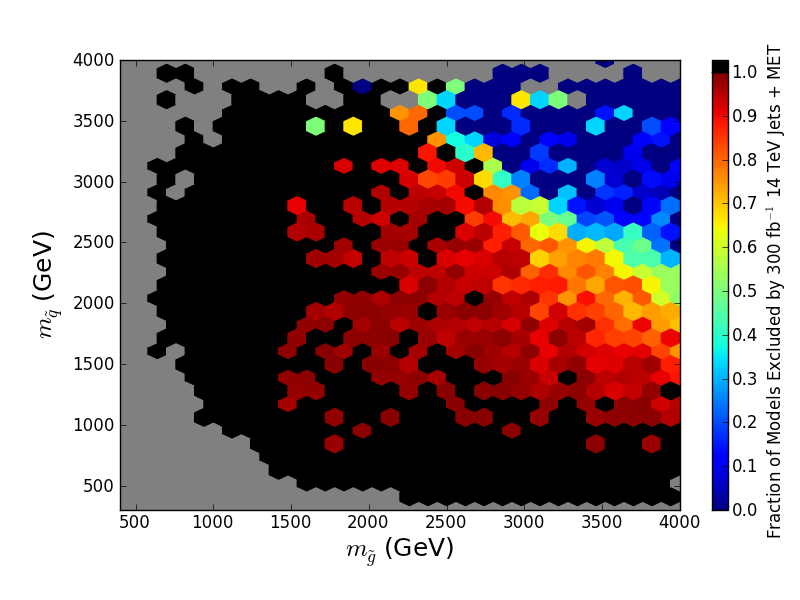}
\hspace{-0.50cm}
\includegraphics[width=3.5in]{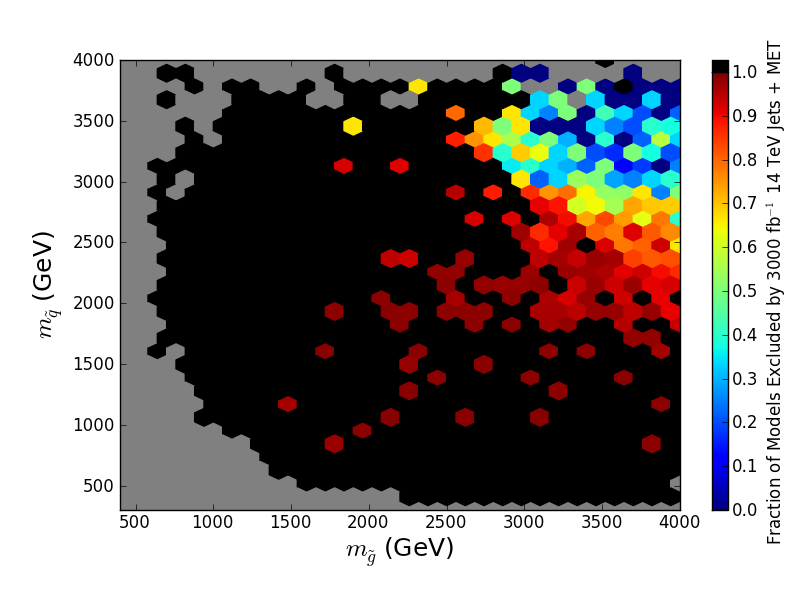}}
\vspace*{0.50cm}
\centerline{\includegraphics[width=3.5in]{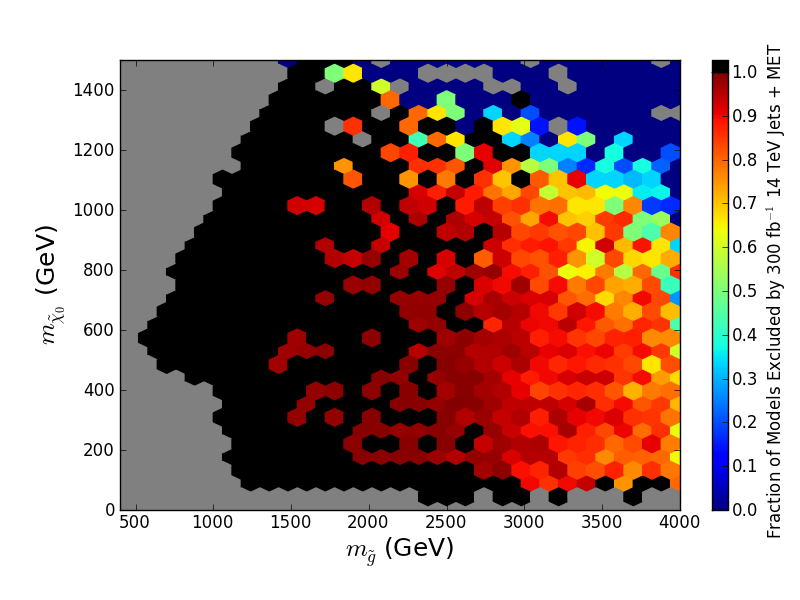}
\hspace{-0.50cm}
\includegraphics[width=3.5in]{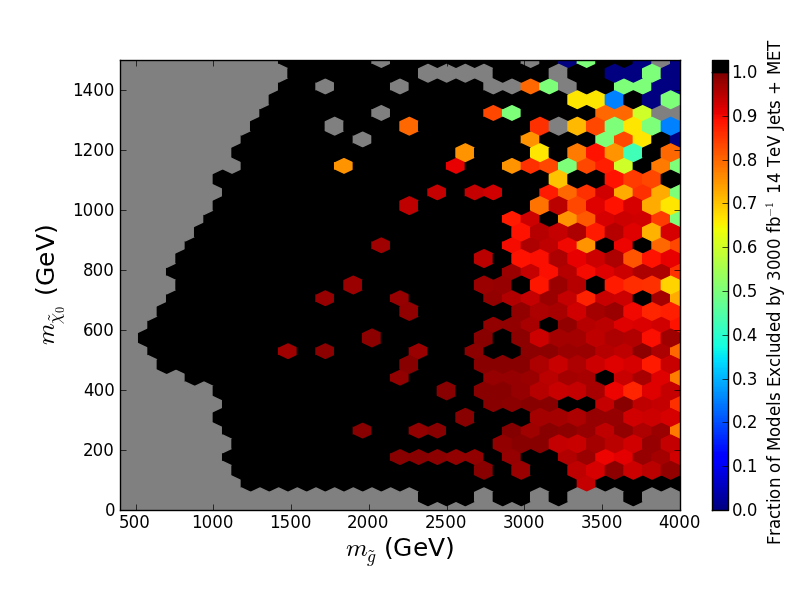}}
\vspace*{-0.10cm}
\caption{Expected results from a jets + MET search at the 14 TeV LHC assuming an integrated luminosity of 300 fb$^{-1}$ (left) and 
3000 fb$^{-1}$ (right), in the lightest squark-gluino and LSP-gluino mass planes.}
\label{figxx1}
\end{figure}

Let us first consider the neutralino LSP model set. 
In Fig.~\ref{figxx1} we see the pMSSM search efficiencies in the lightest squark-gluino and the gluino-LSP mass planes for the general neutralino  model sample at 14 TeV 
arising from the jets + MET analysis with either 300 or 3000 fb$^{-1}$ of integrated luminosity. Here we see that even with the lower value of the integrated 
luminosity this one analysis provides a substantial coverage of these models. Specifically, we find that $92.1 (97.5)\%$ of the models in this subset will be excluded 
by this analysis assuming 300 (3000) fb$^{-1}$ of data, subject to the caveat mentioned above. Given the Higgs mass independence of the SUSY searches as was discussed 
above, we would expect these fractions to be roughly valid for the entire neutralino model set. In particular, in these figures we see that increasing the 
integrated luminosity makes a significant impact on the overall pMSSM model coverage.  Although this coverage is indeed very significant, we observe that models with 
1\textsuperscript{st}/2\textsuperscript{nd} generation squarks as light as $\sim 700-800$ GeV and/or gluinos as light as $\sim 1.5$ TeV still survive this single analysis even at high 
integrated luminosities. Interestingly, surviving models with light squarks and gluinos remain undetected not only because of spectrum compression, but also because of specific decay patterns for the squark and/or gluino which nearly always produce high-$p_T$ leptons. In such cases the models will immediately fail the lepton veto and so remain undetected.  Clearly adding additional analyses, specifically those targeting final state leptons (with possible $b$-tagged jets), will only increase the model coverage and will compensate for the underestimated systematics.

\begin{figure}[htbp]
\centerline{\includegraphics[width=3.5in]{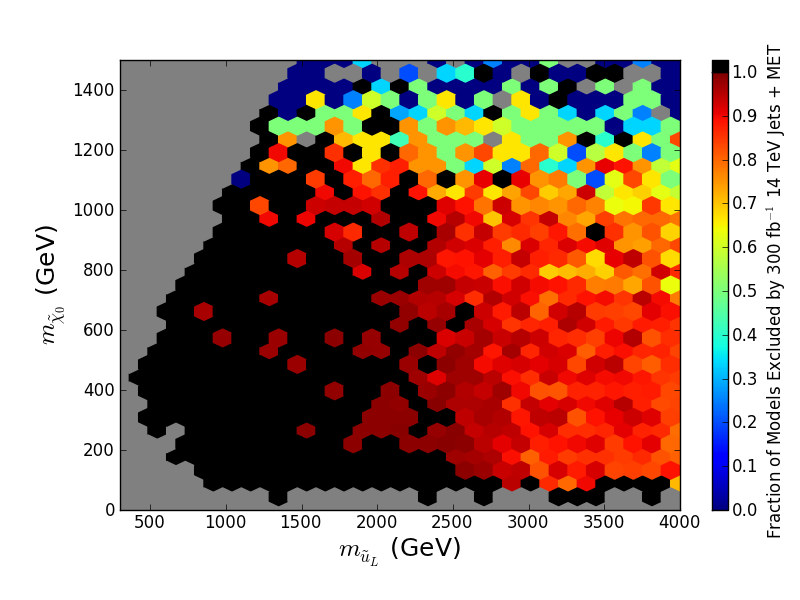}
\hspace{-0.50cm}
\includegraphics[width=3.5in]{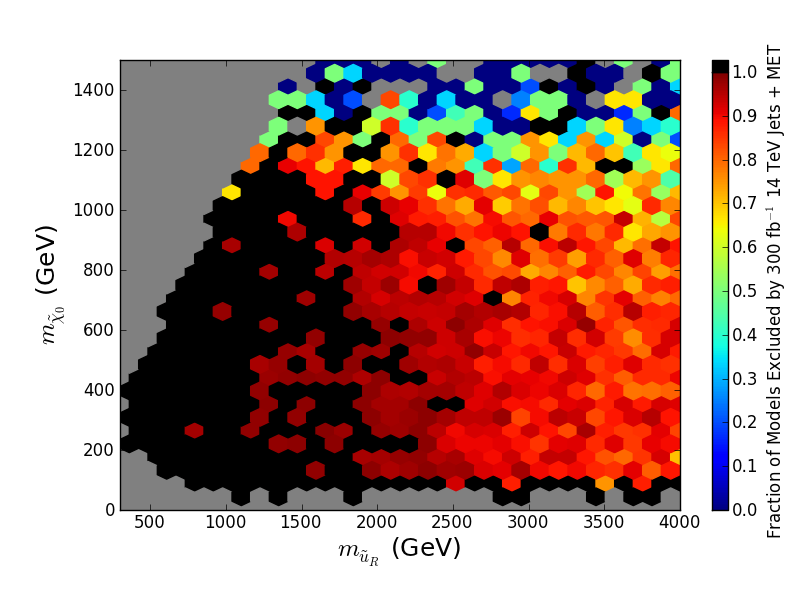}}
\vspace*{0.50cm}
\centerline{\includegraphics[width=3.5in]{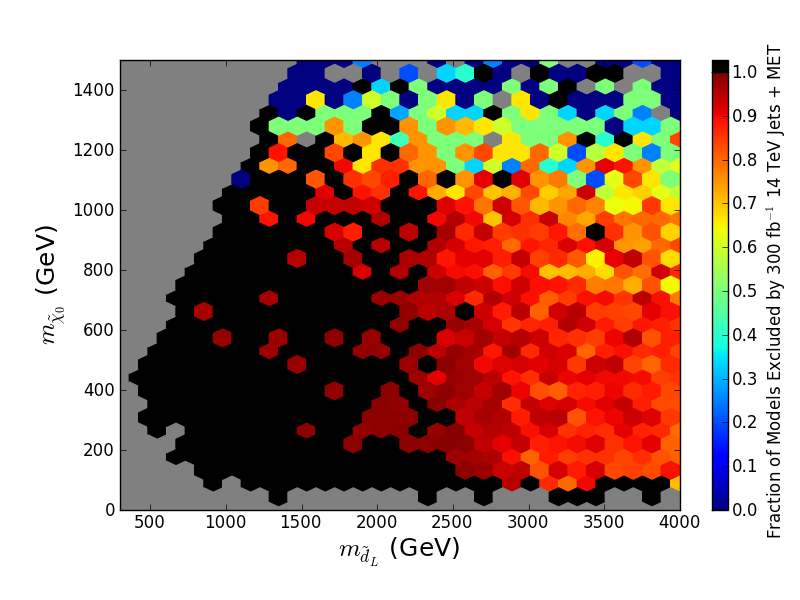}
\hspace{-0.50cm}
\includegraphics[width=3.5in]{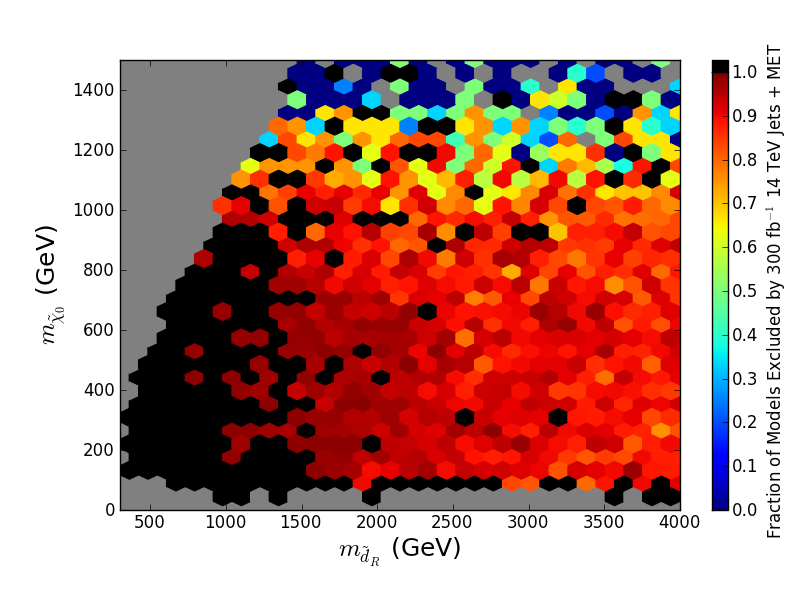}}
\vspace*{-0.10cm}
\caption{Jets + MET search results at a 14 TeV LHC assuming an integrated luminosity of 300 fb$^{-1}$ in the LSP-squark mass plane.}
\label{figxx2}
\end{figure}

\begin{figure}[htbp]
\centerline{\includegraphics[width=3.5in]{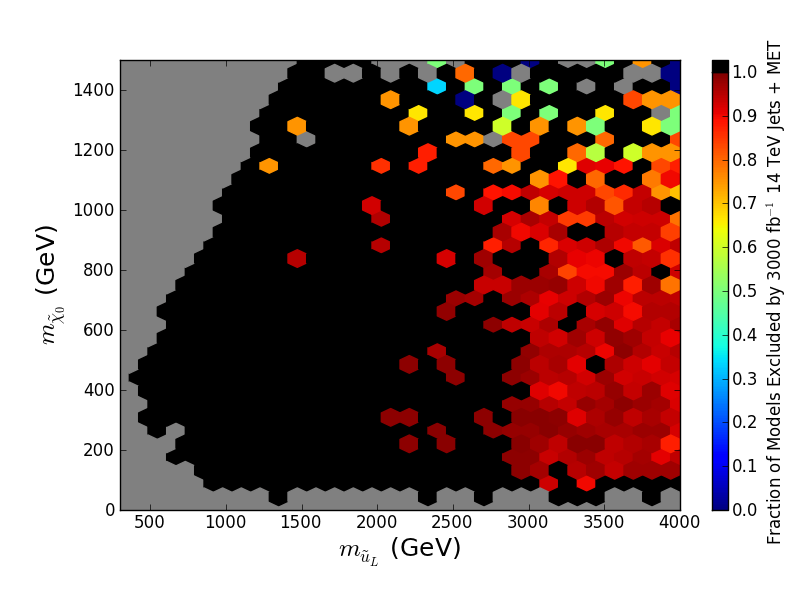}
\hspace{-0.50cm}
\includegraphics[width=3.5in]{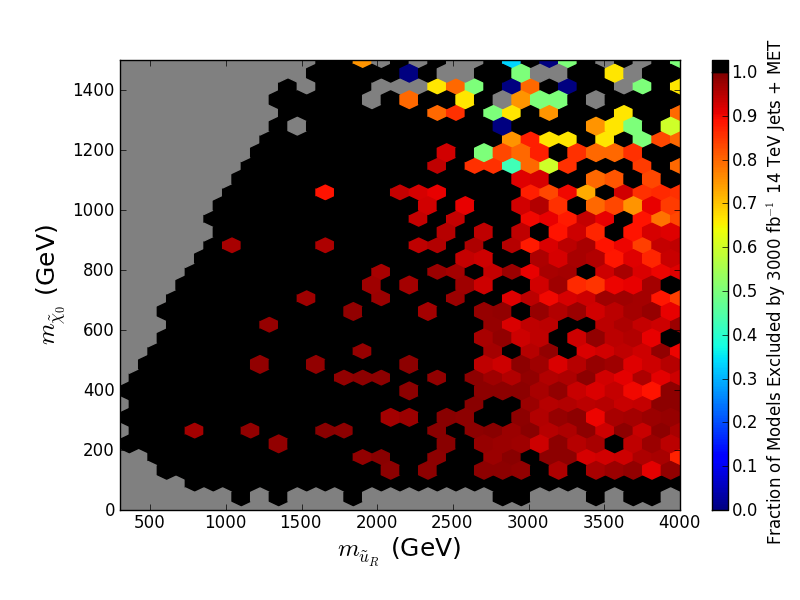}}
\vspace*{0.50cm}
\centerline{\includegraphics[width=3.5in]{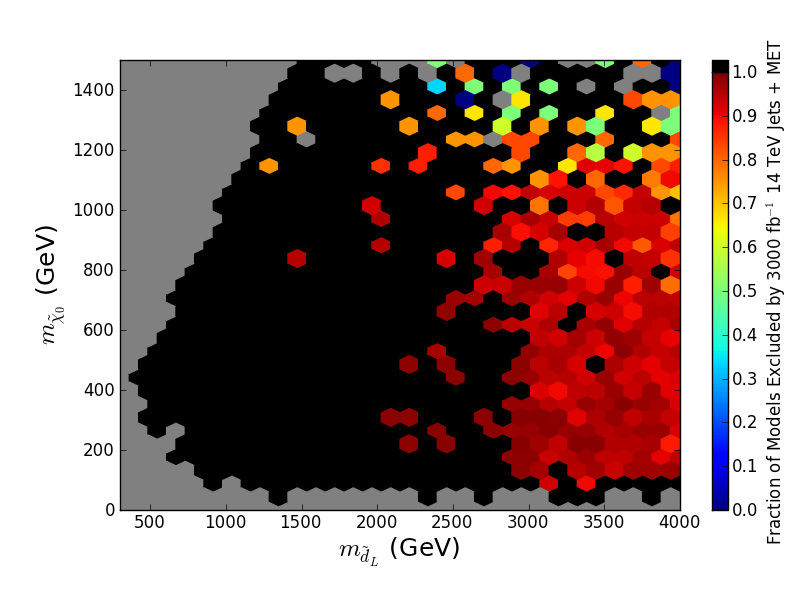}
\hspace{-0.50cm}
\includegraphics[width=3.5in]{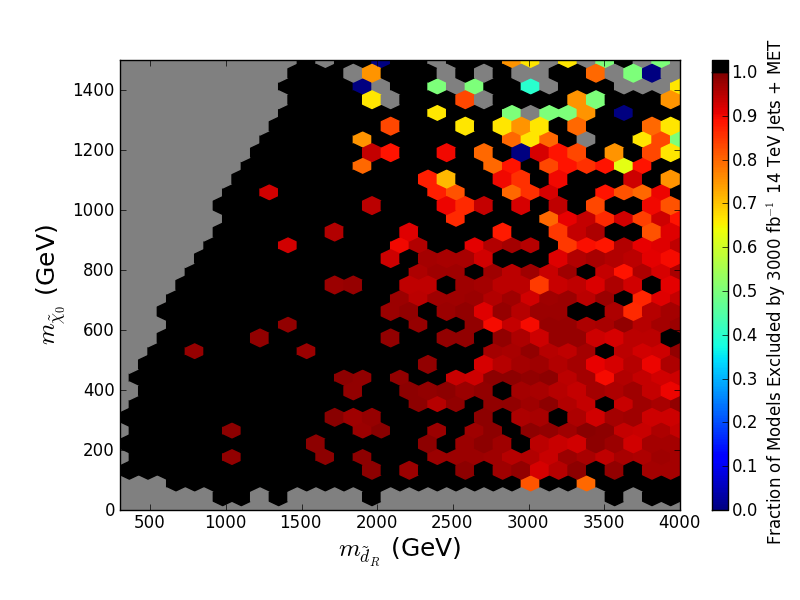}}
\vspace*{-0.10cm}
\caption{Same as the previous figure but now for an integrated luminosity of 3000 fb$^{-1}$.}
\label{figxx3}
\end{figure}

In order to further elucidate the important case of potentially light squarks, Figs.~\ref{figxx2} and ~\ref{figxx3} show the search efficiencies in the squark-LSP 
mass plane separately for the $\tilde u_L,~ \tilde u_R,~ \tilde d_L$ and $\tilde d_R$ squarks at 14 TeV for an integrated luminosity of 300 and 3000 fb$^{-1}$, 
respectively. Here we see a number of things: ($i$) since $\tilde u_L$ and $\tilde d_L$ are similar in mass they are produced together and increase the 
corresponding signal rate as seen before. Thus it is quite rare (but not impossible) for light $\tilde u_L, \tilde d_L$ to still remain after the 14 TeV jets + MET search is 
performed. ($ii$) Since $\tilde u_R$ and $\tilde d_R$ have uncorrelated masses and are iso-singlets, each has a suppressed search reach compared to their 
corresponding left-handed partners. In particular, the $\tilde d_R$ production is also further suppressed by the PDFs and we see that quite light $\tilde d_R$ squarks 
would remain viable after these searches. We note the existence of a model with a rather light LSP, below $\sim 100$ GeV in mass, that remains viable at the HL-LHC 
since all the corresponding squark and gluino masses are in excess of $\sim 3.3$ TeV. It would be interesting to see how models with light squarks would fare when 
additional channels incorporating hard leptons were included in a more complete analysis.

Although the model coverage is seen to be quite significant from just the jets + MET channel alone, further more detailed study of the neutralino pMSSM model set 
at the 14 TeV LHC is certainly warranted.

Models with gravitino LSPs can also be searched for by employing the 14 TeV zero-lepton, jets + MET analysis, but here we begin with a much smaller sample of models, ($\sim 10.2$k), that 
have both survived the 7 and 8 TeV searches and which also predict a Higgs mass of $m_H=126\pm 3$ GeV. A priori, since we had earlier found that the jets + MET searches at 7/8 TeV 
were less effective at covering the parameter space for gravitino LSPs compared with neutralino LSPs, we would expect that this single 14 TeV analysis will exclude a smaller fraction of the remaining gravitino models than was excluded by the same search for the neutralino model set. The main reasons for the relative ineffectiveness of the Jets+MET search are the prevalence of displaced decays (displaced jets will frequently fail quality cuts and cause the jet to be removed or the event to be rejected) and of models in which all SUSY decays produce stable charged particles, instead of missing energy, as their endpoint. Indeed, we find that only 77.7 (87.8)$\%$ of the models in this gravitino subset will be excluded by this 
single analysis assuming 300 (3000) fb$^{-1}$ of data at 14 TeV. Including other searches at 14 TeV, particularly searches for heavy stable charged particles, 
would be most advantageous in covering a larger fraction of the gravitino pMSSM model set and would significantly alter the characteristics of the model coverage as discussed below. Indeed, by analogy with the 7/8 TeV results, we would expect that the consideration of all 14 TeV searches would increase the fraction of models excluded well beyond the fraction of excluded neutralino LSP models.

\begin{figure}[htbp]
\centerline{\includegraphics[width=3.5in]{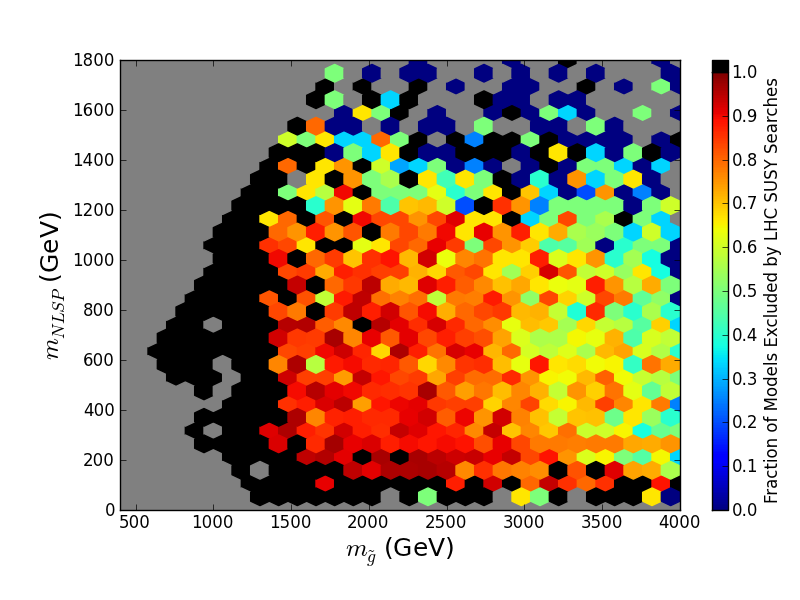}
\hspace{-0.50cm}
\includegraphics[width=3.5in]{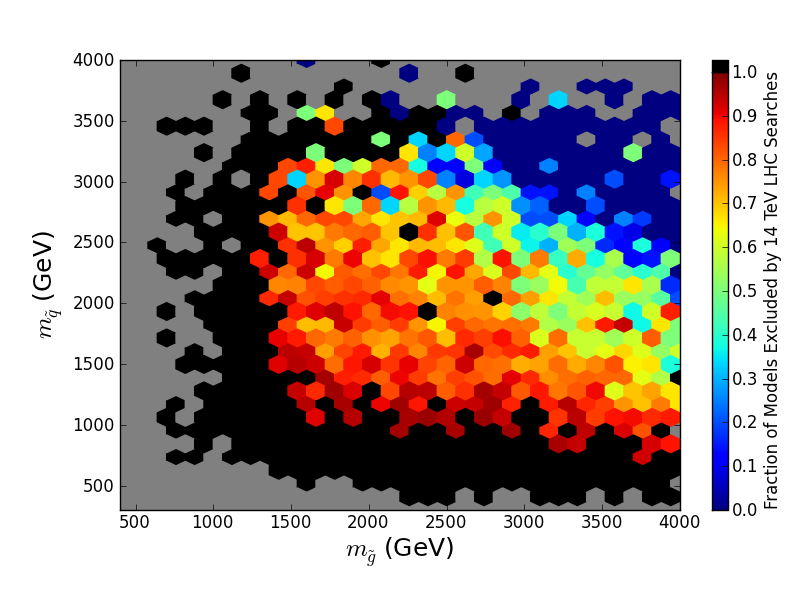}}
\vspace*{0.50cm}
\centerline{\includegraphics[width=3.5in]{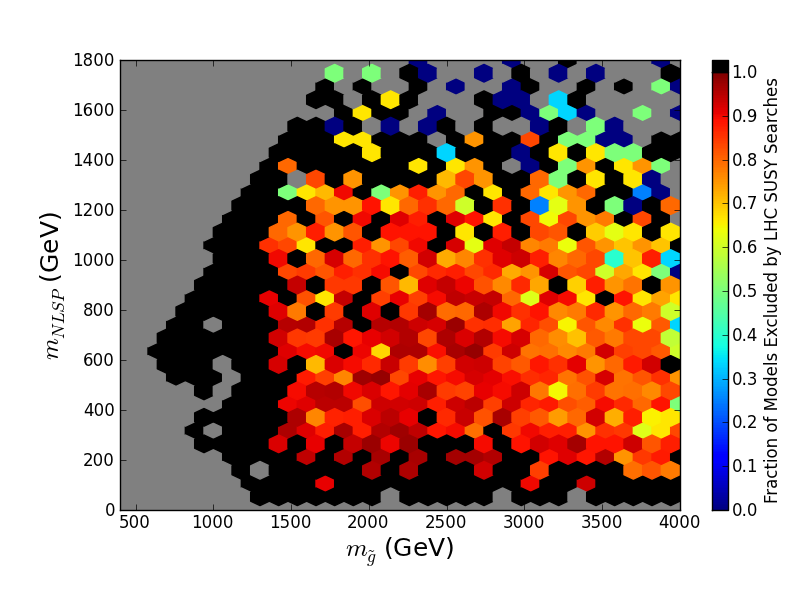}
\hspace{-0.50cm}
\includegraphics[width=3.5in]{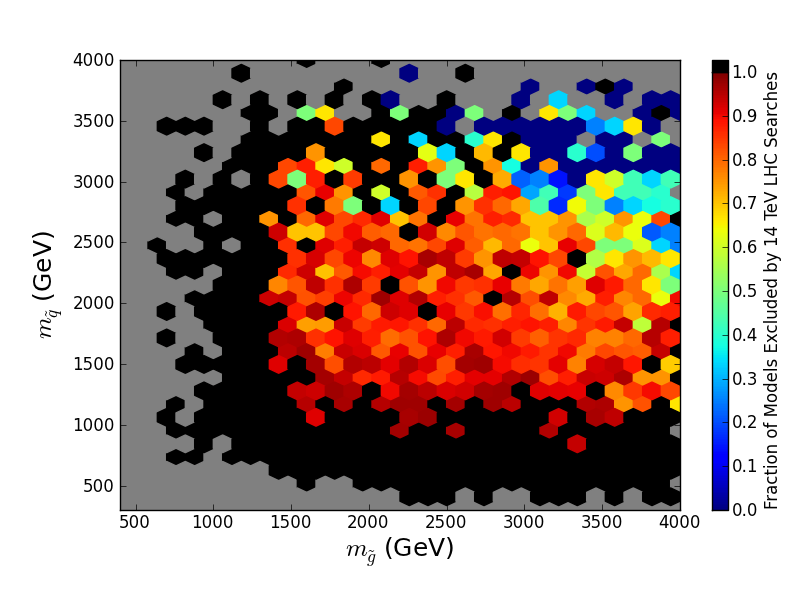}}
\vspace*{-0.10cm}
\caption{Expected results from a jets + MET search at the 14 TeV LHC for the gravitino model set assuming an integrated luminosity of 300 fb$^{-1}$ (top) and 
3000 fb$^{-1}$ (bottom), as shown in both the lightest squark-gluino and NLSP-gluino mass planes.}
\label{figxx1g}
\end{figure}

To further understand the search coverage in the gravitino LSP case, it is useful to examine the results shown in Fig.~\ref{figxx1g} which is the 14 TeV (for both 300 and 3000 fb$^{-1}$) 
analog of Fig.~\ref{fig1g}, shown earlier and based on the results obtained from the 7/8 TeV analyses. We remind the reader that since the 7/8 TeV searches included those for {\it both} 
MET and non-MET final states, the response of the gravitino LSP set to the MET-only search at 14 TeV will be quite different. From these Figures we see that essentially all models 
with gluino masses below $\sim 1.3 (1.4)$ TeV are excluded yet the possibility of squarks below $\sim 1.0 (1.2)$ TeV remains viable with a luminosity of 300(3000) fb$^{-1}$. 
A careful examination reveals that the lower limit on the gluino mass is not much influenced by the addition of the higher luminosity data but one does see that the orange $\sim 80-90\%$ 
exclusion regions are pushed out to larger gluino masses as the luminosity increases. In comparison to the neutralino LSP case, both the lightest squark-gluino mass plane and the 
NLSP-gluino mass plane display a much greater density of allowed models at lower sparticle masses, and the shapes of the remaining allowed regions are quite different for the two model 
sets.

\begin{figure}[htbp]
\centerline{\includegraphics[width=3.5in]{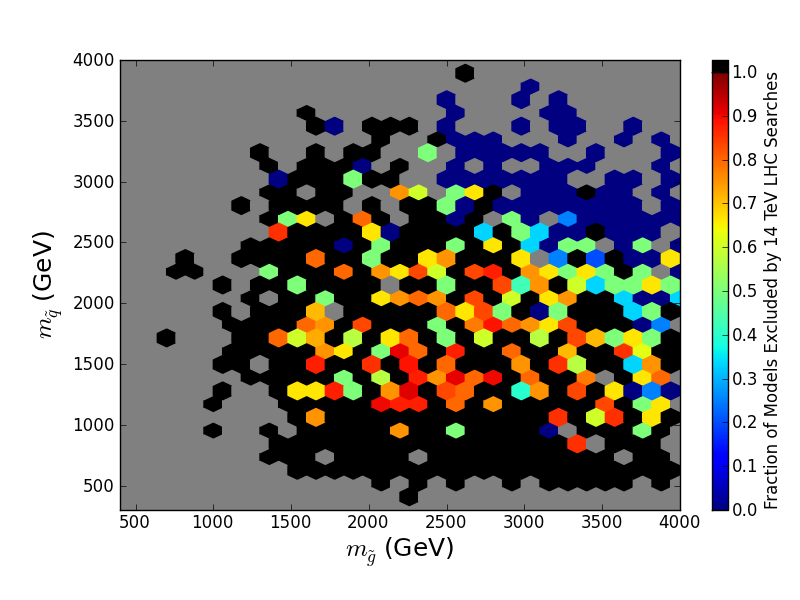}
\hspace{-0.50cm}
\includegraphics[width=3.5in]{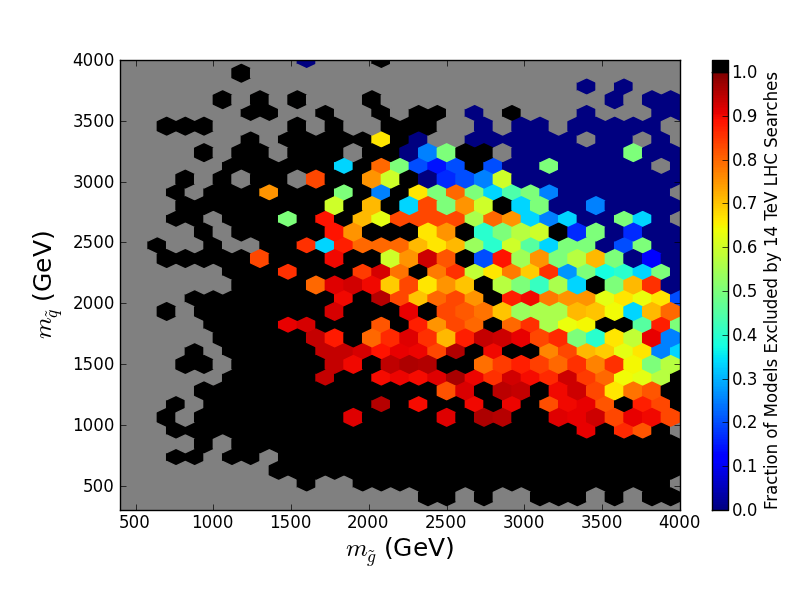}}
\vspace*{0.50cm}
\centerline{\includegraphics[width=3.5in]{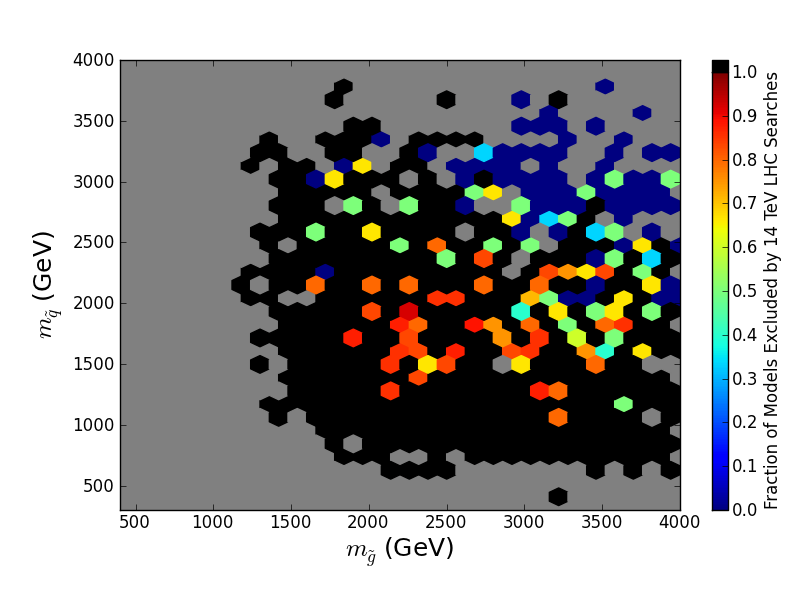}
\hspace{-0.50cm}
\includegraphics[width=3.5in]{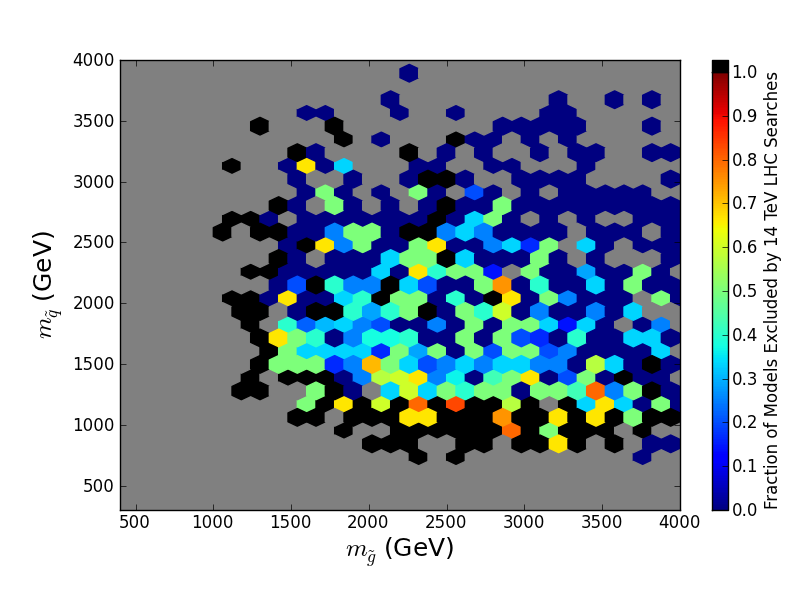}}
\vspace*{-0.10cm}
\caption{Projections of the gravitino pMSSM model coverage efficiencies from the 300 fb$^{-1}$, 14 TeV ATLAS jets plus MET searches shown in the lightest squark-gluino mass plane 
for various model subsets: (top left) models with NLSPs that have displaced decays yielding observable decay products, (top right) models with NLSPs that are detector stable and 
invisible or have invisible decay products, (lower left) models with NLSPs that decay promptly, producing visible decay products and (bottom right) models in which the NLSP is 
detector-stable and charged.}
\label{figxx2g}
\end{figure}

\begin{figure}[htbp]
\centerline{\includegraphics[width=3.5in]{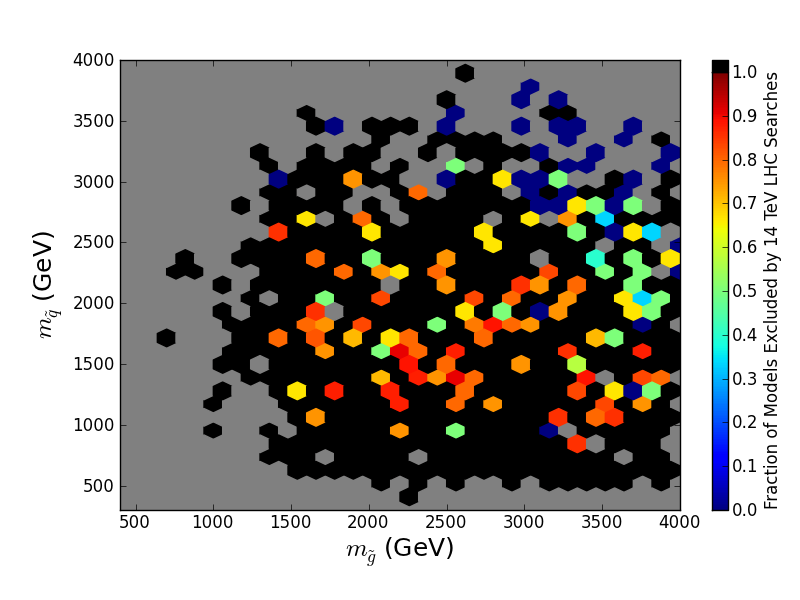}
\hspace{-0.50cm}
\includegraphics[width=3.5in]{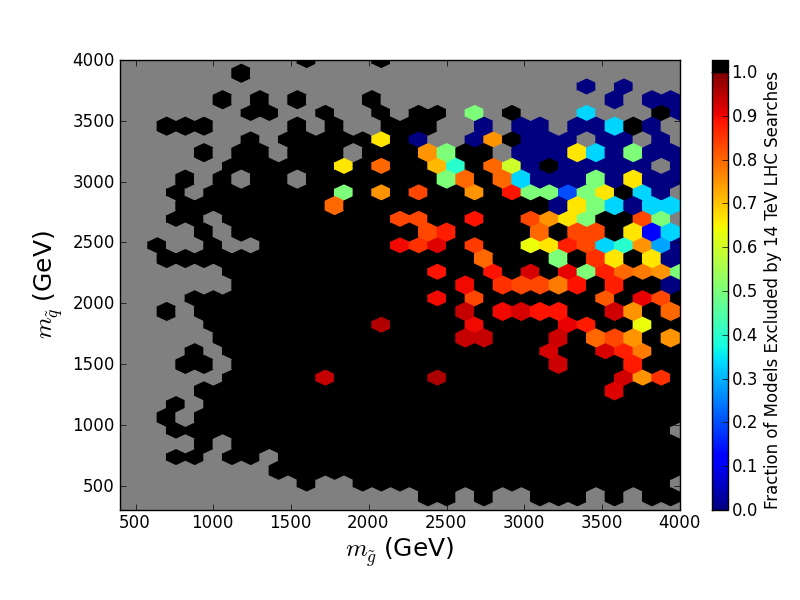}}
\vspace*{0.50cm}
\centerline{\includegraphics[width=3.5in]{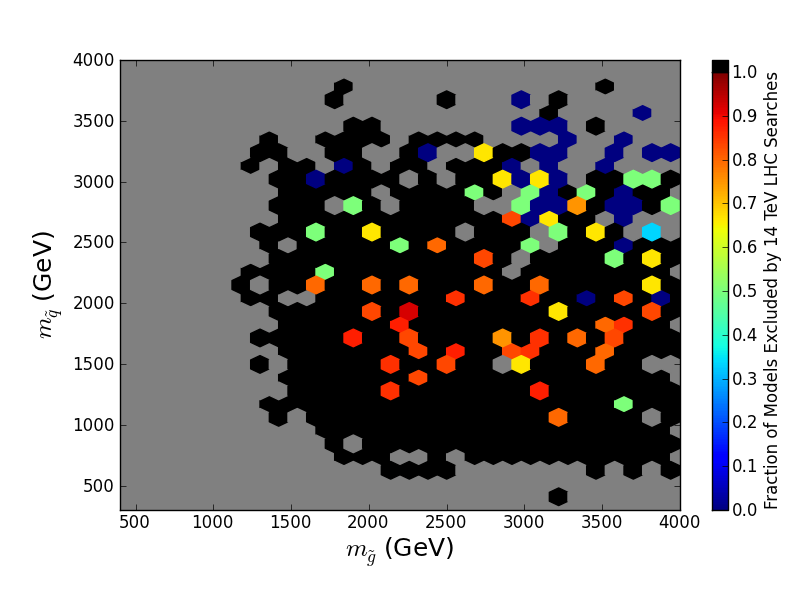}
\hspace{-0.50cm}
\includegraphics[width=3.5in]{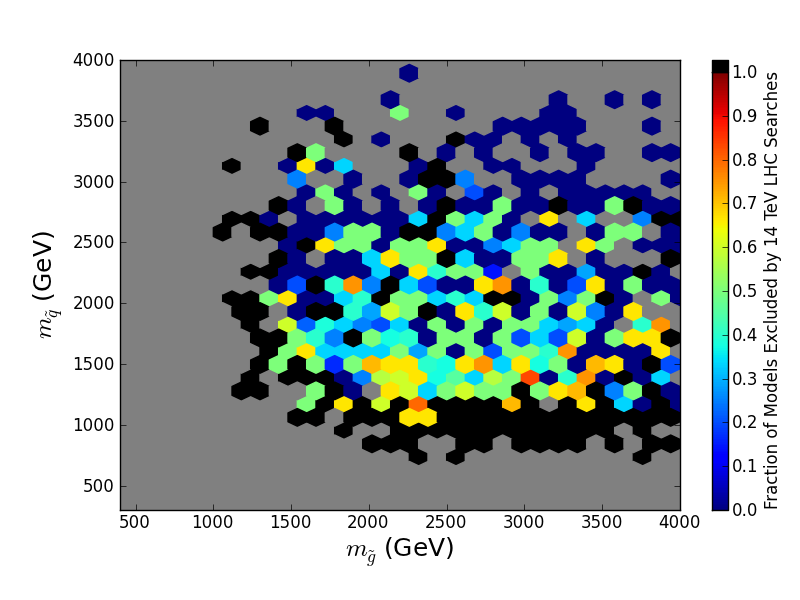}}
\vspace*{-0.10cm}
\caption{Same as the previous Figure but now for 3000 fb$^{-1}$.}
\label{figxx3g}
\end{figure}

As was the case with the 7/8 TeV searches, it is interesting to decompose these gravitino results at 14 TeV into subsets of models which are classified according to the NLSP properties; these decompositions are shown in Figs.~\ref{figxx2g} and ~\ref{figxx3g} for luminosities of 300 and 3000 fb$^{-1}$, respectively. Again, since we do not include any 14 TeV stable 
particle searches we expect the 14 TeV gravitino results and those obtained at 7/8 TeV to behave quite differently when this breakdown is performed as the single 14 TeV search we consider 
requires substantial MET from, \eg, gravitinos or invisible NLSPs. Thus the weak coverage in the lower right panel will improve tremendously with the incorporation of 14 TeV HSCP searches. As we will see, in some cases the ten-fold increase in luminosity is not very useful if only this single MET-based search is employed, since some models simply predict too little MET to be visible. 

In the upper left-hand panel we see the results of the jets plus MET search for the gravitino model subset wherein the NLSP has a displaced decay that yields observable decay products in 
addition to the gravitino. While these decays usually have enough MET to be visible, the events are often vetoed as a result of displaced jets failing quality requirements or displaced muons triggering the cosmic ray veto. In the upper right panel we see the case which is most similar to the neutralino LSP model set, where the NLSP is either invisible and detector-stable or produces invisible products when it does decay, the situation for which the jets plus MET search was designed. 
Unsurprisingly, here we see the strongest coverage of the parameter space which does appear qualitatively similar to the corresponding neutralino model results shown above at both 
integrated luminosities. In the subset of gravitino models where the the NLSP decays promptly with visible decay products (as shown in the lower left panel), the search is slightly more effective than when the decays are displaced, however it still lags behind the invisible NLSP case, in opposition to what was observed for the 7/8 TeV searches. The reason is that all SUSY events will eventually produce two NLSPs; if the production or decay of the NLSP produces leptons, they will cause the event to be rejected by the zero lepton Jets+MET analysis simulated here. However, adding searches for leptonic final states will easily discover or exclude these models with prompt NLSP decays. The last possibility, shown in the lower right panels, is the subset of models where the NLSP is both detector-stable and has electric or color charge. If the NLSP has electric charge, the only MET production will be through neutrinos. The case of colored NLSPs is more complicated, because they can hadronize to create neutral or charged R-hadrons. Neutral R-hadrons will typically deposit only a few GeV in the calorimeter~\cite{Hewett:2004nw}; for simplicity we assume that this energy deposition is negligible, so that large amounts of MET may result if one or both of the R-hadrons is charged. However, several hard jets from ISR or cascade decays are still required to pass the Jets+MET search requirements. We note that the 7/8 TeV results strongly suggest that 14 TeV searches for stable charged particles will exclude any model for which a significant amount of MET is produced by neutral R-hadrons. For the reasons described above, the exclusion efficiency for models with stable charged particles is quite low except for the region near the current exclusion limit, for which the 14 TeV cross-section will be very large.


\begin{figure}[htbp]
\centerline{\includegraphics[width=3.5in]{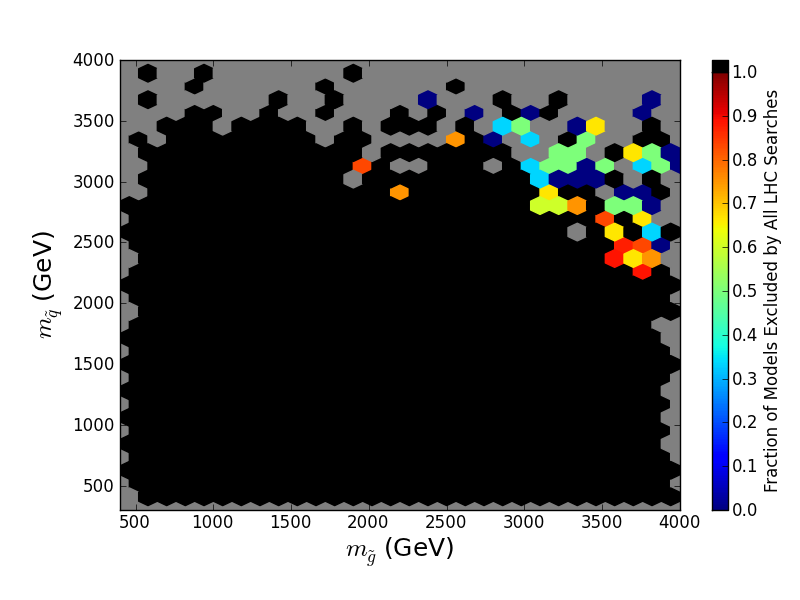}
\hspace{-0.50cm}
\includegraphics[width=3.5in]{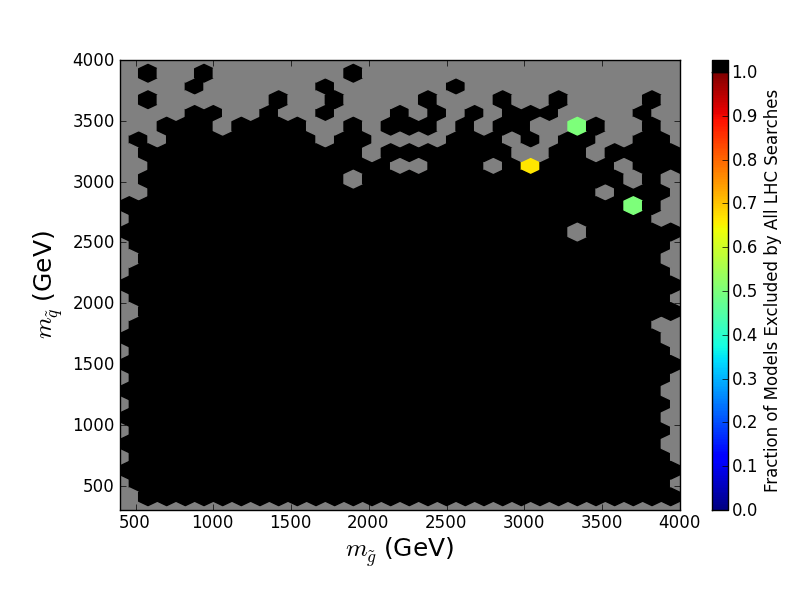}}
\vspace*{0.50cm}
\centerline{\includegraphics[width=3.5in]{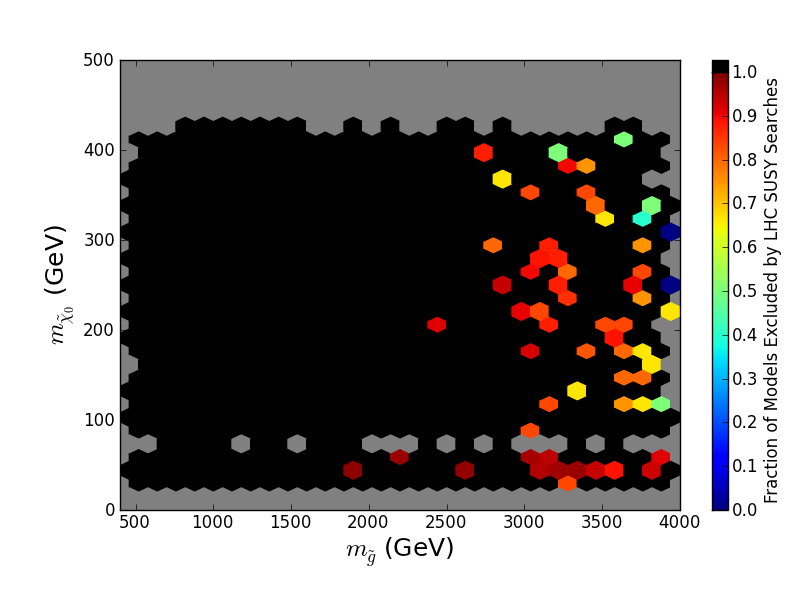}
\hspace{-0.50cm}
\includegraphics[width=3.5in]{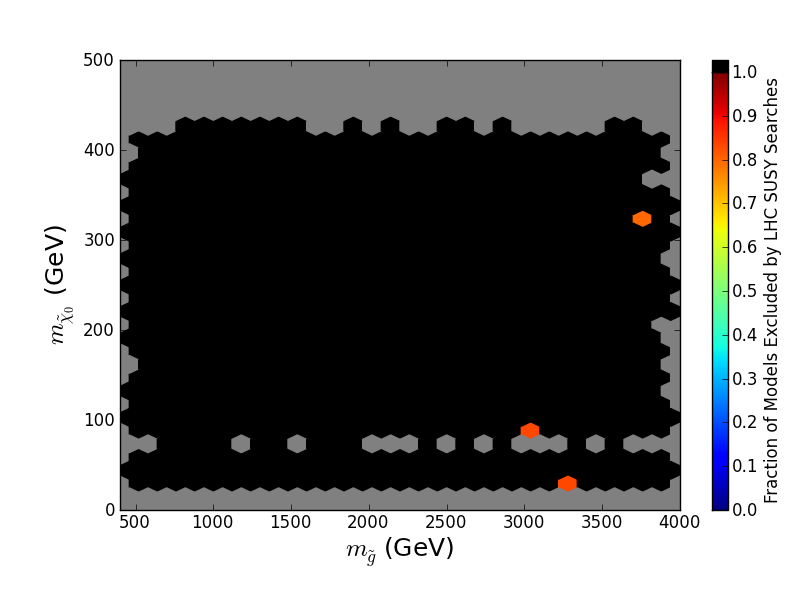}}
\vspace*{-0.10cm}
\caption{Results similar to those as shown in Fig.~\ref{figxx1} above but now for the low-FT model subset as described in the text.}
\label{figxxlowFT}
\end{figure}

Finally, we now briefly turn to the $\sim 3.1$k subset of low-FT models which survive both the 7 and 8 TeV analyses to see how they fare at 14 TeV; the results are shown 
in Fig.~\ref{figxxlowFT}.  Here, except for a relatively few pixels, the panels are found to be almost entirely black, indicating that essentially {\it all} of the remaining low-FT 
model set would be excluded by this single analysis. Indeed, only 74(3) of these models are found to remain after the zero-lepton jets + MET analysis is employed with a luminosity of 
300 (3000) fb$^{-1}$! (This corresponds to a fractional coverage for the remaining low-FT models of $97.6 (99.9)\%$ at these two integrated luminosities.) 
As can be seen from Fig.~\ref{figxxlowFT}, the few surviving models have very heavy squarks and gluinos. The allowed region is smaller than in the general neutralino model set for two reasons. First, the decay patterns that produce high-$p_T$ leptons in the neutralino model generally involve a bino-like intermediate state, which is incompatible with the necessity of having a bino-like LSP in the low-FT model set. Second, the low-FT spectra are required to be relatively uncompressed since fine-tuning places an upper limit of $\sim$ 400 GeV on the LSP mass, in contrast to cases in the general neutralino model set where the LSP can be heavier than a TeV. These effects combine to allow the nearly complete exclusion of the Low-FT model set at the 14 TeV LHC. Since the number of surviving models is so few we learn little additional information by considering the corresponding 
allowed regions in the various 1\textsuperscript{st}/2\textsuperscript{nd} generation squark-LSP results and so for brevity we do not display these results here.

The addition of a 14 TeV jets + MET channel including leptons or jets with $b$-tags would in all probability exclude these few remaining models but such analyses would 
need to be performed to verify these conclusions. In any case, it is clear from this single analysis that the 14 TeV LHC will be able to explore natural MSSM spectra in full detail.

\section{Summary}

The flexibility of the 19/20-parameter pMSSM provides a very powerful way to combine, compare and contrast the various searches for SUSY at the LHC (and elsewhere), 
even those which employ different collision energies.  Here we have examined how the pMSSM parameter space is probed by the suite of ATLAS SUSY searches by `replicating' 
the searches using fast Monte Carlo and then determining how these searches impact three distinct pMSSM model sets: two large models sets with either a ($i$) neutralino or ($ii$) 
gravitino LSP and ($iii$) 
a smaller specialized neutralino LSP set with low-FT and a thermal LSP saturating the relic density. We have shown above that the models in these sets 
generally respond quite differently to the various SUSY searches. However, in all cases, we see that the combination of results obtained from the many LHC searches can 
significantly augment the total coverage of the model space.  Furthermore, not knowing the exact form that the SUSY spectrum might take {\it a priori}, all of the 
searches can play important roles in constraining the pMSSM model parameters. For models in either the neutralino or low-FT sets, we also found that the zero-lepton, 
jets + MET search at the 14 TeV LHC is very likely to be able to exclude (or discover!) the bulk of models that have survived the 7 and 8 TeV searches and do not 
produce high-$p_T$ leptons in their cascades. Augmenting this single search with others at 14 TeV would be of significant interest and something we hope to address in the future.

\section{Acknowledgments}

The authors would like to thank Alan Barr and David C\^{o}t\'{e} for invaluable discussions.  We also thank Richard Dubois and Homer Neal for
computational support with the SLAC PPA batch farm system. This work was supported by the Department of Energy, Contract DE-AC02-76SF00515.

\end{document}